\documentclass[11pt,urlcolor=blue, linkcolor=blue]{article} 
\usepackage{cite}

\usepackage{amsmath, amsthm, amssymb,slashed}
\usepackage{ifpdf}
\ifpdf
  \usepackage[pdftex]{graphicx}
  \usepackage{epstopdf}
\else
  \usepackage[dvips]{graphicx}
\fi
\parskip=\baselineskip

\allowdisplaybreaks[1]

\usepackage{datetime}

\DeclareMathOperator{\lcm}{lcm}
\newcommand{\tI}{\text{I}}
\newcommand{\tII}{\text{II}}

\usepackage{comment}

\usepackage{upgreek}
\usepackage{bbm}

\newcommand{\GSD}{\text{GSD}}

\newenvironment{myfont}[2][]{\csname#2\endcsname[#1]}{}

\newcommand{\bea}{\begin{eqnarray}}
\newcommand{\eea}{\end{eqnarray}}
\def\be{\begin{equation}}
\def\ee{\end{equation}}

\def\RP{{\mathbb{RP}}}

\usepackage{color}
\usepackage[colorlinks,citecolor=blue]{hyperref}
\definecolor{red}{rgb}{1,0,0}
\definecolor{blue}{rgb}{0,0,1}
\definecolor{dblue}{rgb}{0,0,0.4}
\definecolor{green}{rgb}{0,1,0}
\definecolor{black}{rgb}{0,0,0}
\definecolor{white}{rgb}{1,1,1}

\definecolor{brn}{rgb}{.8,.4,.0}
\definecolor{redo}{rgb}{1,.5,.0}
\definecolor{ddgrn}{rgb}{0,0.4,0}
\definecolor{dgrn}{rgb}{0,0.55,0}
\definecolor{dbl}{rgb}{0,0,0.5}

\usepackage[bbgreekl]{mathbbol}
\usepackage{amscd}

\newcommand{\Z}{\mathbb{Z}}
\newcommand{\C}{\mathbb{C}}

\newcommand{\ii}{i}
\newcommand{\dd}{\hspace{1pt}\mathrm{d}}

\newcommand{\eqn}[1]{eqn.~\eqref{#1}}

\newcommand{\Tr}{{\rm Tr}}

\newcommand{\bpm}{\begin{pmatrix}}
\newcommand{\epm}{\end{pmatrix}}
\newcommand{\bmm}{\begin{matrix}}
\newcommand{\emm}{\end{matrix}}

\newcommand{\cC}{ {\cal C} }

\newcommand{\cH}{ {\cal H} }

\newcommand{\cT}{ {\cal T} }

\def\CC{{\cal C}}

\def\CH{{\cal H}}

\def\CM{{\cal M}}
\def\cN{{\cal N}}

\def\CO{{\cal O}}

\def\Z{{\mathbb{Z}}}

\def\C{{\mathbb{C}}}

\def\Tr{{\mathrm{Tr}}}

\newcommand{\tD}{\mathrm{D}}

\numberwithin{equation}{section}

\begin{document}
\begin{titlepage}
\begin{flushright}
\end{flushright}
\vskip .3in

\begin{center}

{\bf\LARGE{
Tunneling Topological Vacua via Extended Operators: \\[2.75mm]
(Spin-)TQFT Spectra and Boundary Deconfinement 
\\[4.1mm]
in Various Dimensions
}}

\vskip0.5cm 
\Large{Juven Wang$^{1,2}$,  Kantaro Ohmori$^{1}$, Pavel Putrov$^{1}$,  \\[2mm] 
Yunqin Zheng$^{3}$, Zheyan Wan$^4$, Meng Guo$^5$,  \\[2mm] 
Hai Lin$^{5,6,2,7}$, Peng Gao$^5$  and Shing-Tung Yau$^{5,2,6,7}$} 
\vskip.5cm
 {\small{\textit{$^1$School of Natural Sciences, Institute for Advanced Study, Princeton, NJ 08540, USA}\\}}
 \vskip.2cm
 {\small{\textit{$^2${Center of Mathematical Sciences and Applications, Harvard University,  Cambridge, MA, USA} \\}}
}
\vskip.2cm
 {\small{\textit{$^3${Department of Physics, Princeton University, Princeton, NJ 08540, USA}\\}}
}
 
 \vskip.2cm
{\small{\textit{$^4${School of Mathematical Sciences, USTC, Hefei 230026, China}\\}}
}

 \vskip.2cm
{\small{\textit{$^5${Department of Mathematics, Harvard University, Cambridge, MA 02138, USA}\\}}
}

\vskip.2cm
{\small{\textit{$^6$
{Yau Mathematical Sciences Center, Tsinghua University, Beijing, 100084, China}\\}}
}

 \vskip.2cm
{\small{\textit{$^7${Department of Physics, Harvard University,  Cambridge, MA 02138, USA} \\}}
}


\end{center}
\vskip.5cm
\baselineskip 16pt
\begin{abstract}

Distinct quantum vacua of topologically ordered states can be tunneled into each other via extended operators.
The possible applications include condensed matter and quantum cosmology.
We present a straightforward approach to calculate the 
partition function on various manifolds and
 ground state degeneracy (GSD), mainly based on continuum/cochain Topological Quantum Field Theories (TQFT), 
 in any dimension. 
This information can be related to the counting of extended operators of bosonic/fermionic TQFT.
On the lattice scale, anyonic particles/strings live at the ends of line/surface operators. 
Certain systems in different dimensions are related to each other through dimensional reduction schemes,
analogous to (de)categorification.
Examples include spin TQFTs derived from gauging the interacting fermionic symmetry protected topological states (with fermion parity $\mathbb{Z}_2^f$) of symmetry group $\mathbb{Z}_4\times \mathbb{Z}_2$ and $(\mathbb{Z}_4)^2$ in 3+1D, also $\mathbb{Z}_2$ and $(\mathbb{Z}_2)^2$ in 2+1D.
Gauging the last three cases begets non-Abelian spin TQFT (fermionic topological order).
We consider situations where a TQFT lives on (1) a closed spacetime
or (2) a spacetime with boundary,
such that the bulk and boundary are fully-gapped 
and short or long-range entangled (SRE/LRE).
Anyonic excitations can be \emph{deconfined} on the boundary.
We introduce new exotic topological interfaces on which \emph{neither particle nor string} excitations alone condensed, but only 
\emph{fuzzy-composite objects of extended operators}
can end (e.g.
a string-like composite object formed by a set of particles can end on a special 2+1D boundary of 3+1D bulk).
We explore 
the relations between group extension constructions and partially 
breaking constructions (e.g. 0-form/higher-form/``\emph{composite}'' breaking)
of topological boundaries, after gauging.
We comment on the implications of entanglement entropy for some of such LRE systems.

%
%

\end{abstract}
\end{titlepage}

\tableofcontents   


\section{Introduction and Summary}

Many-body quantum systems can possess entanglement structures 
--- the entanglement between either neighbor or long-distance quantum degrees of freedom, 
whose property has been pondered by many physicists since Einstein-Podolsky-Rosen's work \cite{PhysRev.47.777-EPR}.
Roughly speaking, there can be short-range or long-range entanglements (See a recent review \cite{1610.03911-Wen-RMP}).
Within the concept of the \emph{locality} in the space (or the spacetime) and the \emph{short-distance cutoff lattice regularization},
the short-range entangled (SRE) state can be deformed to a trivial product state (a trivial vacuum) through local unitary transformations on local sites
by series of local quantum circuits.
The long-range entanglement (LRE) is however much richer.
 
Long-range entangled states 
cannot be deformed to a trivial gapped 
vacuum through local unitary transformations on local sites
by series of local quantum circuits.
Some important signatures of long-range entanglements contain the \emph{subset} or the \emph{full-set} of the following:\\

\noindent
1. Fractionalized excitations and fractionalized quantum statistics: Anyonic particles in 2+1D (See 
\cite{Wilczek:1990ik-Book, freedman2003topological, preskill2004lecture, nayak2008non, 1612.09298PutrovWang} and References therein)
and anyonic strings in 3+1D
(See \cite{Wang1403.7437,Jiang:2014ksa, Wang1404.7854} and \cite{1612.09298PutrovWang}, References therein).\footnote{We denote the spacetime dimensions as
$d+1$D}

\noindent
2. Topological degeneracy: In $d+1$D spacetime dimensions,
the number of (approximate) degenerate ground states on a closed space $M^d$ or an open space $M^d$
with boundary $\Sigma^{d-1}$ (denoted as $\Sigma^{d-1}=\partial M^d$) can depend on the spatial topology.
This is the so-called topological ground state degeneracy (GSD) of zero energy modes.
Although in general for the quantum many-body system, 
both the \emph{gapless} and \emph{gapped} system can have
topological degeneracy, it is easier to extract that for the \emph{gapped} system.
The low energy sector of the \emph{gapped} system can be approximated by a topological quantum field theory(TQFT)\cite{Witten:1988hf}
 (See further discussion in \cite{1612.09298PutrovWang}),
and one can compute GSD from the partition function $Z$ of the TQFT as
\bea
Z(M^{d} \times S^1) = \dim \cH_{M^{d}} \equiv \GSD,
\label{eq:ZequaldimH}
\eea
where $S^1$ is a compact time circle.\footnote{One can also 
consider a generalization of this relation by
turning  on a background flat connection $A^{(G)}$ for a global
symmetry $G$. First, non-trivial holonomies along 1-cycles of $M^d$
will result in replacement $\CH_{M^d}$ by the corresponding twisted
Hilbert space $\CH_{M^d}^\text{tw}$. Second, a non-trivial holonomy
$g\in G$ along the time $S^1$ will result in insertion of $\rho(g)$
into the trace, where $\rho$ is the representation of $G$ on the
Hilbert space:
\begin{equation}
Z(M^d\times S^1;A^{(G)})=\Tr_{\CH_{M^d}^\text{tw}} \rho(g).
\end{equation}
In condensed matter, this is related to the symmetry twist inserted on $M^d$ to probe the 
Symmetry Protected/Enriched Topological states (SPTs/SETs)\cite{1405.7689,1610.03911-Wen-RMP}.
In this work, instead we mainly focus on \eqn{eq:ZequaldimH}.

  }


\noindent
3. Emergent gauge structure: Gauge theory (See \cite{{PhysRevLett.52.2111-Wilczek-Zee-1984},{2005-Levin-Wen-RMP}}, and References therein).

Such long-range entangled states are usually termed as intrinsic \emph{topological orders} \cite{Wen:1990tm}. 
The three particular signatures outlined above are actually closely related. 
For example, the first two signatures must require LRE {topological orders} (e.g. \cite{0506008-Oshikawa-Senthil}).
Other more detailed phenomena are recently reviewed in \cite{1610.03911-Wen-RMP}. 

In this work, we plan to systematically compute the path integral $Z$, namely
GSD$=Z(M^{d} \times S^1)$ for various TQFTs in diverse dimensions.
These GSD computations have merits and applications to distinguish the underlying 
LRE topological phases in condensed matter system, including
quantum Hall states \cite{Wen:1997ce} and quantum spin liquids \cite{2016arXiv160103742S}.
On the other hand,
these GSDs are quantized numbers
obtained by putting a TQFT on a spacetime manifold $M^{d}\times S^1$.
So they are also mathematically rigorous invariants for topological manifolds.
Normally, one defines GSD by putting a TQFT on a \emph{closed} spatial manifold without boundary.
However, recent developments in physics suggests that one can also define
GSD
by putting a TQFT on a \emph{open} spatial manifold with boundary (possibly with multiple components) \cite{1212.4863.WangWen, 1306.4254.Kapustin, 1408.0014HungWan, 1408.6514.LanWW}.
To distinguish the two,
the former, on a \emph{closed} spacetime, is named \emph{bulk topological degeneracy}, 
the latter, on an \emph{open} spacetime, is called \emph{boundary topological degeneracy} \cite{1212.4863.WangWen}.
For the case with boundary the GSD is evaluated as
\bea
Z(M^{d} \times S^1)_{\partial M^d = \Sigma^{d-1}}= \dim \cH_{M^{d}} |_{\partial M^d = \Sigma^{d-1}}\equiv \GSD.
\eea
As already emphasized in \cite{1212.4863.WangWen}, this boundary GSD
encodes both the bulk TQFT data as well as the 
gapped topological boundary conditions data\cite{Haldane, 1008.0654KapustinSaulina, 1104.5047KitaevKong, 1408.6514.LanWW}.
These gapped topological boundary conditions can be viewed also as:
 
\noindent  
$\bullet$ The $(d-1)$-dimensional defect lines/domain walls in the $d$-dimensional space, or \\
\noindent
$\bullet$  The $d$-dimensional defect surfaces in the $(d+1)$-dimensional spacetime.

These topological boundaries/domain walls/interfaces\footnote{
Here a \emph{boundary} generically means the interface between the nontrivial sector 
(TQFT and topological order) and the trivial vacuum (gapped insulator).
A domain wall means the interface between two nontrivial sectors (two different TQFTs).
We will use domain walls and interfaces interchangeably. Although we will only consider the boundaries, and not more general domain walls, 
since the domain walls are related to boundaries by the famous \emph{folding trick}.}
are co-dimension $1$ objects with respect to both the space (in the Hamiltonian picture) or spacetime (in path integral picture). 

We will especially implement the unifying boundary conditions of \emph{symmetry-extension} and \emph{symmetry-breaking}
(of gauge symmetries) 
developed recently by Ref.~\cite{1705.06728WWW}, and will compute GSDs on manifolds with boundaries. 
There in Ref.~\cite{1705.06728WWW}, the computation of path integral is mostly based on discrete cocycle/cochain data of group cohomology on the spacetime lattice, here we will approach from the continuum TQFT viewpoints.

Following the set-up in \cite{1612.09298PutrovWang},
the systems and QFTs of our concern are: (1) Unitary; (2) 
Emergent as the infrared (IR) low energy physics from 
fully-regularized quantum mechanical systems with a ultraviolet (UV) high-energy lattice cutoff 
(This set-up is suitable for condensed matter or quantum information/code);
(3) Anomaly-free for the full $d+1$D. But the $d$D boundary of our QFTs on the open manifold can be anomalous,
with gauge or gravitational 't Hooft anomalies (e.g. \cite{1303.1803-Wen}).

\subsection{Tunneling topological vacua, counting GSD and extended operators}

\label{Sec:tunnel-1}
Using these GSDs, one can characterize and count the discrete vacuum sectors
of QFTs and gauge theories.
In 2+1D or higher dimensions, the distinct vacuum sectors for topological order
are robustly separated against local perturbations. 
Distinct vacuum sectors cannot be tunneled into each other by \emph{local operator} probes. 
In other words, the correlators of local probes should be zero or exponentially decaying:
\bea
\langle \CO_1(x_1) \CO_2(x_2) \rangle|_{|x_1-x_2|\to \infty} = \langle \text{g.s.} | \CO_1(x_1) \CO_2(x_2) |\text{g.s.}\rangle|_{|x_1-x_2|\to \infty} \simeq 0.
\eea
Here
$|\text{g.s.} \rangle$ means one of the ground states, and sometimes denoted as $|\text{g.s.} \rangle=| 0\rangle$.

\begin{figure}[!h]
\centering
\includegraphics[scale=.95]{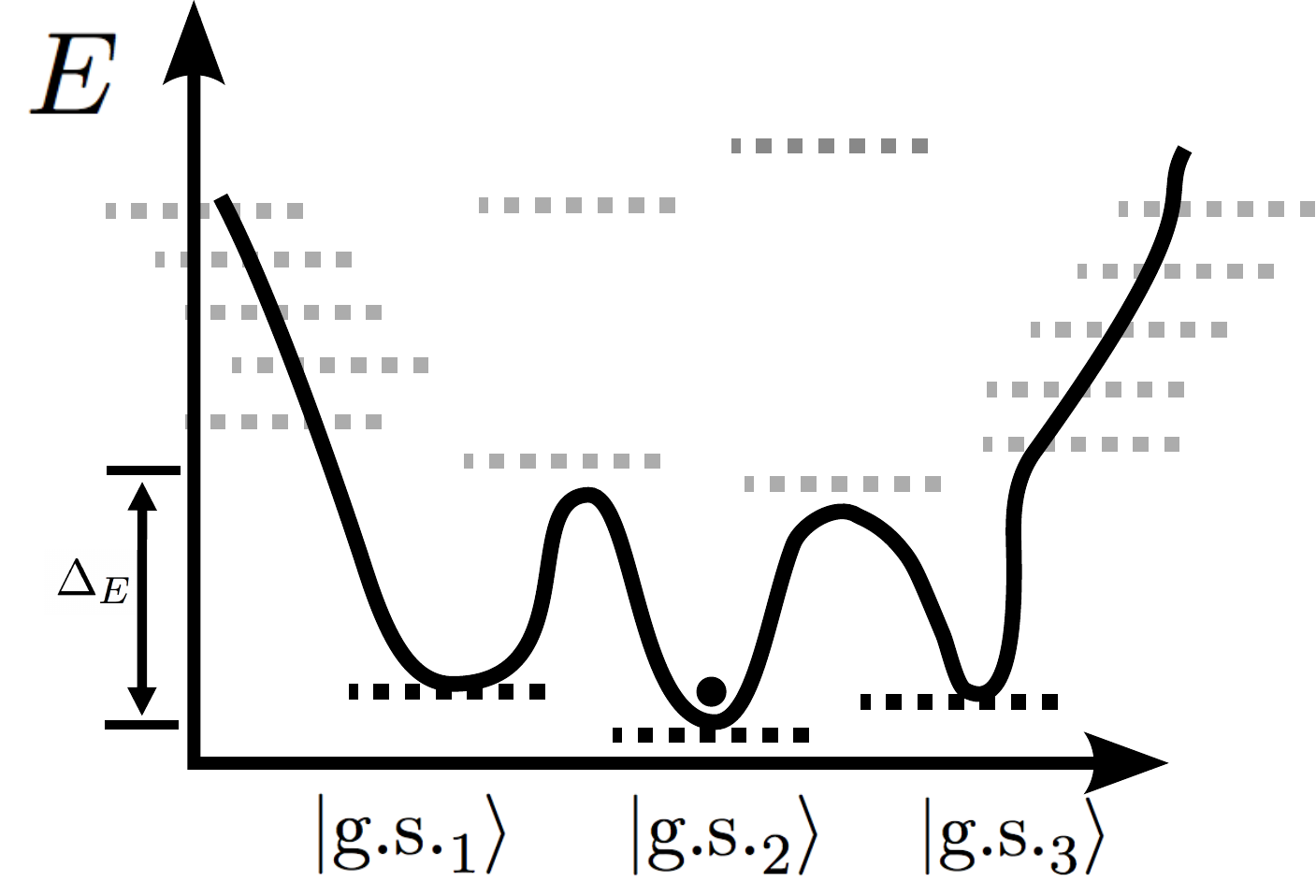}
\caption{We show the \emph{quantum energy spectrum} as several discrete energy levels in terms of horizontal dashed lines (- - -).
The approximate \emph{semi-classical energy potential} are drawn in terms of the continuous solid black curve.
The vertical axis shows the energy value $E$.
The horizontal axis illustrates their different quantum numbers, which can be, for example, (1) different
eigenvectors spanning different subspaces in the Hilbert space;
or (2) different spin/angular/spacetime momenta, etc.
This figure shows 3 topological degenerate ground states $|\text{g.s.}_1\rangle$, $|\text{g.s.}_2\rangle$ and $|\text{g.s.}_3\rangle$ with the
dark gray horizontal dashed lines (- - -) for their energy levels
--- Their energy levels only need to be \emph{approximately} the same (within the order of $e^{-\# V}$ where $V$ is the system size), but they remain \emph{topologically robust}.
Namely, only via the insertion of the extended operator shown in \eqn{eq:tunnel-GSD-1} winding around a non-contractible cycle can
 the $|\text{g.s.}_2\rangle$ tunnel to the other sectors, even though their energy levels are nearly the same.
The energy barrier is proportional to the cost of creating two anyonic excitations 
at the end of extended operators $W$ in \eqn{eq:tunnel-GSD-1}.
This energy barrier $\Delta_E$ naively seems to be infinite in TQFT, 
but it is actually of a finite order $\Delta_E \simeq 4J$ where $J$ is the lattice coupling constant in the UV complete lattice (e.g. in Kitaev's toric code \cite{Kitaev2003}
or more general twisted quantum double models \cite{Wan1211.3695, Wan:2014woa}).
{In reality, as an example in 2+1D, the 3 topological degenerate ground states
on a $T^2_{\text{space}} \times S^1_{\text{time}}$ can be induced from the filling fraction $\nu=\frac{1}{3}$-Laughlin fractional Quantum Hall states from electrons, or a $U(1)_3$-Chern-Simons theory at the deep IR.
Further illustration is shown in Fig.~\ref{fig:Tunnel}}
}
\label{fig:vacua}
\end{figure}

However, distinct vacuum sectors can be 
unitarily deformed into each other only through \emph{extended operators} $W$ (line and surface operators, etc.) winding nontrivial cycles (1-cycle, 2-cycle, etc.) along compact directions of space.
In the case that {extended operator} $W$ is a line operator, the insertion of $W$ can be understood as the process of 
creation and annihilation of a pair of anyonic excitations.
Namely, a certain well-designed extended operator $W$ can indeed \emph{connect} two different ground states/vacua,
$|\text{g.s.}_\alpha \rangle$ and $|\text{g.s.}_\beta \rangle$, inducing nontrivial correlators: 
\bea \label{eq:tunnel-GSD-1}
 \langle \text{g.s.}_\alpha | W(\gamma) |\text{g.s.}_\beta\rangle \to \text{finite} \neq 0.
\eea
Again $|\text{g.s.}_\alpha \rangle$ means the ground state $\alpha$ among the total GSD sector, and $\gamma$ is a nontrivial cycle in the space.
 Therefore, computing GSD also serves us as important data
 for  \emph{counting extended operators}, thus counting distinct types of
 \emph{anyonic particles or anyonic strings, etc.}

Different degenerate ground states can also be regarded as different approximate vacuum sectors in particle physics or in cosmology,
see Fig.~\ref{fig:vacua} for further explanations and analogies.
Therefore, in summary, our results might be of general interests to the condensed matter, mathematical physics, high-energy particle theory and quantum gravity/cosmology community. 

\subsection{The plan of the article and a short summary}
\label{sec:1.2}

First, in Sec.\ref{Sec:Categorification}, 
we describe how formal mathematical idea of decategorification can be helpful to organize the topological data.
In down-to-earth terms, we can decompose GSD data read from $d+1$D into a direct sum of several sub-dimensional GSD sectors in $d$D,
by compactifying one of the spatial dimensions on a small circle.

Then in Sec.~\ref{Sec:GSD-bTQFT} and Sec.~\ref{Sec:GSD-bTQFT-higher},
we start from more familiar discrete gauge theories. For example, 
the {$\Z_N$ gauge theory\cite{Wegner:1971jf, 1978FradkinSusskindPRD}}.
More generally, we can consider the twisted discrete gauge theories, known as
Dijkgraaf-Witten (DW) gauge theories\cite{Wittencohomology}.
These are bosonic TQFTs that can be (1) realized at the UV lattice cutoff through purely bosonic degrees of freedom,
and (2) defined on both non-spin and spin manifold.
We will study the GSD for these bosonic TQFTs.

There has been a lot of recent progress on understanding bosonic Dijkgraaf-Witten (DW) gauge theories in terms of continuum TQFTs.
However, to our best knowledge, so far there are no explicit calculations of GSD 
from the continuum field theories for the proposed \emph{non-Abelian} DW 
gauge theories.\footnote{By \emph{non-Abelian} DW 
gauge theories, we do not mean the gauge group is non-Abelian. 
Some \emph{non-Abelian} DW theories
can be obtained from certain Abelian gauge group with additional cocycle twists.}
\footnote{In an unpublished article \cite{1507.Wang} in 2015, some of the current authors had computed these {non-Abelian} GSD.
Part of the current work is based on the extension of that previously unpublished work.  We wish to thank Edward Witten for firstly suggesting this
continuum 
QFT method for computing {non-Abelian} GSD in June 2015.
In contrast, for computation of GSD for Abelian TQFTs, it has been done in
\cite{Gaiotto:2014kfa} and other related work.}
Our work will fill in this gap for better analytical understanding, 
by computing non-Abelian GSD using continuum TQFTs,
\footnote{
The simplest continuum bosonic TQFTs of discrete gauge theories, 
have the following form $\int \frac{ N_I}{2\pi}{B^I  d A^I} + 
{ \frac{N_1 N_2 \dots N_n\;
p_{}}{{(2 \pi)^{n-1} } N_{123 \dots n}}} A^1  A^2   \dots A^n $.
See details in later sections, we will show their GSD computations
in Sec.~\ref{Sec:GSD-bTQFT} for 1+1D to 3+1D, and Sec.~\ref{Sec:GSD-bTQFT-higher} for any dimension.
For all the $N_I=1$, the GSD$=1$ is computed earlier in \cite{Gu:2015lfa}. And there is only a trivial ground state, thus suitable for describing Symmetry-Protected Topological states (SPTs) without
intrinsic topological order.
}
that matches precisely to the predictions of GSD computed from the original Dijkgraaf-Witten group cohomology data: Discrete cocycle path integrals.
We present these results in Sec.~\ref{Sec:GSD-bTQFT} and Sec.~\ref{Sec:GSD-bTQFT-higher}.

By non-Abelian topological orders, 
we mean that some of the following properties are matched:
\begin{itemize}  
\item  The GSD $= Z(S^{d} \times S^1; \sigma_1, \sigma_2, \sigma_3, \dots)= \dim \cH_{S^{d}; \sigma_1, \sigma_2, \sigma_3, \dots} $
computed on a sphere $S^{d}$ with operator insertions (or the insertions of anyonic particle/string excitations on $S^{d}$)
have the following behavior:
(1) $Z(S^{d} \times S^1; \sigma_1, \sigma_2, \sigma_3, \dots)= \dim \cH_{S^{d}; \sigma_1, \sigma_2, \sigma_3, \dots} $
will grow exponentially as $k^n$ for a certain set of large $n$ number of insertions, for some number $k$.
The anyonic particle causes this behavior is called non-Abelian anyon or non-Abelian particle.
The anyonic string causes this behavior can be called non-Abelian string \cite{Wang1404.7854, CWangMLevin1412.1781}.
(2)  $ Z(S^{d} \times S^1; \sigma_1, \sigma_2, \sigma_3)= \dim \cH_{S^{d}; \sigma_1, \sigma_2, \sigma_3}>1$
for a certain set of three insertions.
\item  The Lie algebra of underling Chern-Simons theory is non-Abelian, if such a Chern-Simons theory exists.
\item  The GSD
$Z(T^{d} \times S^1)= \dim \cH_{T^{d}}$ for a discrete gauge theory of a gauge group $G$ on $T^{d}$ spatial torus, behaves as GSD$<|G|^d$, i.e. reduced to a smaller number than Abelian GSD. 
This criterion however works only for 1-form gauge theory.\footnote{We will see that examples like higher-form gauge theories, e.g. $\int BdA+BB$, have
GSD reduced compared to $|G|^d$, but they are still Abelian in a sense that they are free theories (have quadratic action).
In additional, its GSD $= Z(S^{3} \times S^1; \sigma_1, \sigma_2, \sigma_3, \dots)= \dim \cH_{S^{d}; \sigma_1, \sigma_2, \sigma_3, \dots} =1$ which means 
an Abelian topological order.} 
\end{itemize}  
Demonstrating that $Z(M^d\times S^1)$ effectively counts the dimensions of Hilbert space on $M^d$,
provides a more convincing quantum mechanical understanding of continuum/cochain TQFTs.
By computing the following data, independently without using particular triangulations of spacetime, 
\begin{enumerate}
\item
 GSD data, counting dimensions of Hilbert space,
\item Various braiding statistics and link invariants derived in \cite{1612.09298PutrovWang}, 
\end{enumerate}
for Abelian or non-Abelian cases, we
solidify and justify their continuum/cochain field theory descriptions of both Abelian or non-Abelian Dijkgraaf-Witten theories, 
as we show that the data are matched with the calculations based on triangulations \cite{Wang1404.7854,1612.01418Wen}.
Our present results combined together with Ref.~\cite{1612.09298PutrovWang}  positively support the previous attempts based on continuum TQFTs 
\cite{deWildPropitius:1995cf, WangSantosWen1403.5256,  Kapustin1404.3230, 1405.7689, {Gu:2015lfa}, {Ye1508.05689}, RyuChenTiwari1509.04266,
1602.05951, RyuTiwariChen1603.08429, He1608.05393, NingLiuPengYe1609.00985, Ye1610.08645, 1612.01418Wen, Ye1703.01926, 1710.04730-Tiwari-Ryu, 1801.01638NingYe}. 
Various data derived from continuum TQFTs 
can be checked and compared through the discrete cocycle and lattice formulations 
\cite{Wan1211.3695, 1212.0835Ran, 1212.1827Hung,
Wang1403.7437, Jiang:2014ksa, Wang1404.7854, Wan:2014woa,
CWangMLevin1412.1781, Bridgeman:2017etx, 1706.09769-Tantivasadakarn, 1710.01747YZhengBernevig, 1710.11168.Xueda}.

In Sec.~\ref{Sec:fTQFT-GSD}, we study fermion TQFTs (the so-called spin-TQFTs) and their GSD.
These fermion spin TQFTs are much subtler. They are obtained from dynamically gauging the global symmetry of
fermionic SPTs \cite{1612.09298PutrovWang}.
Although the original fermionic SPTs and the gauged fermionic spin TQFTs have the UV completion on the lattice,
the effective IR field theory may not necessarily guarantee good \emph{local} action descriptions.
These somehow \emph{non-local} topological invariants include, for example,
Arf-Brown-Kervaire (ABK) and $\eta$ invariants, intrinsic to the fermionic nature of systems.
Nevertheless, there are still well-defined partition functions/path integrals and we can compute explicit physical observables.
 Our examples include intrinsically interacting 3+1D and 2+1D fermionic SPTs (fSPTs) as short-range entangled (SRE) states, and their dynamically gauged spin-TQFTs as long-range entangled (LRE) states.
Recently,
Ref.~\cite{1701.08264Kapustin,1703.10937WangGu,1705.08911-Tantivasadakarn, 2018Fidkowski}
also explore the related interacting 3+1D fSPTs protected by the symmetry of finite groups. In Sec.~\ref{Sec:fTQFT-GSD},  we will briefly comment the relations between
our work and Ref.~\cite{1701.08264Kapustin,1703.10937WangGu,1705.08911-Tantivasadakarn, 2018Fidkowski}.

In Sec.~\ref{Sec:Dim-Reduce},
we explore {dimensional reduction scheme of partition functions}.
This section is based on the 
abstract and general thinking in Sec.~\ref{Sec:Categorification}
on (de)categorification. We
implement it on explicit examples, in
Sec.~\ref{Sec:GSD-bTQFT} and \ref{Sec:GSD-bTQFT-higher} on bosonic TQFTs
and in Sec.~\ref{Sec:fTQFT-GSD} on fermionic TQFTs.

In Sec.~\ref{Sec:LRE-bulk-brdy},
we mainly consider 
the long-range entangled (LRE) topologically-ordered bulk and boundary systems, denoted as
LRE/LRE bulk/boundary for brevity.
The LRE/LRE bulk/boundary systems can be 
obtained from dynamically gauging 
the bulk and 
unifying boundary conditions of \emph{symmetry-extension} and \emph{symmetry-breaking}
introduced in 
Ref.~\cite{1705.06728WWW}.\footnote{  
For LRE/LRE bulk/boundary topologically ordered system,
    \emph{symmetry-breaking/extension} really means
    the gauge symmetry-breaking/extension. The symmetry usage here is slightly abused to include the gauge symmetry.}
In contrast, we will also compare 
the systems of
LRE/LRE bulk/boundary to those of
SRE/SRE bulk/boundary 
and
SRE/LRE bulk/boundary.

In Sec.~\ref{Sec:conclude}, we conclude with various remarks 
on long-range entanglements and entanglement entropy,
and implications for the studied systems in various dimensions.

\subsection{Topological Boundary Conditions: 
Old Anyonic Condensation v.s. New Condensation of Composite Extended Operators}

Sec.~\ref{Sec:LRE-bulk-brdy} offers a mysterious and exotic new topological boundary mechanism, 
worthwhile enough for us to summarize its message in Introduction first.
An important feature of LRE/LRE bulk/boundary is that both the bulk and boundary can have deconfined anyonic excitations.
The anyonic excitations
are 0D particles, 1D strings, etc., which can be regarded as the energetic excitations
at the ends of extended operators supported on 1D lines, 2D surfaces, etc.

In contrast to the past conventional wisdom which suggests that the LRE topological gapped boundary 
is defined through the \emph{condensation} of certain \emph{anyonic excitations},
we emphasize that there are some additional subtleties and modifications needed.
The previously established folklore that suggests the topological gapped boundary conditions are given
by \emph{anyon condensation} (\cite{BM07, BS09, BSH09, 1104.5047KitaevKong, 1307.8244Kong, 1308.4673HungWan, 1408.0014HungWan}, also References therein 
a recent review \cite{1706.04940.Burnell}),  
\emph{Lagrangian subgroups} or their generalization 
\cite{1008.0654KapustinSaulina, 1212.4863.WangWen, Levin:2013gaa, BJQ13, BJQ13a, 1408.6514.LanWW}.
For example, in 1+1D boundary of 2+1D bulk (say $\Sigma^2 = \partial M^3$ is the boundary), 
the condensation of anyons suggest their line operators can end on the boundary $\Sigma^2$. 
Formally, we have boundary conditions of the following type:
\bea
\left. \sum_i  q_i A_{i} \right\vert_{\Sigma^2}=0,
\eea
or similar, that is certain linear combinations of line operators (with coefficients $q_i$) can end on ${\Sigma^2}$. Here and below $A_i$ denote 1-form gauge fields.

For example, if we consider a $\Z_N$ gauge theory of action 
$\int {\frac{N}{2\pi}{B  \wedge d A}}$
(i.e. $\Z_N$ toric code /topological order) on any $d+1$D $M^{d+1}$,
we can determine \emph{two} types of conventional topological gapped boundary conditions on $\Sigma^d = \partial M^{d+1}$:\footnote{In 2+1D, 
given the $\Z_N$-gauge bulk theory as
$\int {\frac{N}{2\pi}{B  \wedge d A}}$,
 we can gap the boundary by a cosine term of vortex field $ \phi_1$ of $A$, via
$$g_1 \int  dt dx \cos(N  \phi_1)$$ at the strong $g_1$ coupling, which corresponds to the $A=0$ boundary condition\cite{1212.4863.WangWen}.
We can also gap the boundary by another cosine term of vortex field $ \phi_2$ of $B$, via
$$g_2 \int  dt dx \cos(N  \phi_2)$$ at the strong $g_2$ coupling, which corresponds to the $B=0$ boundary condition\cite{1212.4863.WangWen}.
These two boundaries correspond to the rough $e$ and the smooth $m$ boundaries in the lattice Hamiltonian formulation of Bravyi-Kitaev's \cite{9811052BravyiKitaev}.
} 

\noindent
1.  By condensations of 
$\Z_N$ charge (i.e. the electric $e$ particle attached to the ends of $\Z_N$ Wilson worldline $\int A$ of 1-form gauge field), 
set by:
\bea \label{eq:e-bdry}
 \left. A_{} \right\vert_{\Sigma^{d}}=0, \;\;\;\;\;\; \text{  as $\Z_N$ charge $e$ condensed on ${\Sigma^{d}}$}.
\eea

\noindent
2.  By condensations of 
$\Z_N$ flux (i.e. the magnetic $m$ flux attached to the ends of  $\Z_N$ 't Hooft worldvolume $\int B$ of $(d-1)$-form gauge field), with the $m$-condensed 
boundary set by:
\bea \label{eq:m-bdry}
 \left. B_{} \right\vert_{\Sigma^{d}}=0, \;\;\;\;\;\; \text{  as $\Z_N$ flux $m$ condensed on ${\Sigma^{d}}$}.
\eea
The UV lattice realization of above two boundary conditions are constructed in the Kitaev's toric code \cite{9811052BravyiKitaev} as well as Levin-Wen string-net \cite{1104.5047KitaevKong}.
The two boundary conditions in \eqn{eq:e-bdry} and \eqn{eq:m-bdry} are incompatible. Namely, each given physical boundary segment can 
choose either one of them, either $e$ or $m$ condensed but not the other.

However, in Sec.~\ref{Sec:LRE-bulk-brdy}, we find that the usual anyon condensations like $\left. \sum_i  q_i A_{i} \right\vert_{\Sigma^d}=0$ 
(including \eqn{eq:e-bdry} and \eqn{eq:m-bdry}) are
\emph{not} sufficient. We find that there are certain \emph{exotic, unfamiliar, new} topological boundary conditions on 2+1D boundary of 3+1D bulk, such that neither $\left. A_{i} \right\vert_{\Sigma^3}=0$ nor $\left. A_{j} \right\vert_{\Sigma^3}=0$,
but \emph{only} the composite of extended operators can end on the boundary,
\bea \label{eq:composite-bdry}
\left. {A_{i} \cup  A_{j}} \right\vert_{\Sigma^3}=0.
\eea
Here $\cup$ is a cup product.
Heuristically, we interpret these types of topological boundary conditions as the condensation of
\emph{composite objects of extended operators}.
Here on a 2+1D boundary of a certain 3+1D bulk,
we have  a string-like composite object formed by a set of particles.
The 1D string-like composite object is at the ends of 2D worldsheet $A_{i} \cup  A_{j}$.
The set of 0D particles we refer to are the ends of 1D worldlines $A_{i}$ and $A_{j}$.
The boundary condition
$\left. {A_{i} \cup  A_{j}} \right\vert_{\Sigma^2}=0$
is achieved  \emph{neither by intrinsic 0D particle nor by intrinsic 1D string} excitation condensation alone.
We suggest, this exotic topological deconfined boundary condition 
may be interpreted as condensing certain \emph{composite} 1D string formed by 0D particles.

In summary, in Sec.~\ref{Sec:LRE-bulk-brdy}, we find that \emph{gauge symmetry-breaking} boundary conditions
are indeed related to the usual \emph{anyon condensation} of particles/strings/etc. The \emph{gauge symmetry-extension} of LRE/LRE bulk/boundary in Ref.~\cite{1705.06728WWW} 
sometimes can be reduced to  the usual \emph{anyon condensation} story
(e.g. for 2+1D bulk),
while other times, instead of the {condensations} of a set of anyonic excitations,
one has to consider {condensations} of certain composite objects of extended operators (e.g. for certain 3+1D bulk).

\subsection{Tunneling topological vacua in LRE/LRE bulk/boundary/interface systems}

\label{Sec:tunnel-2}

\begin{figure}[!h]
\centering
(a) \includegraphics[scale=1.15]{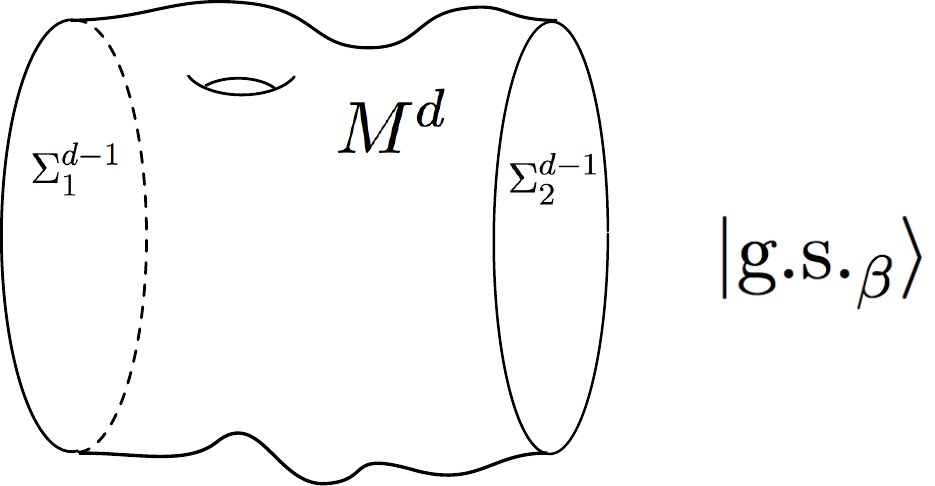}\\
 (b) \includegraphics[scale=1.15]{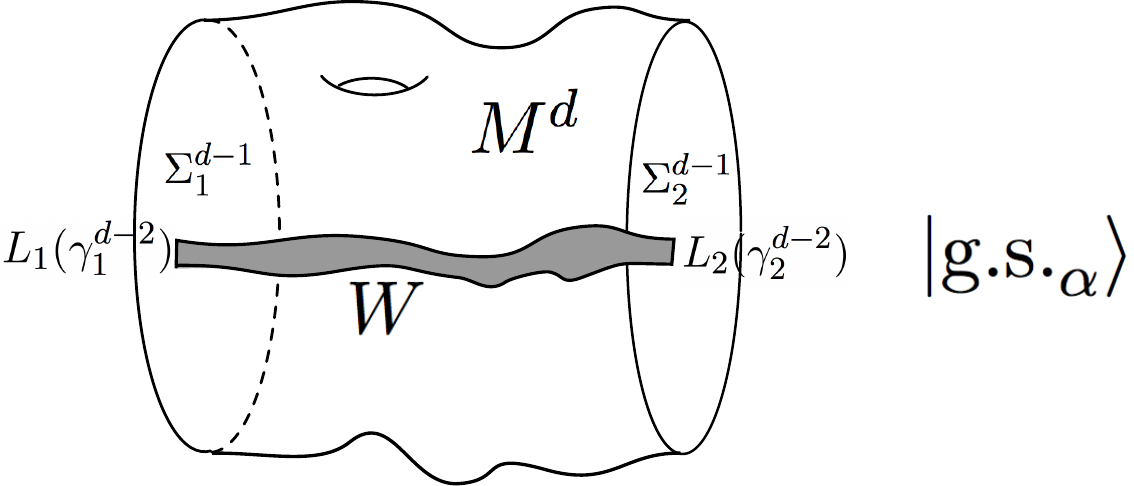}
\caption{Illustration of 
tunneling between topological vacua, from $|\text{g.s.}_\beta\rangle$ to $|\text{g.s.}_\alpha\rangle$,
 via an extended operator $W$.
In fig.(a), we see the topological vacuum in an original ground state $|\text{g.s.}_\beta\rangle$, where the spatial manifold $M^d$ is shown.
On top of $M^d$, there are
LRE/LRE bulk/boundary with topologically orders (TQFTs).
In fig.(b), after inserting certain extended operator $W$ connecting two boundary components ($\Sigma_1^{d-1}$ and $\Sigma_2^{d-1}$), usually by an adiabatic process,
we switch or \emph{tunnel} to another   
topological vacuum $|\text{g.s.}_\alpha\rangle$.
In the case of a closed manifold, the extended $W$ goes along a non-contractible cycle (representing a nontrivial element of the homology group of $M^d$).
}
\label{fig:Tunnel}
\end{figure}

We offer one last remark before moving on to the main text in Sec.~\ref{Sec:Categorification}.
Similarly to \eqn{eq:tunnel-GSD-1}, we can also interpret switching the topological sectors of gapped boundary/interface systems of Sec.~\ref{Sec:LRE-bulk-brdy},
in terms of {tunneling topological vacua}
by using extended operators $W$.
The equation \eqn{eq:tunnel-GSD-1} still holds when when the operator $W$ has a support with two boundary components $\gamma_1^{d-2}$ and $\gamma_2^{d-2}$ that, in turn, support $(d-2)$D operators $L_1/L_2$ and lie in two different boundary components/interfaces  $\Sigma_1^{d-1}$ and $\Sigma_2^{d-1}$ of the spatial manifold\footnote{More generally, one can consider a configuration where the support of $W$ has a boundary $\gamma^{d-2}$ (possibly with multiple connected components) that coincides with a non-trivial cycle in the boundary $\Sigma^{d-1}$ (also possibly with multiple components) of the spatial manifold $M^d$. Each connected component of $\gamma^{d-2}$ supports a certain $(d-1)$-dimensional operator.}:
\bea \label{eq:tunnel-GSD-2}
 \left.
 \langle \text{g.s.}_\alpha | L_1(\gamma_1) W(\gamma_1, \gamma_2) L_2(\gamma_2) |\text{g.s.}_\beta\rangle  
 \right\vert_{\gamma_j^{d-2} \subset \Sigma_j^{d-1} }
 \to \text{finite} \neq 0.
\eea
Here the open spacetime manifold $\partial M^d$ has two or more  boundary components 
$$\partial M^d=\Sigma_1^{d-1} \sqcup \Sigma_2^{d-1} \sqcup \dots.$$
As usual, $\sqcup$ means disjoint union.
Physically, by moving certain (anyonic) excitations of either the usual extended operators or the \emph{composite} extended operators,
from one boundary component $\Sigma_1^{d-1}$ to another boundary component $\Sigma_2^{d-1}$,
we have switched the ground state between $ | \text{g.s.}_\alpha \rangle$ and $|\text{g.s.}_\beta\rangle$, as
\eqn{eq:tunnel-GSD-2} suggested, shown in Fig.~\ref{fig:Tunnel}.

This idea is deeply related to Laughlin's thought experiment in condensed matter \cite{Laughlin1981PhysRevB.23.5632}:
Adiabatically dragging fractionalized quasiparticle between two edges of the annulus via threading a background magnetic flux through the
 hole of annulus --- this would change the ground state sector.
This also lays the foundation of Kitaev's fault-tolerant quantum computation in 2+1D by anyons \cite{Kitaev2003}.
Various applications can be found in \cite{Santos:2013uda, 1212.4863.WangWen, 1403.6902CTHsiehRyu, 1408.0014HungWan}
and references therein.
In our work, we generalize the idea to \emph{any} dimensions. 
This idea in some sense also helps us to the counting of GSD and extended operators for LRE/LRE bulk/boundary systems.



\section{Strategy: (De)Categorification, Dimensional Decomposition and Intuitive Physical Picture} 

\label{Sec:Categorification}

\begin{figure}[!h]
\centering
 \includegraphics[scale=.5]{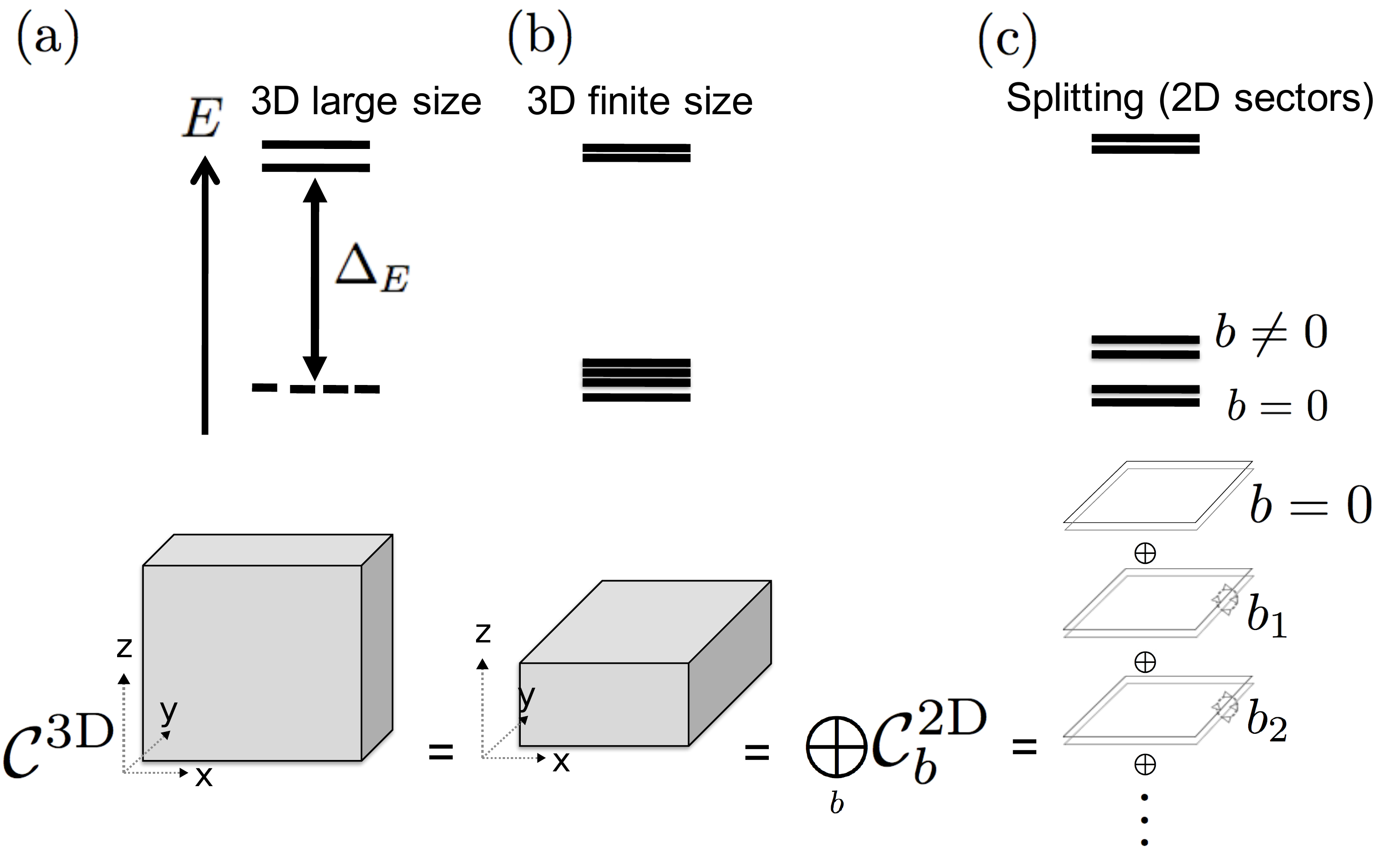}
\caption{Relating
the dimensional reduction and 
(de)categorification
to
measurable physical quantum phenomena
in the laboratory.
The top part of the subfigure (a) shows the bulk energy spectrum $E$ with energy gap $\Delta_E$,
in the large 3+1D size limit. The bottom part shows in grey color
a 3D spatial sample on $T^3$ torus with large compact circles in all $x,y,z$ directions. The degenerate zero modes in the
energy spectrum are due to the non-trivial topological order (described by a TQFT) of the quantum system.
The subfigure (b) shows that the energy spectrum slightly splits due to finite size effect, 
but its approximate GSD is still topologically robust.
The subfigure (c) shows the system on $T^3$ torus in the limit of small circle in the $z$ direction. The energy spectrum
forms several sectors, that can be labeled by a quantum number $b$ associated to the holonomy $\oint_{S^1_z} A$ of gauge field $A$
along $z$ (or a background flux through the compact circle) as $b \sim \oint A$.
See more detailed explanation in the main text.}
\label{fig:dim-reduce}
\end{figure}

In this section, we address physical ideas of dimensional reduction/extension of partition functions and topological vacua (GSD),
and their relations to formal mathematical ideas of decategorification/categorification.
These ideas are actually relevant to  physical phenomena measurable
in a laboratory, see Fig.~\ref{fig:dim-reduce}.
In condensed matter, the related idea of dimensional reduction
was first studied in \cite{Moradi:2014cfa} and \cite{Wang1404.7854} for 3+1D bulk theories. Here we apply the idea to an arbitrary dimension.
Later, gathering the concrete calculations in
Sec.~\ref{Sec:GSD-bTQFT}, \ref{Sec:GSD-bTQFT-higher} and \ref{Sec:fTQFT-GSD},
we will implement the strategy outlined here on those examples in Sec.~\ref{Sec:Dim-Reduce}.

Fig.~\ref{fig:dim-reduce} shows how the energy spectrum of a topologically ordered sample (shown as a cuboid in grey color) effectively described by an underlying TQFT
gets affected by the system size and by the holonomies of gauge fields through the compact circles.
The topologically ordered cuboid is displayed in the real space.
The energy eigenstates live in the Hilbert space.  
The energy spectrum can be solved from a quantum mechanical Hamiltonian system.

Fig.~\ref{fig:dim-reduce} (a) shows the system at a large or infinite size limit in the real space (in the case when the spatial manifold is $M^{d}=T^d$, $d=3$ with every $S^1$ circle size $\to \infty$), when
the topological degeneracy of zero energy modes becomes almost exact. The zero modes are separated 
from higher excitations by a finite energy gap $\Delta_E$.

Fig.~\ref{fig:dim-reduce} (b) shows the system at finite size in real space. The GSD becomes 
approximate but still topologically robust.

Fig.~\ref{fig:dim-reduce} (c) shows that, when $M^{d}=M^{d-1}\times S^1_z$ (in the case $M^d=T^3$) and  the compact $z$ direction's $S^1_z$ circle becomes small,
the approximate zero energy modes form several sectors, labeled
by quantum number $b$ associated to the holonomy $b \sim \oint_{S^1_z} A$ of a gauge field $A$
along $z$ (or a background flux threading via the compact circle).
In $d+1$-dimensions, this means that
\bea \label{eq:GSD-decompose}
\GSD_{T^d,\; d+1\text{D-TQFT}}= \sum_{b} \GSD_{T^{d-1},\; (d-1)+1\text{D-TQFT}'(b)}.
\eea
The energy levels within each sector of $\GSD_{T^d, (d-1)+1\text{D-TQFT}'(b)}$ are approximately grouped together.
However, energy levels of different sectors, labeled by different $b$, can be shifted upward/downward differently due to tuning the quantum number
$b \sim \oint_{S^1_z} A$. 
This energy level shifting is due to Aharonov-Bohm type of effect. 
This provides the physical and experimental meanings of this decomposition.

More generally, one can consider the decomposition of the (zero mode part of the) Hilbert space of a $d+1$-dimensional TQFT on $M^{d-1}\times S^1$ into the Hilbert spaces of $(d-1)+1$-dimensional TQFTs on $M^d$:
\begin{equation}
\CH_{d+1\text{D-TQFT}}(M^{d-1}\times S^1)=
\bigoplus_{b} \CH_{(d-1)+1\text{D-TQFT}'(b)}(M^{d-1})
\label{eq:Hilbert-decompose}
\end{equation}
Note that (\ref{eq:Hilbert-decompose}) in principle contains more information than just the decomposition of the GSDs (as in (\ref{eq:GSD-decompose}) for $M^{d-1}=T^{d-1}$). This is because the Hilbert space $\CH_{M^d}$ of a $d+1$-dimensional TQFT on $M^d$ forms a representation of the mapping class group of $M^d$, MCG($M^d$). Therefore (\ref{eq:Hilbert-decompose}) should be understood as the direct sum decomposition of representations of MCG($M^{d-1}$). This generalizes the relation of MCG to the dimensional decomposition scheme proposed in \cite{1401.0518Moradi}.
Examples in \cite{Wang1404.7854, Moradi:2014cfa} show that
for a 3+1D to 2+1D decomposition,
we indeed have the modular $S$ and $T$ representation of $\text{MCG}(T^n)=SL(n, \Z)$ data decomposition:
$S^{xy,3\text{D}}= \bigoplus_{b} S^{xy,2\text{D}}_b$
and $T^{xy,3\text{D}}= \bigoplus_{b} T^{xy,2\text{D}}_b$ on a 2D spatial torus $T^{2}_{xy}$. 

The statement can be made more precise in the case when the $d+1$-dimensional TQFT is realized by gauging a certain SPT with finite abelian (0-form) symmetry $G$. Suppose that other (ungauged) symmetries of the theory are contained in $H$, an extension of $SO$ (or $O$ when there is time-reversal symmetry) structure group of a space-time manifold. For example, when there is $\Z_2^f$ fermionic parity, one considers manifolds with Spin-structure. The corresponding SPT state then are classified by\footnote{Suppose for simplicity $\Omega^H_{d+1}(BG)$ contains only \emph{torsion} elements Tor. Otherwise we redefine it by replacing $\Omega^H_{d+1}(BG)$ with $\text{Tor}\,\Omega^H_{d+1}(BG)$ in the formulas below. Throughout this work, we focus on the \emph{torsion} Tor part.} 
\begin{equation}
\Omega^{d+1}_H(BG):= \text{Hom}(\Omega^H_{d+1}(BG),U(1)),	
\end{equation}
where $\Omega^H_{d+1}(BG)$ is the bordism group of manifolds with $H$-structure (e.g. $H=\text{Spin}$) \cite{Kapustin1403.1467,Kapustin:2014dxa,1604.06527FH} and equipped with maps to $BG$ (the classifying space of $G$).

 Then, for an SPT state corresponding to a choice
 \begin{equation}
	 \mu \in \Omega^{d+1}_H(BG),
	\end{equation}
the partition function of the corresponding gauged theory on a closed $d+1$-manifold is given by
 \begin{equation}
	 Z_\mu (M^{d+1}):=\frac{1}{|G|^{|\pi_0(M^{d+1})|}}
	 \sum_{a_{d+1}\in H^1(M^{d+1},G) }\mu([(M^{d+1},a_{d+1})])
	 \label{gauged-SPT-def}
	\end{equation}
where the pair $(M^{d+1},a_{d+1})$, a $(d+1)$-manifold with $H$-structure and a map $M_{d+1}\rightarrow BG$, represents an element in $\Omega^H_{d+1}(BG)$. Here and below we use one-to-one correspondence between homotopy classes of maps $M^{d+1}\rightarrow BG$ and elements of $H^1(M^{d+1},G)$. Note that the choice of $\mu$ can be understood as the choice of the action for finite group gauge theory: $\mu([(M^{d+1},a)])\equiv e^{iS(M^{d+1},a_{d+1})}\in U(1)$.

Suppose $H$-structures on $M^d$ and $S^1$ define $H$-structure on $M^d\times S^1$ (this is true for $H$=Spin example). Then one can consider the following map for a given element $b\in G$:
\begin{equation}
	\begin{array}{cccc}
	\phi_b: & \Omega^H_d(BG) & \longrightarrow & \Omega^H_{d+1}(BG) \\
	& [(M^{d},a_{d})]& \longmapsto & [(M^d\times S^1, a_d\oplus (\underbrace{b\oplus \ldots \oplus b}_{|\pi_0(M^d)|}))]
	\end{array}
\end{equation}
where we used that $H^1(M^d\times S^1)\cong H^1(M^d)\oplus G^{|\pi_0(M^d)|}$. It is easy to see that the map above is well defined, that is the image of $[(M^d,a_d)]$ does not depend on the choice of the representative $(M^d,a_d)$ in $\Omega^H_d(BG)$. 

From the definition of gauged SPTs (\ref{gauged-SPT-def}), it follows that for connected closed manifolds we have the following relation
\begin{equation}
	\left. Z_\mu(M^d\times S^1)\right|_{|\pi_0(M^d)|=1} = 
	\sum_{b\in G} Z_{\phi_b^*(\mu)}(M^d).
	\label{partition-function-decomp}
\end{equation}
That is the partition function of the gauged $d+1$-dimensional SPT labeled by $\mu\in \Omega^{d+1}_H(BG)$ is given by the sum of gauged $d$-dimensional SPTs labeled by $\phi^*_b(\mu)\in \Omega^{d}_H(BG)$.

Similarly, for Hilbert spaces of the corresponding TQFTs we have
\begin{equation}
	\left. \CH_\mu(M^{d-1}\times S^1)\right|_{|\pi_0(M^{d-1})|=1} = 
	\bigoplus_{b\in G} \CH_{\phi_b^*(\mu)}(M^{d-1}).
	\label{hilbert-space-decomp}
\end{equation}
For a connected bordism $N^{d}$, $\partial N^d =\big(\sqcup_i(-M_i^{d-1}) \big) \sqcup \,\big(\sqcup_j \tilde{M}_j^{d-1} \big)$ we then have
\begin{equation}
	\left.Z_\mu(N^d\times S^1)\right|_{|\pi_0(N^d)|=1} =\mathfrak{i}_\text{diag}\circ \left(\bigoplus_{b\in G} Z_{\phi^*_b(\mu)}(N^d)\right) \circ \mathfrak{pr}_\text{diag}
	\label{morphism-decomp}
\end{equation}
where
\begin{equation}
	\begin{array}{cccc}
		Z_\mu(N^d\times S^1): & \otimes_i \CH_\mu(M_i^{d-1}) & \longrightarrow &
		\otimes_j \CH_\mu (\tilde{M}_j^{d-1}) \\
		Z_{\phi^*_b(\mu)}(N^d \times S^1): & \otimes_i \CH_{\phi^*_b(\mu)}(M_i^{d-1}) & \longrightarrow &
		\otimes_j \CH_{\phi^*_b(\mu)} (\tilde{M}_j^{d-1}) 
	\end{array}
\end{equation}
and
\begin{equation}
	\begin{array}{cccc}
		\mathfrak{i}_\text{diag}: & \bigoplus\limits_{b\in G}\bigotimes\limits_j \CH_{\phi^*_b(\mu)}(\tilde{M}_j^{d-1}) & \longrightarrow &
		\bigotimes\limits_j \bigoplus\limits_{b_j\in G}\CH_{\phi^*_{b_j}(\mu)} (\tilde{M}_j^{d-1}) = \bigotimes\limits_j \CH_\mu (\tilde{M}_j^{d-1}) \\
		\mathfrak{pr}_\text{diag}: & 
		\bigotimes\limits_i \CH_\mu ({M}_i^{d-1}) = \bigotimes\limits_i \bigoplus\limits_{b_i\in G}\CH_{\phi^*_{b_i}(\mu)} ({M}_i^{d-1}) 
		& \longrightarrow &
		\bigoplus\limits_{b\in G}\bigotimes\limits_i \CH_{\phi^*_b(\mu)}({M}_i^{d-1}) 
		\\
	\end{array}
\end{equation}
are inclusion of the diagonal and projection onto the diagonal.

Let us denote the TQFT functor (in the usual Atiyah's meaning) for the (d+1)-dimensional gauged SPT labeled by $\mu$ as $\cC^{d\text{D}}_\mu$ (so that its value on objects is given by $\CH_\mu(\bullet)$ and its value on morphisms is $Z_\mu(\bullet)$). Then throughout the paper we will often write simply
\begin{equation}
	\cC^{d\text{D}}_\mu  = \bigoplus_{b\in G} \cC^{(d-1)\text{D}}_{\phi^*_b(\mu)} 
	\label{eq:C-decompose}
\end{equation}
by which we actually mean that the functors satisfy relations\footnote{Relation (\ref{partition-function-decomp}) can be understood as a special case of (\ref{morphism-decomp}) with $\mathfrak{i}_\text{diag}:\C^{|G|}\stackrel{\sum}{\rightarrow}\C$, $\mathfrak{pr}_\text{diag}:\C\stackrel{\text{diag}}{\rightarrow}\C^{|G|}$} (\ref{partition-function-decomp}), (\ref{hilbert-space-decomp}), (\ref{morphism-decomp}). We thus decompose a $(d+1)${D-TQFT} to many sectors of
$((d-1)+1)${D-TQFT}$'$ labeled by $b$, in the topological vacua subspace within the nearby lowest energy Hilbert space. We mark that the related ideas of dimensional decomposition scheme are explored in \cite{2017DelcampDittrich, Lan1704.04221, 1709.04924Delcamp, Lan1801.08530}.

Relation (\ref{partition-function-decomp}) can also be interpreted as follows:
\begin{equation}
\sum_b Z_{\phi^*_b(\mu)}(M^d)=Z_{\mu}(M^d\times S^1) 
=\Tr_{\CH_\mu(M^{d})}1.
\end{equation}
That is, the trace over the Hilbert space of $d+1$-dimensional TQFT has interpretation of sum over the partition functions of $d$-dimensional TQFTs. This is an example of the general notion of decategorification in mathematics, where the vectors space are replaced by numbers. The inverse, that is a lift of numbers to vector spaces is known as categorification. Note that even though the partition function of a single $d$-dimensional TQFT${'}$ in the sum above cannot be categorified (i.e. interpreted as a trace over some Hilbert space), a particular sum of them can be. The notion of (de)-categorification can be extented to the level of the extended TQFT functors. In particular, (\ref{hilbert-space-decomp}) can be interpreted as
\begin{equation}
	K^0(\text{BCond}_\mu(M^{d-1}))=\bigoplus_{b\in G} \CH_{\phi_b^*(\mu)}(M^{d-1})
\end{equation}
where $\text{BCond}_\mu(M^{d-1})$ is the category of boundary conditions of the $(d+1)$-dimensional TQFT (obtained by gauging SPT labeled by $\mu$) on $M^{d-1}$ and $K^0$ is the Grothendieck group.

Note that in the case of fermionic theories the Hilbert spaces in (\ref{eq:Hilbert-decompose}) have an additional structure: $\Z_2^f$-grading (see section \ref{Sec:fTQFT-GSD} for details).

\section{Bosonic TQFTs and Ground State Degeneracy}

\label{Sec:GSD-bTQFT}

In this section we compute the ground state degeneracy (GSD, or, equivalently, the vacuum degeneracy) of some topological field theories, 
using the strategy and the set up similar to the one in \cite{1612.09298PutrovWang}.
We will consider TQFTs with a continuum field description in terms of $n$-form gauge fields. The level-quantization constraint for such theories is derived and given in \cite{1405.7689}.
Below we compute the GSD on a spatial manifold ${M}^{d}$ via the absolute value of the partition function ${Z}$ 
on a spacetime manifold ${M}^d \times S^1$  based on its relation of to the dimension of Hilbert space $\cH$:
$$
\text{GSD}_{M^{d}}=\dim \cH_{M^{d}} = Z(M^{d} \times S^1).
$$
As a warm-up, we start with (untwisted)
{$\Z_N$ gauge theory\cite{Wegner:1971jf}}, also known as 
{$\Z_N$ spin liquid\cite{ReadSachdevPRL66.1773},}
{$\Z_N$ topological order\cite{WenPRBZ2TO44.2664}},
 or
{$\Z_N$ toric code\cite{Kitaev2003}}.
Then we proceed to more general twisted discrete gauge theories:
bosonic Dijkgraaf-Witten (DW) gauge theories.
In most of the cases 
we consider the torus as the spatial manifold for simplicity:
$$
\text{GSD}_{T^{d}}=\dim \cH_{T^{d}} = |Z(T^{d+1})|.
 $$
 Below we use the notation
$N_{ijk\dots}\equiv \gcd(N_i,N_j,N_k,\dots)$.
We will always use $A$ to denote a 1-form gauge field,
while $B$ can be a higher-form gauge field.
In most cases, without introducing ambiguity, we omit the explicit wedge product $\wedge$
between differential forms. We will also often omit the explicit summations over the indices $I,J,K,\ldots$ in the formulas.
We note that related calculations of bosonic 
GSD are also derived based on independent and different methods in \cite{Wang1404.7854, 1612.01418Wen, 1703.03266Wan}.
Some of the main results of this section are briefly summarized in Table \ref{Table:bTQFT}.

\begin{table}[h!]
\footnotesize
\begin{center}
    \begin{tabular}{|c|c|c|c|c|c|c|}
    \hline
   Dim & gauge group $G$ & $\begin{array}{c} \text{Action} \\ (\text{Local}) \end{array}$  &  $\GSD_{T^{d}}$  & $\GSD_{S^{d-2}\times S^1}$\\ \hline\hline
any D & $\prod_I \Z_{N_I}$ & 
$\int \frac{ N_I}{2\pi}{B^I  d A^I}$ &  $|G|^{d}$  & --
\\ \hline  \hline     
2+1D & $U(1)^n$ (level $K$) &
$ \int  \frac{K_{IJ}}{4 \pi} A^I  d A^J$
 & $|\det K|$& --
 \\ \hline  
2+1D & $\prod_I \Z_{N_I}$ &
$\int \frac{N_I}{2\pi}{B^I   d A^I} + { \frac{ p_{IJ}}{4 \pi}} A^I  d A^J$
 & $|G|^{2}$  & --
 \\ \hline   
2+1D & $\underset{{I=1,2,3}}{\prod} \Z_{N_I}$ &
$ \int \frac{ N_I}{2\pi}{B^I   d A^I}+{ \frac{N_1 N_2 N_3\;
p_{}}{{(2 \pi)^2 } N_{123}}} A^1  A^2   A^3$
 & $N^4+ N^3-N$  & --
 \\ \hline  \hline  
3+1D & $\underset{{I=1,2,3}}{\prod} \Z_{N_I}$ & 
$\int \frac{ N_I}{2\pi}{B^I   d A^I} {{+}}  
\frac{ N_I N_J \; p_{IJK}}{{(2 \pi)^2 } N_{IJ}}   
A^I  A^J  d A^K $
& $|G|^{3}$  & $|G|$
\\ \hline    
3+1D & $\underset{{I=1,2,3,4}}{\prod} \Z_{N_I}$ & 
$\int \frac{ N_I}{2\pi}{B^I  d A^I} + 
{ \frac{N_1 N_2 N_3 N_4\;
p_{}}{{(2 \pi)^3 } N_{1234}}} A^1  A^2  A^3  A^4 $
& $\begin{matrix}N^{10}+N^9+N^8-N^7\\
-N^6 -N^5+ N^3
\end{matrix}$  & $|G|$
\\ \hline    
3+1D & $\underset{{I}}{\prod} \Z_{N_I}$ & 
$\int \frac{N_I}{2\pi}B^I dA^I+\frac{p_{IJ}N_IN_J}{4\pi N_{IJ}}\,B^I B^J $
& $\gcd(p,N)^3$  & $\gcd(p,N)$
\\ \hline
\hline    
4+1D & $\underset{{I=1,\dots,5}}{\prod} \Z_{N_I}$ & 
$\int \frac{ N_I}{2\pi}{B^I  d A^I} + 
{ \frac{N_1 N_2 \dots N_5\;
p_{}}{{(2 \pi)^4 } N_{12345}}} A^1  A^2  A^3  A^4  A^5 $
& \eqn{eq:GSD-BdA+5A}, \eqn{eq:GSD-BdA+5A-p}
  & $|G|$
\\ \hline
\hline    
{$d$D} 
& $\underset{{I=1,\dots,{d}}}{\prod} \Z_{N_I}$ & 
$\int \frac{ N_I}{2\pi}{B^I  d A^I} + 
{ \frac{N_1 N_2 \dots N_{d}\;
p_{}}{{(2 \pi)^{d-1} } N_{1\dots d}}} A^1  \dots A^{d} $
& \eqn{eq:GSD-BdA+dA-p}
  & $|G|$
\\ \hline    
 \hline   
    \end{tabular}
    \end{center}
\caption{Table of TQFTs and GSDs. 
For twisted gauge theories of Dijkgraaf-Witten (DW) theory, 
 we will sometimes restrict to the case $\Z_{N_1}=\Z_{N_2}=\Z_{N_3}=\Z_{N_4}=\dots \equiv \Z_{N}$ where $N$ is prime. 
Here $p$ is nontrivial and $\gcd(p, N)=1$ for those non-Abelian theories within DW theories; $G$ denotes the total finite gauge group in DW setup. 
Our derivations are based on continuum field descriptions. These results can be independently compared with
discrete cocycle/cochain lattice path integral method in \cite{Wang1404.7854, 1612.01418Wen}. 
}
\label{Table:bTQFT}
\end{table}

\subsection{$\int {B  d A}$ in any dimension} 
\label{sec:BdA-any-dim}
To warm up, we evaluate the ground state degeneracy of the untwisted $\Z_N$ gauge theory in $d+1$D on torus $T^d$ as the partition function on
$M^{d+1}=T^{d+1}$ spacetime in two different ways. %
In the first approach, we  integrate out a $(d-1)$-form $B$ field which yields a condition of $A$ being flat together with quantization of its holonomies. 
We evaluate\footnote{Sometimes we may make the wedge product ($\wedge$) implicitly without writing it down.}
\bea
\text{GSD}_{{T}^{d}}
&=&\int [D B][DA] 
\exp[ \int_{{T}^d \times S^1} {\frac{\ii N}{2\pi}{B  \wedge d A}} ]\\
&=&\int [DA] 
\;1 \vert_{dA=0, \;\;\; \oint_{S^1\subset T^{d+1}} A= \frac{2 \pi n_\mu}{N},  \;\;\; n_\mu \in \Z_{N} }=\cN^{-1}\sum_{a\in H^1(M^{d+1},G)} 1 \nonumber\\
&=&\cN^{-1}  \cdot {|H^1(M^{d+1},G)|}= \frac{|H^1(M^{d+1},G)|}{|H^0(M^{d+1},G)|}=\frac{N^{d+1}}{N}=N^d. \nonumber
\eea
The $G=\Z_N$ is the gauge group.  
The $\cN^{-1}={|H^0(M^{d+1},G)|}^{-1}$ is the normalization factor that takes into account gauge redundancy of 1-form gauge field.

In the second approach, we integrate out a 1-form $A$ field which yields a flat $B$ condition together with quantization of its flux through any codimension-2 cycle $M^{d-1}\subset T^{d+1}$. We evaluate
\bea
\text{GSD}_{{T}^d}
&=&\int [D B][DA] 
\exp[ \int_{{T}^{d+1}} {\frac{\ii N}{2\pi}{B  \wedge d A}} ] \\
&=&\int [DB] 
\;1 \vert_{dB=0, \;\;\; \oint_{M^{d-1}} B= \frac{2 \pi n}{N},  \;\;\; n \in \Z_{N} } \nonumber\\
&=& \cN^{-1}  \cdot {|H^{d-1}(M^{d+1},G)|}=(\prod_{j=0}^{d-2} {|H^{d-2-j}(M^{d+1},G)|}^{(-1)^{j}})^{-1} {|H^{d-1}(M^{d+1},G)|} \nonumber\\
&=& (N^{(-)^{d+1}{ d+1 \choose 0}} \dots N^{{ d+1 \choose d-3}} N^{-{ d+1 \choose d-2}}) N^{{ d+1 \choose d-1}}
= N^d.  \nonumber
\eea
The $\cN^{-1}$ factor again takes into account the gauge redundancy of $(d-1)$-form gauge field $B$.
The gauge transformation of $B \to B+ d \lambda^{(d-2)}$ contains the $(d-2)$-form gauge parameter $\lambda^{(d-2)}$, whose
gauge transformation allows $ \lambda^{(d-2)} \to \lambda^{(d-2)} + d \lambda^{(d-3)}$ change with further lower form redundancy.
Considering the gauge redundancy layer by layer, we obtain the $\cN^{-1}$ factor in the third line in the above equation.
The last equality uses $(1-s)^{d+1}={ d+1 \choose d+1} 1^{d+1} - { d+1 \choose d} s 
+ { d+1 \choose d-1} s^2 + \dots + (-)^{d+1} { d+1 \choose 0} s^{d+1}$ with  $s=1$.
The results of the above first and second approach match indeed, $\text{GSD}_{{T}^d}=|G|^d$.
\footnote{Partition function for $\mathbb{Z}_N$ gauge theory with 1-form and $d-1$ form gauge fields in $d+1$D match only up to the gravitational counter term $N^{\chi(M^{d+1})}$ where $\chi$ is the Euler number, if we use the normalization factors $\mathcal{N}$ explained in the main text. However, when $M^{d+1}=S^1\times M^{d}$, $\chi=0$ and the partition function agrees, which is consistent with the fact that the GSD itself is observable quantity. See (B.23) of \cite{Gaiotto:2014kfa}.}

\subsection{$\int K_{IJ} A_I   d A_J$, $\int {B  d A}{{+} } A   d A$  in 2+1D, $\int {B d A}{{+} } A  A d A$  in 3+1D and $\int BdA+A^{d-1}dA$ in any dimension}


First we compute the GSD of {$\int \sum_{I=1}^3\frac{ N_I}{2\pi}{B^I  \wedge d A^I}{{+} c_{123}} A^1 \wedge A^2 \wedge d A^3 $} (where $c_{123}=\frac{p_{123}N_1N_2}{(2\pi)^2N_{12}},\;p_{123}\in \Z$) theory 
 on
a torus. Other details of the theory are studied in \cite{1612.09298PutrovWang},
with the level-quantization constraint derived/given in \cite{1405.7689}.
\bea
\text{GSD}_{{T}^3}
&=&\int [D B][DA] 
\exp[ \int_{{T}^3 \times S^1} {\frac{\ii N_I}{2\pi}{B^I  \wedge d A^I}{{+\ii} c_{123}} A^1 \wedge A^2 \wedge d A^3} ]\\
&=&\int [DA] 
\exp[ \int_{{T}^3 \times S^1} {{{\ii} c_{123}} A^1 \wedge A^2 \wedge dA^3 } ] \vert_{dA^I=0, \;\;\; \oint_{S^1} A^I= \frac{2 \pi n_I}{N_I},  \;\;\; n_I \in \Z_{N_I} } \nonumber\\
&=& \sum_{n_{I,x},n_{I,y},n_{I,z} \in \Z_{N_I}} 1 
= ({N_1}{N_2}{N_3})^3 = |G|^3. \nonumber
\eea
We have used that $A$ satisfies the flatness condition in the second line, so all the configurations weigh with
$\exp[\int{{\ii} c_{123}} A^1 \wedge A^2 \wedge d A^3 ]=1$. 
To sum over $\int [DA]$ in the partition function, we simply need to sum over all
the possible holonomies $\oint_{S^1} A^I= \frac{2 \pi n_I}{N_I}$ around every non-contractible directions.
Similarly, for
{$\int \sum_{I=1}^2\frac{ N_I}{2\pi}{B^I  \wedge d A^I}{{+} c_{122}} A^1 \wedge A^2 \wedge d A^2$} theory,
from the flatness condition on $A$ on the torus it follows that the partition function is given by
\bea
\text{GSD}_{{T}^3}
&=& \sum_{n_{I,x},n_{I,y},n_{I,z} \in \Z_{N_I}} 1 
= ({N_1}{N_2})^3 = |G|^3. 
\eea
In 2+1D, the same strategy allows us to evaluate the GSD for $\int\sum_I\frac{ N_I}{2\pi}{B^I  \wedge d A^I}
+ \sum_{IJ}{ c_{IJ}} A^I \wedge \dd A^J $ ( $c_{IJ}=\frac{p_{IJ}}{4\pi},\;p_{IJ}\in \Z$) theory on a torus: 
\bea
\text{GSD}_{{T}^2}
&=&\int [D B][DA] 
\exp[ \int_{{T}^2 \times S^1} {\frac{\ii N_I}{2\pi}{B^I  \wedge d A^I} {{+\ii} c_{IJ}} A^I \wedge \dd A^J } ]\\
&=&\int [DA] 
\exp[ \int_{{T}^2 \times S^1} {{{\ii} c_{IJ}} A^I \wedge \dd A^J} ] \vert_{dA^I=0, \;\;\; \oint_{S^1} A^I= \frac{2 \pi n_I}{N_I},  \;\;\; n_I \in \Z_{N_I} } \nonumber\\
&=& \sum_{n_{I,x},n_{I,y} \in \Z_{N_I}} 1 
= \prod_{I}({N_I})^2 = |G|^2. \nonumber
\eea
The result can be interpreted as the volume of the rectangular polyhedron with edges of sizes $N_I$ (each appearing twice). More generally, for an abelian Chern-Simons theory with matrix level $K$ \cite{WenKmatrix}, 
that is with the action\footnote{If there is
an odd entry along the diagonal of $K_{II}$, then it requires a spin structure, otherwise it is non-spin.} $\int\frac{ K_{IJ}}{4\pi}{A^I  \wedge d A^J}$,
the  flatness condition is modified to $\sum_J K_{IJ} d A^J=0$. The result is then given by the volume of the polyhedron with edges given by column vectors of the matrix $K$:
\bea
\text{GSD}_{{T}^2}=|\det K|.
\eea

The calculation above can be easily generalized to the case of $d+1$-dimensional theory with the action of the form $\int BdA+A^{d-1}dA$. The result is $\text{GSD}_{T^d}=|G|^{d}$. This is in line with the fact that these theories are of abelian nature.

One can also obtain the GSD of the above theories based on the cochain path integral, see Ref.~\cite{1612.01418Wen} on these Abelian TQFTs.

\subsection{$\int {B   d A}+ \int B  B$  in 3+1D}

\label{sec:BB-GSD}

\subsubsection{Twisted $\Z_{N}$ theory with a $B \wedge B$ term}
\label{sec:BB-GSD-1}

We first consider a 3+1D action
{$\int \frac{N}{2\pi}{B  \wedge d A}{{+} \frac{N p}{2\pi}} B \wedge B$. This theory has been considered in detail in \cite{Gaiotto:2014kfa} (appendix B). In the action above we chose a less refined level quantization, which is valid for any manifold possibly without a spin structure.
For a non-spin bosonic TQFT, the level quantization can be easily derived based on \cite{1405.7689}.
For a spin fermionic TQFT, Ref.\cite{Gaiotto:2014kfa} provides a refined level quantization on a spin manifold, 
where the $p$ can take half integer values, namely
we can redefine $p = p'/2$ with an integer $p'$. In short, we get 
$\int \frac{ N}{2\pi}{B  \wedge d A}{{+} \frac{ N p'}{4\pi}} B \wedge B$ where now $p' \in \Z$.
It is a spin TQFT when both $N$ and $p'$ are odd.
The gauge transformation is
$B \to B + d \lambda$,
$A \to A - 2p \lambda +dg = A - p' \lambda +dg$.

Using the approach similar to the one in the second part of section \ref{sec:BdA-any-dim} we can evaluate its GSD on a 3-torus:
\bea
\text{GSD}_{{T}^3}
&=&\int [D B][DA] 
\exp[ \int_{{T}^3 \times S^1} {\frac{\ii N}{2\pi}{B  \wedge d A}{{+} \frac{\ii N p}{2\pi}} B \wedge B} ]\\
&=&\int [DB] 
\exp[ \int_{{T}^3 \times S^1} { \frac{\ii N p}{2\pi}} B \wedge B ] \vert_{dB=0, \;\;\; \oint_{M^2} B= \frac{2 \pi n}{N},  \;\;\; n \in \Z_{N} }\nonumber\\
&=& \cN^{-1} \sum_{n_{\alpha\beta} \in \Z_{N}}  \exp[  {{\ii}}  \frac{2 \pi (2 p_{})}{N}(n_{xy} n_{zt} -n_{xz} n_{yt}+n_{yz} n_{xt})  ] \nonumber\\
&=& \cN^{-1} \sum_{n_{\alpha\beta} \in \Z_{N}} \exp[  {{\ii} }  \frac{2 \pi ( {2}p/\gcd( {2}p,N))}{(N/\gcd( {2}p,N))} (n_{xy} n_{zt} -n_{xz} n_{yt}+n_{yz} n_{xt})
] \nonumber\\   
&=& \cN^{-1} (\frac{N \cdot N}{(N/\gcd( {2}p,N))})^3=(\frac{N^4}{N})^{-1} (\frac{N \cdot N}{(N/\gcd( {2}p,N))})^3=\gcd( {2} p,N)^3
=\gcd( p',N)^3 . \nonumber
\eea
Where $n_{\alpha\beta}$ are fluxes of the field $B$ through 2-tori in the directions $\alpha,\,\beta$.
The sum over $n_{\alpha\beta}$ factorizes into the product of sums
over the pairs $n_{xy},n_{zt}$, $n_{xz},n_{yt}$ and $n_{yz},n_{xt}$. These sums can be interpreted as sums over integral points inside squares of size $N$. Each square has $N\cdot N$ area. We divide this area by  $(N/\gcd(p',N))$, since a summation of $(N/\gcd(p',N))$ number of exponential factors gives one.
We used the fact that $(p'/\gcd(p',N))$ and $(N/\gcd(p',N))$ are relatively prime. 
The $\cN^{-1}$ factor is derived from dividing by the number of 1-form gauge symmetries, $|H^1(M,G)|$ which is equal to $N^4$ on the ${{T}^3 \times S^1}$, and then multiplying by the order of the gauge group, $|H^0(M,G)|=N$.
This gives the normalization factor $\cN^{-1}=1/(N^4/N)=1/N^3$, which accounts for the redundancy
of ``gauge symmetries'' and ``gauge symmetries of gauge symmetries.''
Overall, we obtain $\text{GSD}_{{T}^3}=\gcd(p',N)^3$ which is consistent with Ref.\cite{Gaiotto:2014kfa}. 
See \cite{Gaiotto:2014kfa} for the evaluation of partition function on other manifolds.

We can also use another independent argument based on Ref.\cite{Gaiotto:2014kfa} to verify the GSD obtained above.
In Ref.\cite{Gaiotto:2014kfa}, it was found that $\int {\frac{ N}{2\pi}{B  \wedge d A}{{+} \frac{ N p'}{4\pi}} B \wedge B}$ theory
has a similar GSD as $\Z_{\gcd(N,p')}$ gauge theory at the low energy. First, we know that the 
commutator between conjugate field and momentum operators is 
$[A(x),B(x')]=\frac{\ii 2\pi}{N} \delta{(x-x')}$. 
At $p=0$,
there is a 2-form global symmetry $\Z_N$ and a 1-form global symmetry\footnote{Recall that in general a generator of $q$-form symmetry is realized by an operator supported on a submanifold codimension $q+1$ (that is of dimension $d-q$ for a $d+1$ dimensional spacetime).} $\Z_N$
generated by:
$$U=e^{\ii \oint_\gamma A}, \;\;\;\; V=e^{\ii \oint_{\Sigma} B}.$$ 
At $p=0$, the symmetry transformation gives,
\bea
B  &\to& UBU^{-1}=B-\frac{1}{N} \xi_B, \\
A &\to& VAV^{-1}=A+\frac{1}{N} \zeta_A,
\eea
with $\xi_B$ and $\zeta_A$ are flat and satisfying $\oint_{\Sigma} \xi_B = 2 \pi$ and $\oint_{\gamma} \zeta_A = 2 \pi$,
so that $U^N=V^N=1$ and $UV=e^{\frac{\ii 2\pi}{N}} VU$.
The operators $U$ and $V$ can be referred to as the clock and the shift operators (like the angle and angular momentum operators). They generate
$N$ distinct ground states along each non-contractible loop. 
On the other hand, when $p' \neq 0$, we can consider
an open cylindrical surface ($\Sigma$) operator with two ends on closed loops $\gamma$ and $\gamma'$:
$$W=\exp[\ii \oint_{\gamma} A + \ii   p' \int_{\Sigma} B  -\ii \oint_{\gamma'} A  ].$$
The boundary components of $\Sigma$ are $\gamma$ and $\gamma'$, which makes the operator gauge invariant under the gauge transformation.
The closed line operator with $\exp[\ii \oint_{\gamma} A]$ can be defined
whenever the contribution from the open surface part becomes trivial. Let
$\Sigma_1$ and $\Sigma_2$ be two distinct surfaces bounded by $\gamma$
and $\gamma'$ (i.e., $\partial \Sigma_1=\partial \Sigma_2=\gamma_1\cup (-\gamma_2)$ where the minus sign indicates the opposite orientation),
then $\Sigma_1-\Sigma_2$ is a closed surface, and we have
$\int_{\Sigma_1} B-\int_{\Sigma_2} B= \frac{2 \pi}{N}n$ with some $n
\in \Z_{N}$,
since
$\oint_{\Sigma_1 -\Sigma_2} B= \frac{2 \pi}{N}n$.
The minimum integer $I$ enforcing $I p' \int_{\Sigma_1-\Sigma_2} B={2 \pi}$ is
$I=\frac{N}{\gcd(p',N)}$.
This means that
$\exp[  \ii ( I \oint_{\gamma} A + I   p' \int_{\Sigma} B  -I
\oint_{\gamma'} A )]
$ does not depend on the choice of the open surface $\Sigma$, and we can view $\exp[  \ii ( I \oint_{\gamma} A + I   p' \int_{\Sigma} B  -I
\oint_{\gamma'} A )]
$ as two deconfined line operators formally as $\exp[  \ii ( I \oint_{\gamma} A   -I
\oint_{\gamma'} A )]
$. 
Thus we can define the line operator alone as:
\begin{eqnarray}
U=\exp[ \ii \; I\oint_{\gamma} A]=e^{\ii \frac{N}{\gcd(p',N)} \oint
A}, \;\;\;\; \text{with} \;\;\;\;  U^{\gcd(p',N)}=1.
\end{eqnarray}
The reasoning is, again, that since $\oint_{\gamma} A= \frac{2
\pi}{N}n$ with some $n \in \Z_{N}$,
then we have $U=\exp[\ii  \frac{2 \pi n}{\gcd(p',N)}]$ satisfying
$U^{\gcd(p',N)}=1$.

The closed surface operator alone can be defined as:
\bea
V=\exp[\ii  \oint_{\Sigma} B], \;\;\;\; \text{ while } V^N=1 \text{ and } V^{p'}=1,  \;\;\;\; \text{ so } V^{\gcd(p',N)}=1.
\eea 
Here $V^N=1$ is due to $\oint_{\Sigma} B= \frac{2 \pi}{N}n$.
On the other hand, we can close the open surface
by letting two closed curves ${\gamma}$ and ${\gamma'}$ coincide, then
the open surface operator $\exp[\ii \oint_{\gamma} A + \ii   p' \int_{\Sigma} B  -\ii \oint_{\gamma'} A  ]$
becomes the surface operator
$\exp[ \ii   p' \oint_{\Sigma} B]$. But the original open surface operator must be trivial (inside correlation functions) because the theory describes topological and gapped systems.
This implies $W=1$ and thus $\exp[ \ii   p' \oint_{\Sigma} B]=1 \Rightarrow V^{p'}=1$.
The superposed conditions of $V^N=1$ {and} $V^{p'}=1$,  give the final finest constraint $V^{\gcd(p',N)}=1$.
Finally we obtain:
$$UV=e^{\frac{\ii 2\pi}{\gcd(p',N)}} VU,  \;\;\;\; \text{ because of } [\frac{N}{\gcd(p',N)} A(x),B(x')]=\frac{\ii 2\pi}{\gcd(p',N)} \delta{(x-x')}.$$
Thus the new clock and shift operators generate
${\gcd(p',N)}$ distinct ground states along each non-contractible loop. 
For $\text{GSD}_{{T}^3}$ on a ${{T}^3}$ spatial torus with three spatial non-contractible loops,
we obtain $\text{GSD}_{{T}^3}={\gcd(p',N)}^3$ as in \cite{Gaiotto:2014kfa}.


\subsubsection{More general theory}

We can also consider a more generic action $\int \sum_{I=1}^2\frac{ N_I}{2\pi}{B^I  \wedge d A^I}
+{ \frac{p_{12} N_{1} N_{2}}{2\pi N_{12} }} B^1 \wedge B^2 $} where
$N_{12}\equiv{\gcd(N_1,N_2)}$ and $p_{12}$ can be a half-integer.
Again we choose a less refined level quantization, which is true for any generic manifold without a spin structure.
The gauge transformation is the following:
\bea
&&B^I \to B^I + d \lambda^I, \;\;\; 
A^1 \to A^1 - p_{12} \frac{N_1}{N_{12}}\lambda +dg^1, \;\;\; 
A^2 \to A^2 - p_{12} \frac{N_2}{N_{12}}\lambda +dg^2.
\eea
Again, we derive $\text{GSD}_{{T}^3}$ on a ${{T}^3}$ spatial torus:
\bea
\text{GSD}_{{T}^3}
&=&\int [D B][DA] 
\exp[ \int_{{T}^3 \times S^1} { \frac{\ii N_I}{2\pi}{B^I  \wedge d A^I}
{{+} \frac{\ii p_{12} N_{1} N_{2}}{2\pi N_{12} }} B^1 \wedge B^2 } ]\\
&=&\int [DB] 
\exp[ \int_{{T}^3 \times S^1} 
{+} \frac{\ii p_{12} N_{1} N_{2}}{2\pi N_{12} } B^1 \wedge B^2
 ] \vert_{dB^I=0, \;\;\; \oint_{M^2} B^I= \frac{2 \pi n_I}{N_I},  \;\;\; n_I \in \Z_{N} } \nonumber \\
&=& \cN^{-1} \sum_{n^I_{\alpha\beta} \in \Z_{N_I}} \exp[  {{\ii} } \frac{2 \pi (\frac{2 p_{12}}{\gcd(2 p_{12},N_{12})})}{(\frac{N_{12}}{\gcd(2 p_{12},N_{12})})} (n_{xy}^1 n_{zt}^2 -n_{xz}^1 n_{yt}^2+n_{yz}^1 n_{xt}^2+
 n_{zt}^1 n_{xy}^2 - n_{yt}^1n_{xz}^2+n_{xt}^1 n_{yz}^2)
] \nonumber\\   
&=& \cN^{-1} (\frac{N_{1} \cdot N_{2}}{(N_{12}/\gcd(2 p_{12},N_{12}))})^6=
(\frac{|H^1(M,G)|}{|H^0(M,G)|})^{-1} \cdot (\frac{N_{1} \cdot N_{2}}{(N_{12}/\gcd(2 p_{12},N_{12}))})^6 \nonumber\\
&=&\frac{N_{1} N_{2}}{N_{1}^4 N_{2}^4} \cdot (\frac{N_{1} \cdot N_{2}}{(N_{12}/\gcd(2 p_{12},N_{12}))})^6=
(\frac{\lcm(N_1,N_2)}{\gcd(N_1,N_2)})^3 \gcd({2} p_{12},N_{12})^6. \nonumber 
\eea
Similarly to Sec.~\ref{sec:BB-GSD-1}, 
the $\cN^{-1}$ factor is derived from dividing by the number of 1-form gauge symmetries, $|H^1(M,G)|$, and
then multiplying by the order of the gauge group, $|H^0(M,G)|$.
This accounts for the redundancy of ``gauge symmetries'' and ``gauge symmetries of gauge symmetries.''

\subsection{$\int {B   d A}+ \int A  A$  in 1+1D}

We consider the 1+1D TQFT with the action $\int \frac{ N_I}{2\pi}{B^I  \wedge d A^I}+\frac{ p_{12}N_1N_2}{2\pi\,N_{12}} A^1 \wedge A^2$. 
Locally B is a 0-form field and $A$ is a 1-form field.
The level quantization is described in \cite{1405.7689, 1612.09298PutrovWang}.
This theory can be obtained by dynamically gauging an SPTs with the symmetry group $G_s=\prod_{I=1}^2 \Z_{N_I}$ \cite{Gu:2015lfa}.
Its dimension of Hilbert space on ${S}^1$
is computed as a discrete sum, after integrating out $B$ field:
\begin{equation}
\text{GSD}_{{S}^1}
= \cN^{-1} \sum_{\vec{n}_I \in \Z_{N_I}^2} \exp[ - {{\ii} } p_{12} \frac{2 \pi }{N_{12}} \det(\vec{n}_1,   \vec{n}_2 )].
\end{equation}
Consider a specific example $N_1=N_2=N$ which is 
a prime number, so that $\gcd(p_{12}, N)=1$. In this case,
\bea
\text{GSD}_{{S}^1}
&=& \cN^{-1} \sum_{\vec{n}_I \in \Z_{N}^2} \exp[ - {{\ii} p_{12}} \frac{2 \pi }{N_{}} \det(\vec{n}_1,   \vec{n}_2) ] 
=\sum_{n_{2,x},n_{2,t} \in \Z_{N}}\delta( {n}_{2,x})\delta( {n}_{2,t})=1.\;\;\;\;\;\;\;\;\;\;
\eea
There is a unique ground state degeneracy without robust topological order in this case.

For more generic $N_1$ and $N_2$, 
the normalization factor $\cN^{-1}$ is $\frac{1}{N_1 N_2}$.
We can rewrite $\frac{p_{12}}{N_{12}}$ as
$\frac{p_{12}/{\gcd(N_{12}, p_{12})}}{N_{12}/{\gcd(N_{12}, p_{12})}}$ for generic non-coprime $N_{12}$ and $p_{12}$.
A direct computation shows
\bea \label{Eq.GSDAA}
\text{GSD}_{{S}^1}&=&
\frac{1}{N_1 N_2}\sum_{\vec{n}_I \in \Z_{N_I}^2} \exp[ - {{\ii} } p_{12} \frac{2 \pi }{N_{12}} \det(\vec{n}_1,   \vec{n}_2 )] =
\frac{1}{N_1 N_2}(\frac{N_1 N_2}{N_{12}/{\gcd(N_{12}, p_{12})}})^2 \nonumber\\
&=&\frac{\lcm(N_1,N_2)}{\gcd(N_1,N_2)} {\gcd(N_{12}, p_{12})}^2.\;\;\;\;
\eea 
The GSD depends on the level/class index $p_{12}$. Note that ${\gcd(N_{12}, p_{12})}={\gcd(N_1,N_2, p_{12})}$.

Some numerical evidences, such as the tensor network renormalization group method \cite{GuSPT}, suggest that there is \emph{no robust intrinsic topological order in 1+1D}.
We can show that ``no robust topological order in 1+1D'' can be already seen in terms of the fact that 
\emph{local non-extended operator}, such as the 0D vortex operator $B$, can lift the degeneracy. 
Thus this GSD is accidentally degenerate, not topologically robust.

From eq. (\ref{Eq.GSDAA}) it follows that for the level-1 action (i.e. $N_I=1$) we have GSD$=1$ and no intrinsic topological order, which is
consistent with the use of this level-1 theory for SPTs \cite{Gu:2015lfa}.

\subsection{$\int {B   d A}+ \int A  A A$ in 2+1D}

\label{sec:A3-GSD}


We can also consider the 2+1D TQFT with the action $\int \frac{ N_I}{2\pi}{B^I  \wedge d A^I}+c_{123}A^1 \wedge A^2 \wedge A^3$ (where $c_{123}=\frac{p_{123}N_1N_2N_3}{(2\pi)^2N_{123}},\;p_{123}\in \Z$). It can be obtained from dynamically gauging some SPTs with the symmetry group $G_s=\prod_{I=1}^3 \Z_{N_I}$ \cite{Gu:2015lfa}. The level quantization is discussed in \cite{1405.7689, 1612.09298PutrovWang}.
One can confirm that it is equivalent to 
Dijkgraaf-Witten topological gauge theory with the gauge group $G=\prod_{I=1}^3 \Z_{N_I}$ with the type-III cocycle twists
by computing its dimension of Hilbert space on a torus.
In the first step, we integrate out $B$ to get a flat $A$ constraint and obtain the following expression for $\text{GSD}_{{T}^2}$:
\begin{eqnarray}
\begin{split}
\int [D B][DA]\exp[ \int_{{T}^2 \times S^1} {\frac{\ii N_I}{2\pi}{B^I  \wedge d A^I}+{{\ii} c_{123}} A^1 \wedge A^2 \wedge A^3} ]\\
= \cN^{-1} \sum_{\vec{n}_I \in \Z_{N_I}^3} \exp[  {{\ii} } p_{123} \frac{2 \pi }{N_{123}} \det(\vec{n}_1,   \vec{n}_2,   \vec{n}_3 )].
\end{split}
\end{eqnarray}
The above formula is general but we take a specific example $N_1=N_2=N_3=N$ where $N$ is a prime number, so that $\gcd(p_{123}, N)=1$ below.
The calculation of $\text{GSD}_{{T}^2}$ reduces to a calculation of the following discrete Fourier sum. 
\bea
\text{GSD}_{{T}^2}
&=& \cN^{-1} \sum_{\vec{n}_I \in \Z_{N}^3} \exp[  {{\ii} p_{123}} \frac{2 \pi }{N_{}} \det(\vec{n}_1,   \vec{n}_2,   \vec{n}_3) ] 
= \sum_{\vec{n}_2, \vec{n}_3 \in \Z_{N}^3} \prod_{j} \delta( \det(\text{minor}(\vec{n}_2,   \vec{n}_3)_{1,j})). \;\;\;\;\; \;\;\;\;\;
\eea
We first sum over the vector $\vec{n}_1$, and this gives us the product of discrete delta functions of the determinants of the minors $\text{minor}(\vec{n}_2,   \vec{n}_3)_{1,j}$.
Case by case, there are a few choices of $\vec{n}_2,\vec{n}_3$ when the delta function is non-zero: 
(1) $\vec{n}_2$ is a zero vector, then $ \vec{n}_3$ can be arbitrary.
Each of this choices gives one distinct ground state configuration for $\text{GSD}_{{T}^2}$.
We have in total $1 \cdot N^3 $ such choices.
(2) $\vec{n}_2$ is not a zero vector, then as long as $\vec{n}_3$ is parallel to the $\vec{n}_2$, 
namely $\vec{n}_2 = C  \; \vec{n}_3 \pmod N$ for some factor $C$, the determinants of the minor
matrices are zero. 
The number of such configurations is $(N^3-1) \cdot N $.
The total ground state sectors are the sum of contribution from (1) and (2):
\bea \label{Eq.GSDAAA}
\text{GSD}_{{T}^2}=1 \cdot N^3+ (N^3-1) \cdot N
=N^4+N^3-N. \;\;\;\;\;
\eea
Our continuum field-theory derivation here independently reproduces the result from the discrete spacetime lattice formulation of 2+1D 
Dijkgraaf-Witten topological gauge theory computed in Sec. IV C of Ref.\cite{Wang1404.7854} and Ref.\cite{deWildPropitius:1995cf}. The agreement of the Hilbert space dimension (thus GSD) 
together with the braiding statistics/link invariants\cite{CWangMLevin1412.1781, 1612.09298PutrovWang}
confirms that
the field-theory can be regarded as the low-energy long-wave-length continuous field description of Dijkgraaf-Witten theory with the gauge group
$G=\prod_{I=1}^3 \Z_{N_I}$
with the type-III 3-cocycle twists.


\subsection{$\int {B   d A}+ \int A  A A A$ in 3+1D}

\label{sec:A4-GSD}


Below we consider the 3+1D TQFT action $\int \frac{ N_I}{2\pi}{B^I  \wedge d A^I}+{ c_{1234}} A^1 \wedge A^2 \wedge A^3 \wedge A^4$ (where $c_{1234}=\frac{p_{1234}N_1N_2N_3N_4}{(2\pi)^3N_{1234}},\;p_{1234}\in \Z$)
obtained from dynamically gauging some SPTs with the symmetry group $G_s=\prod_{I=1}^4 \Z_{N_I}$. 
See the level quantization in \cite{1405.7689, 1612.09298PutrovWang}.
It is equivalent to 
Dijkgraaf-Witten topological gauge theory at the low-energy 
of the gauge group $G=\prod_{I=1}^4 \Z_{N_I}$ with the type-IV 4-cocycle twists\cite{1405.7689}.
First, we verify it by computing its dimension of Hilbert space on a torus. 
%
%
\bea
\text{GSD}_{{T}^3}
&=&\int [D B][DA] 
\exp[ \int_{{T}^3 \times S^1} {\frac{\ii N_I}{2\pi}{B^I  \wedge d A^I}+{{\ii} c_{1234}} A^1 \wedge A^2 \wedge A^3 \wedge A^4} ]\\
&=&\int [DA] 
\exp[ \int_{{T}^3 \times S^1} {{{\ii} c_{1234}} A^1 \wedge A^2 \wedge A^3 \wedge A^4} ] \vert_{dA^I=0, \;\;\; \oint_{S^1} A^I= \frac{2 \pi n_I}{N_I},  \;\;\; n_I \in \Z_{N_I} } \nonumber \\
&=& \cN^{-1} \sum_{\vec{n}_I \in \Z_{N_I}^4} \exp[  {{\ii} } p_{1234} \frac{2 \pi }{N_{1234}} \det(\vec{n}_1,   \vec{n}_2,   \vec{n}_3,  \vec{n}_4 )
].   \nonumber
\eea
Here we assume the four non-contractible $S^1$ in ${T}^3 \times S^1$ have  coordinates $x,y,z,t$.
\be
\det(\vec{n}_1,   \vec{n}_2,   \vec{n}_3,  \vec{n}_4 )
\equiv\begin{vmatrix}
n_{1,x} & n_{1,y} & n_{1,z} & n_{1,t} \\
n_{2,x} & n_{2,y} & n_{2,z} & n_{2,t} \\
n_{3,x} & n_{3,y} & n_{3,z} & n_{3,t} \\
n_{4,x} & n_{4,y} & n_{4,z} & n_{4,t} 
\end{vmatrix}=\sum_{j} (-1)^{1+j} n_{1,j} \cdot \det(\text{minor}(\vec{n}_2,   \vec{n}_3,  \vec{n}_4)_{1,j}).
\ee
Here the minor sub-matrix
$\text{minor}(\vec{n}_2,   \vec{n}_3,  \vec{n}_4)_{1,j}$ of the remaining vectors $\vec{n}_2,   \vec{n}_3,  \vec{n}_4$
excludes the row and the column of $n_{1,j}$.
Also $\cN^{-1}$ is the proper normalization factor that takes into account the gauge redundancy. Namely, $\cN^{-1} =|G|^{-1}$ is the inverse of the order of the gauge group 
so that $|{Z}|$ have the proper integer value.
Without losing the generality of our approach, we take a specific example $N_1=N_2=N_3=N_4=N$ where $N$ is a prime number.
Hence we use the fact that $\gcd(p_{1234}, N)=1$ below.
The calculation of $\text{GSD}_{{T}^3}$ reduces to a calculation of the discrete Fourier summation. 
$$
\text{GSD}_{{T}^3}
= \cN^{-1}\sum_{\vec{n}_I \in \Z_{N}^4} \exp[  {{\ii} p_{1234}} \frac{2 \pi }{N_{}} \det(\vec{n}_1,   \vec{n}_2,   \vec{n}_3,  \vec{n}_4 ) ] 
= \sum_{\vec{n}_2, \vec{n}_3, \vec{n}_4 \in \Z_{N}^4} \prod_{j} \delta( \det(\text{minor}(\vec{n}_2,   \vec{n}_3,  \vec{n}_4)_{1,j})).
$$
We first sum over the vector $\vec{n}_1$, and this gives us discrete delta functions on the $\text{minor}(\vec{n}_2,   \vec{n}_3,  \vec{n}_4)_{1,j}$.
Case by case, there are a few choices when the product of delta functions does not vanish: 
(1) $\vec{n}_2$ is a zero vector, then $ \vec{n}_3, \vec{n}_4$ can be arbitrary.
Each of these gives us a distinct ground state configuration for $\text{GSD}_{{T}^3}$ with weight one.
All together this countributes $1 \cdot N^4 \cdot N^4$.
(2) $\vec{n}_2$ is not a zero vector, then as long as $\vec{n}_3$ is parallel to  $\vec{n}_2$,
namely $\vec{n}_2 \parallel  \vec{n}_3$ and $\vec{n}_2 = C  \; \vec{n}_3 \pmod N$, for some factor $C$, then the product of the determinants of the minor
matrices is zero. Here $\vec{n}_4$ can be arbitrary.
This gives  $(N^4-1)\cdot N\cdot N^4$ distinct ground state configurations.
(3) $\vec{n}_2$ is not a zero vector, and $\vec{n}_3$ is not parallel to  $\vec{n}_2$,
namely $\vec{n}_2 \neq C  \; \vec{n}_3 \pmod N$ for any $C$, then the determinant of the minor
matrices is zero if $\vec{n}_4$ is a linear combination of $\vec{n}_2$ and $\vec{n}_3$.
Namely, $\vec{n}_4 = C_1 \vec{n}_2 + C_2 \vec{n}_3$ for some integers $C_1,C_2 \in \Z_{N}$.
This gives  $(N^4-1) \cdot (N^4-N) \cdot N \cdot N$ distinct ground state configurations.
The total number topological vacua  is the sum of the contributions from (1), (2) and (3):
\bea \label{Eq.GSDAAAA}
\text{GSD}_{{T}^3}&=&1 \cdot N^4 \cdot N^4+(N^4-1) \cdot N \cdot N^4+(N^4-1) \cdot (N^4-N) \cdot N \cdot N \\
&=&N^8+N^9-N^5+N^{10}-N^7-N^6 + N^3. \;\;\;\;\; \nonumber
\eea
Our continuous field-theory derivation here independently reproduces the result from the discrete spacetime lattice formulation of DW topological gauge theory computed in Sec. IV C of Ref.\cite{Wang1404.7854}. The agreement of the Hilbert space dimension (thus GSD) 
together with the braiding statistics/link invariants\cite{CWangMLevin1412.1781, 1612.09298PutrovWang}
imply that
the field-theory can be regarded as the low-energy long-wave-length continuum field description of DW theory of the gauge group
$G=\prod_I \Z_{N_I}$
with the type-IV 4-cocycle twists.
%
%

\section{Higher Dimensional Non-Abelian TQFTs}

\label{Sec:GSD-bTQFT-higher}

\subsection{$\int {B   d A}+ \int A^5$ in 4+1D}


Consider continuum field theory which describes twisted Dijkgraaf-Witten (DW) theory with the gauge group $G=\mathbb{Z}_N^{5}$ with type V 5-cocycle twist
in $4+1$ dimensions:
$
\int \frac{ N_I}{2\pi}{B^I  \wedge d A^I}+{ c_{12345}} A^1 \wedge A^2 \wedge A^3 \wedge A^4 \wedge A^5,
$ (where $c_{12345}=\frac{pN_1N_2N_3N_4N_5}{(2\pi)^3N_{12345}},\;p\in \Z$).
The level quantization is described in \cite{1405.7689}. We would like to compute the GSD on a torus. Integrating over $B^I$ restricts $A^I$ to be  flat, and the only degree of freedom is the holonomies around cycles of the spacetime torus. We denote the holonomy of $A^I$ around the cycle $\gamma^{\mu}$ (which wrap around the $\mu$ direction of $T^4 \times S^1$) to be $\oint_{\gamma^{\mu}}A_i=2\pi n_i^{\mu}/N$, $\mu=0, 1, 2, 3, 4$.  Following the method in Sec.~\ref{sec:A4-GSD}, the partition function reduces to 
\begin{eqnarray}
\GSD_{T^4}=\sum_{n_1^{\mu}, n_2^{\mu}, n_3^{\mu}, n_4^{\mu}, n_5^{\mu}=0}^{N-1}\bigg[\frac{1}{N^5}\bigg] \exp\bigg[i \frac{2\pi p}{N}\sum_{\mu, \nu, \rho, \sigma, \lambda=0}^{4}\epsilon^{\mu\nu\rho\sigma\lambda}n_1^{\mu}n_2^{\nu}n_3^{\rho}n_4^{\sigma}n_5^{\lambda}\bigg].
\end{eqnarray}
We further sum over $n_1^\mu$ using the discrete Fourier transformation, 
\begin{eqnarray}
\sum_{n \in \Z_{N}}\exp\bigg[\frac{i 2\pi p \,\alpha\, n }{N}\bigg]=N\delta\bigg[\alpha=0\mod \frac{N}{\gcd(N,p)}\bigg]
\end{eqnarray}
which yields
\begin{eqnarray}
\GSD_{T^4}=\sum_{n_I^\mu \in \Z_{N}}^{}\prod_{\mu=0}^{4}\delta\bigg[\sum_{\nu, \rho, \sigma, \lambda=0}^{4}\epsilon^{\mu\nu\rho\sigma\lambda}n_2^{\nu}n_3^{\rho}n_4^{\sigma}n_5^{\lambda}=0\mod \frac{N}{\gcd(N,p)}\bigg].
\end{eqnarray}
The product of the delta functions imposes the constraint that 
$
\vec{n}_2, \vec{n}_3, \vec{n}_4, \vec{n}_5
$
are linearly independent $\mod \frac{N}{\gcd(N,p)}$.
and the partition function counts the number of configurations which satisfy such constraint. There are a few cases:
(1) We first consider the case when $p=1$. 
If $\vec{n}_2=0\mod N$, the other vectors $\vec{n}_3, \vec{n}_4, \vec{n}_5$ can be chosen at will. Hence there are 
$
1\cdot N^5\cdot N^5\cdot N^5
$
configurations in this case. 
(2)
If $\vec{n}_2\neq 0\mod N$ and $\vec{n}_3=C \vec{n}_2$, 
the other vectors $\vec{n}_4, \vec{n}_5$ can be chosen at will. There are $(N^5-1)$ choices of $\vec{n}_2$, 
$N$ choices of $\vec{n}_3$, and $N^5$ choices of $\vec{n}_4$ and $\vec{n}_5$ separately. Hence there are
$
(N^5-1)\cdot N\cdot N^5\cdot N^5
$
configurations in this case. 
(3)
If $\vec{n}_2\neq 0\mod N$, $\vec{n}_3\neq C \vec{n}_2$ and $\vec{n}_4=C_1\vec{n}_2+C_2\vec{n}_3$,  $\vec{n}_5$ can be chosen at will. There are $N^5-1$ choices of $\vec{n}_2$, $N^5-N$ choices of $\vec{n}_3$, $N\cdot N$ choices of $\vec{n}_4$ (there are $N$ choices of $C_1$ and $C_2$ respectively) and $N^5$ choices of $\vec{n}_5$. Hence there are 
$
(N^5-1)\cdot (N^5-N)\cdot (N\cdot N)\cdot N^5
$
configurations in this case. 
(4)
If $\vec{n}_2\neq 0\mod N$, $\vec{n}_3\neq C \vec{n}_2$,  $\vec{n}_4\neq C_1\vec{n}_2+C_2\vec{n}_3$ and $\vec{n}_5= C_3\vec{n}_2+C_4\vec{n}_3+C_5 \vec{n}_4$, there are $N^5-1$ choices of $\vec{n}_2$, 
$N^5-N$ choices of $\vec{n}_3$, 
$N^5-N^2$ choices of $\vec{n}_4$ 
and $N\cdot N\cdot N$ choices of $\vec{n}_5$. 
Hence there are
$
(N^5-1)\cdot (N^5-N)\cdot (N^5-N^2)\cdot (N\cdot N \cdot N)
$
configurations in this case. 
In summary, the 
GSD with the $\gcd(N,p)=1$ is
\begin{eqnarray}
\label{eq:GSD-BdA+5A}
\GSD_{T^4} \mid_{\gcd(N,p)=1}= N^6\Big[-1+N^2+N^3+N^4-N^6-2N^7-N^8+N^{10}+N^{11}+N^{12}\Big].
\end{eqnarray}
For a generic level $p$, the configurations for each $n_i^{\mu}$ split into $\gcd(N,p)$ sectors, and we need to sum over all the sectors in the partition function. For instance, when $\vec{n}_2=0\mod \frac{N}{\gcd(N,p)}$, there are $(\gcd(N,p))^5$ choices of $\vec{n}_2$, and $N^5$ choices of $\vec{n}_3, \vec{n}_4, \vec{n}_5$ separately. Hence there are 
$
\gcd(N,p)^5\cdot N^5\cdot N^5\cdot N^5
$
configurations in this case. 
It is clear that this result can be obtained from the $p=1$ case by replacing $N$ with $\frac{N}{\gcd(N,p)}$, and multiplying by the number of sectors $\gcd(N,p)^5$ for each $\vec{n}_i$. Specifically, $\gcd(N,p)^5\cdot N^5\cdot N^5\cdot N^5$ can be rewritten as 
$
\Big[\gcd(N,p)^5\Big]^4 \cdot \Big[\Big(\frac{N}{\gcd(N,p)}\Big)^5\Big]\cdot \Big[\Big(\frac{N}{\gcd(N,p)}\Big)^5\Big]\cdot \Big[\Big(\frac{N}{\gcd(N,p)}\Big)^5\Big].
$
For the other cases, we can count similarly. Generalizing the ground state degeneracy to generic $p$, one obtains the following expression
\begin{eqnarray}
\label{eq:GSD-BdA+5A-p}
\begin{split}
\GSD_{T^4} =& \bigg[\gcd(N, p)^5\bigg]^4\bigg[\frac{N}{\gcd(N, p)}\bigg]^6\bigg\{-1+\bigg[\frac{N}{\gcd(N, p)}\bigg]^2+\bigg[\frac{N}{\gcd(N, p)}\bigg]^3\\&+\bigg[\frac{N}{\gcd(N, p)}\bigg]^4-\bigg[\frac{N}{\gcd(N, p)}\bigg]^6-2\bigg[\frac{N}{\gcd(N, p)}\bigg]^7-\bigg[\frac{N}{\gcd(N, p)}\bigg]^8\\&+\bigg[\frac{N}{\gcd(N, p)}\bigg]^{10}+\bigg[\frac{N}{\gcd(N, p)}\bigg]^{11}+\bigg[\frac{N}{\gcd(N, p)}\bigg]^{12}\bigg\}.
\end{split}
\end{eqnarray}
In particular, when $p=0$, $\gcd(N,0)=N$, the partition function is reduced to 
$
Z(T^4 \times S^1)_{p=0}=\bigg[\gcd(N, p)^5\bigg]^{4}= (N^5)^4=|G|^4
$
as expected.

\subsection{Counting Vacua in Any Dimension for Non-Abelian $\int {B   d A}+ \int A ^{d}$}

We can discuss such non-Abelian TQFTs in any general dimensions.
We first consider $p=1$ theories, and the pattern is obvious, 
\begin{eqnarray}
\begin{split}
\GSD_{T^{d-1}} \mid_{p=1}=&1\cdot \underbrace{N^d\cdots N^d}_{d-2}+ (N^d-1)\cdot N\cdot \underbrace{N^d\cdots N^d}_{d-3}
+(N^d-1)\cdot(N^d-N)\cdot N^2\cdot \underbrace{N^d\cdots N^d}_{d-4}\\&+\cdots +(N^d-1)\cdot(N^d-N)\cdots (N^d-N^{d-3})\cdot N^{d-2}\\=& (N^d)^{d-2}+\sum_{k=0}^{d-3}\prod_{i=0}^{k}(N^d-N^i)N^{k+1}(N^d)^{d-(k+2)-1}. 
\end{split}
\end{eqnarray}
For general $p$, the pattern can be generalized, we have 
\begin{eqnarray}
\label{eq:GSD-BdA+dA-p}
\begin{split}
\GSD_{T^{d-1}}=& Z(T^{d-1} \times S^1)=
\big[ \gcd(N, p)^d \big]^{d-1}\Bigg\{\bigg[\frac{N}{\gcd(N, p)}\bigg]^{d(d-2)}\\&+\sum_{k=0}^{d-3}\prod_{i=0}^{k}\Bigg(\bigg[\frac{N}{\gcd(N, p)}\bigg]^d-\bigg[\frac{N}{\gcd(N, p)}\bigg]^i\Bigg)\bigg[\frac{N}{\gcd(N, p)}\bigg]^{d^2-(k+3)d+k+1}  \Bigg\}.
\end{split}
\end{eqnarray}
When $p=0$, we have 
$
Z(T^{d-1} \times S^1)_{p=0}=|G|^{d-1}
$
as expected. 

All these examples, including Sec.~\ref{sec:A3-GSD},~\ref{sec:A4-GSD}, and \ref{Sec:GSD-bTQFT-higher} of
$\int {B   d A}+ \int A ^{d}$ type, 
are non-abelian TQFTs due to the GSD reduction from $|G|^{d-1}$ to a smaller value.
This can be understood as the statement that the \emph{quantum dimensions} $d_\alpha$ of some anyonic excitations are not equal to, but greater than 1,
i.e. $d_\alpha>1$ \cite{Wang1404.7854}.

By the same calculation, we obtain that the GSD on ${{T}^{d-1}}$ of the theory with the action 
in Eq.(\ref{Eq.GSDAAAA}) with $N=1$ is GSD$=1$ (no intrinsic topological order in this case).

\section{Fermionic Spin TQFTs from Gauged Fermionic SPTs and Ground State Degeneracy}

\label{Sec:fTQFT-GSD}
The gapped theories in  $d+1$ dimensions with fermionic degrees of freedom can be effectively described in terms $d+1$-dimensional spin-TQFTs. Unlike in the bosonic case, the partition function of a spin-TQFT on a $(d+1)$-manifold $M^{d+1}$ depends not just on topology of $M^{d+1}$, but also on a choice of spin-structure. If a spin-structure exists, there are $H^1(M^{d+1},\Z_2)$ different choices. Similarly, the Hilbert space $\CH_{M^d}$ depends on the choice of spin structure on the spatial manifold $M^d$. Moreover, $\CH_{M^d}$ can be decomposed into fermionic ($f$) and bosonic ($b$) parts:
\begin{equation}
\CH_{M^d}=\CH_{M^d}^f\oplus \CH_{M^d}^b.
\end{equation}
Equivalently, $\CH_{M^d}$ is a $\Z_2^f$-graded vector space. When we state results about $\CH_{M^d}$ in particular examples we will use the following condensed notation:
\begin{equation}
\text{GSD}_{M^d}=n_ff + n_bb,\qquad n_{f,b}\in \Z_+
\end{equation}
where $n_{f,b}\equiv \dim \CH_{M^d}^{f,b}$. In general the fermionic and bosonic GSDs $n_f$ and $n_b$ can be determined from the following partition functions of the spin-TQFT:
\begin{equation}
Z(M^d\times S^1_\text{A})=\Tr_{\CH_{M^d}}1=n_b+n_f,
\end{equation}
\begin{equation}
Z(M^d\times S^1_\text{P})=\Tr_{\CH_{M^d}}(-1)^F=n_b-n_f,
\end{equation}
where A/P denote anti-periodic/periodic boundary conditions on fermions along the time circle $S^1$ (i.e. even/odd spin structure on $S^1$).

\subsection{Examples of fermionic SPTs and spin TQFTs: 
$\Z_2 \times \Z_2^f$ and $(\Z_2)^2 \times \Z_2^f$ in 2+1D. 
$\Z_4\times \Z_2 \times \Z_2^f$ and $(\Z_4)^2 \times \Z_2^f$
in 3+1D
}

In this section, we consider fermionic spin-TQFTs arising from gauging a unitary global symmetry of fermionic SPTs (fSPTs), set up in \cite{1612.09298PutrovWang} (with some corrections and improvements, the basic idea remaining the same). More 2+1D/3+1D spin TQFTs are given later in Sec.~\ref{Sec:Dim-Reduce-F}. A systematic study 
(using the cobordism approach) of fermionic SPTs with finite group symmetries and the corresponding fermionic gauged theories will be given in \cite{finite-group-fSPT-cobordisms}, here we will just use some of the results from that work. 

Previously Ref.~\cite{1501.01313Cheng,1610.08478Wang} (and References therein)
study the classifications of 2+1 interacting fSPTs involving finite groups.
Recently, Ref.~\cite{1701.08264Kapustin,1703.10937WangGu,1705.08911-Tantivasadakarn, 2018Fidkowski} study pertinent issues of 3+1 interacting fSPTs.
Ref.~\cite{1701.08264Kapustin} provides explicit 3+1D fSPTs and their bosonized TQFTs.
The bosonization is performed by dynamically gauging the fermion parity $\Z_2^f$, which results in bosonic TQFTs (or non-spin TQFTs). In contrast, in our work,
we \emph{only} dynamically gauge the (finite unitary onsite) symmetry group but leave the $\Z_2^f$ global symmetry intact, which results in fermionic spin TQFTs.
Ref.~\cite{1703.10937WangGu} provides fixed-point  fSPT wavefunctions for 3+1D interacting fermion systems and generalized group super-cohomology theory.
Ref.~\cite{1705.08911-Tantivasadakarn} uses the gauged fSPTs and their braiding statistics to detect underlying nontrivial fSPTs and propose their tentative classifications.
Ref.~\cite{2018Fidkowski} studies the surface TQFTs for 3+1D fSPTs.

We briefly summarize results about examples considered in the current paper in Table \ref{Table:fTQFT}. The precise meaning of the expressions for the actions is explained below (see \cite{finite-group-fSPT-cobordisms} for details). Note that the theories considered here do not have time-reversal symmetry, so the space-time manifold $M^{d+1}$ is considered to be oriented.

\begin{table}[h!]
\footnotesize
\begin{center}
    \begin{tabular}{|c| c | c|c | c | c|c|c|}
    \hline
  & Dim & Group & $\begin{array}{c} \text{spin TQFTs}\\ 
   \text{from gauging fSPTs}:\\ \text{Action } (\text{Formal notation}) \end{array}$  &  ${T^{d}}$ (all P) & ${T^{d}}$ (other) & $\RP^{d}$\\ \hline\hline
1) & 2+1D & $\Z_2 \times \Z_2^f $ & 
$\frac{\pi}{4}\int a\cup \text{ABK}$ &  $3f$ & $3b$ & --
\\ \hline   
2)
 & 2+1D & $ (\Z_2)^2 \times \Z_2^f$ &
$\frac{\pi}{2} \int a_1 \cup a_2 \cup \tilde{\eta}$
 & $6f + 1b$ & $7b$ & --
 \\ \hline  \hline  
3)
 & 3+1D & $\Z_4\times \Z_2 \times \Z_2^f $ & 
$\frac{\pi}{2}\int (a_1\mod 2) \cup (a_2\cup \text{ABK})$ 
& $512b$  & $512b$ & $3b$
\\ \hline   
4)
 & 3+1D & $(\Z_4)^2 \times \Z_2^f $  & 
$\pi\int (a_1\mod 2)\cup (a_2 \mod 2) \cup\text{Arf}$ 
 & $64\cdot (42f+1b)$ & $64\cdot 43b$ & $4b$ 
\\ \hline   
    \end{tabular}
    \end{center}
\caption[Text excluding the matrix]{Table of $d+1$-dimensional ($d=$2 or 3) spin TQFTs, and their corresponding GSDs. The expressions in the fourth column are actions for cochain spin-TQFTs with $a_i\in H^1(M^{d+1},\Z_{n_i})$. Note that there seems to be no purely fermionic SPTs in (3+1)-dimensions with just $\Z_2^n$ global symmetry, due to the fact that $\Omega^4_\text{Spin}(B\Z_2^n)=\Omega^4_{SO}(B\Z_2^n)=H^4(B\Z_2^n, U(1))$. 
This also suggests \emph{no} gauged version of intrinsic fermionic spin TQFTs of a gauge group $\Z_2^n$, thus
their gauged TQFTs are Dijkgraaf-Witten theories and their continuum bosonic TQFTs discussed in Sec.~\ref{Sec:GSD-bTQFT}.
In contrast, in \cite{finite-group-fSPT-cobordisms},
we obtain torsion parts of above cobordism groups as
$
\left\{\begin{array}{l} 
\Omega^3_{\text{Spin}}(B\Z_2)\cong \Z_8 \\  
\Omega^3_{\text{Spin}}(B(\Z_2^2))\cong \Z_8^2\times \Z_4\\
\Omega^4_{\text{Spin}}(B(\Z_2\times \Z_4))\cong \Z_2\times \Z_4\\
\Omega^4_{\text{Spin}}(B(\Z_4^2))\cong \Z_4^2\times \Z_2
\end{array}\right.
$, which introduce additional new intrinsic interacting fermionic SPTs, 
beyond group cohomology and bosonic SPTs.
The fourth column shows their topological terms (SPT invariants)
which generate the intrinsic interacting fermionic SPTs.
We gauge their global symmetries (leaving $\Z_2^f$ remained ungauged) to study their fermionic spin TQFTs, 
and GSDs in the last three columns.
In particular, the odd generators for the 1), 2) and 4) gauged spin TQFTs
are indeed \emph{non-Abelian} fermionic topological orders.
The reason is that,  Abelian topological orders, say 3),
yield $\GSD_{T^{d}}=|G|^d$. However,
 \emph{non-Abelian} topological orders yield the reduction $\GSD_{T^{d}}<|G|^d$,
 see Sec.~\ref{sec:1.2}.
 In \cite{finite-group-fSPT-cobordisms}, we also compute
 cobordism groups of
 $\Omega^4_{\text{Spin}}(B(\Z_2^2 \times \Z_4))$, $\Omega^4_{\text{Spin}}(B(\Z_2 \times \Z_4^2))$ and $\Omega^4_{\text{Spin}}(B(\Z_4^3))$, etc.,
 which suggest the classification of fSPTs and new additional topological terms.
}
\label{Table:fTQFT}
\end{table}

1) 
\begin{equation} \label{eq:ABK-def}
 \int_{M^3} a\cup \text{ABK} \equiv \text{ABK} [\text{PD}(a)] \in \Z_8
\end{equation}
where $\text{PD}(a)$ is a smooth, possibly non-orientable submanifold in $M^3$ representing Poincar\'e dual to $a\in H^1(M^3,\Z^2)$ (it always exist in codimension 1 case). Given a spin structure on $M^3$ the submanifold $\text{PD}(a)$ can be given a natural induced $\text{Pin}^-$ structure, see\footnote{The idea is that the normal bundle to the submanifold $\text{PD}(a)\equiv N^2\subset M^3$ for oriented $M^3$ can be realized as 
determinant line bundle
$\det T{N^2}$, so that $TM^3|_{N^2}=TN^2\oplus \det TN^2$. For a general vector bundle $V$, there is a natural bijection between Pin$^-$- structures on $V$ and Spin-structures on $V\oplus \det V$.} \cite{kirby1990pin}. ABK denotes $\Z_8$ valued Arf-Brown-Kervaire invariant of surfaces with Pin$^-$ structure (the invariant which provides explicit isomorphism $\Omega^{\text{Pin}^-}_2\cong \Z_8$). 

2)\footnote{The corresponding cobordism group is $\Omega^3_{\text{Spin}}(B(\Z_2^2))\cong \Z_8^2\times \Z_4$. The presented action corresponds to the generator of $\Z_4$ factor.
More generally in \cite{finite-group-fSPT-cobordisms},
we obtain torsion parts of cobordism groups as,
$\Omega^3_{\text{Spin}}(B\Z_2^n)=\Z_8^n\oplus\Z_4^{\frac{n^2-n}{2}}\oplus\Z_2^{\frac{n^3-3n^2+2n}{6}}$ and
$\Omega^4_{\text{Spin}}(B\Z_2^n)=\Z_2^{\frac{n^4+2n^3+11n^2-14n}{24}}.$
The later coincides with group cohomology result \cite{1405.7689}.}
\begin{equation}
\int_{M^3} a_1 \cup a_2 \cup \tilde{\eta} \equiv 
\tilde{\eta}(\text{PD}(a_1)\cap \text{PD}(a_2)) \equiv f_{M^3}(a_1,a_2) \in \Z_4
\end{equation}
As before by $\text{PD}(a_i)$ we mean smooth submanifold representing Poincar\'e dual to $a_i$. By $\tilde{\eta}$ we denote a $\Z_4$ valued invariant associated to a 1-dimensional submanifold equipped with an additional structure (induced by Spin structure on $M^3$ as well as by embedded surfaces $\text{PD}(a_i)$). Its value, $f_{M^3}(a_1,a_2)$ with 
\begin{equation}
 f_{M^3}: H^1(M^3,\Z_2)\times H^1(M^3,\Z_2) \rightarrow \Z_4
\end{equation} 
can be concretely defined as follows \cite{kirby1990pin}. Take $a_{1,2}\in H^1(M_3,\Z_2)$. As was already discussed in case 1), Spin structure on $M_3$ induces Pin$^-$-structures $q_{i}$ on $\text{PD}(a_{i})$. Pin$^-$- structures on Riemann surfaces can be understood as quadratic enhancments of intersection form on $H_1(\text{PD}(a_i),\Z_2)$:
\begin{equation}
 q_{i}:H_1(\text{PD}(a_i),\Z_2) \longrightarrow \Z_4
\end{equation} 
Then
\begin{equation}
 f_{M^3}: (a_1,a_2)\longmapsto q_{1}([\text{PD}(a_1)\cap \text{PD}(a_2)])
\end{equation} 
Note that in fact $f$ is symmetric under $a_1\leftrightarrow a_2$. More geometrically the value $f_{M^3}(a_1,a_2)$ can be understood as follows. Choose a trivialization of the normal bundle to $\text{PD}(a_1)\cap \text{PD}(a_2)$ such that the induced Spin-structure makes this 1-dimensional manifold a Spin-boundary. Then $f_{M^3}(a_1,a_2)$ is the number of half-twists modulo 4 (only mod 4 value is independent of choices made) of the section of the normal bundle inward to $\text{PD}(a_1)$ or $\text{PD}(a_2)$. Note that 
\begin{equation}
 f_{M^3}(a_1,a_2) = \int_{M^3}a_1\cup a_1 \cup a_2 = \int_{M^3}a_1\cup a_2 \cup a_2 \mod 2. 
\end{equation}

3)\footnote{The corresponding cobordism group is $\Omega^4_{\text{Spin}}(B(\Z_2\times \Z_4))\cong \Z_2\times \Z_4$. The presented action corresponds to the generator of $\Z_4$ factor.}
\begin{equation}
 \int_{M^4}(a_1\mod 2) \cup (a_2\cup \text{ABK}) \equiv  \int_{\text{PD}(a_1 \mod 2)} a_2|_{\text{PD}(a_1 \mod 2)} \cup \text{ABK} \mod 4\qquad \in \Z_4
 \label{fSPT-Z2Z4-action}
\end{equation}
Where $\text{PD}(a_1 \mod 2)$ is a 3-dimensional smooth submanifold in $M^4$ representing Poincar\'e dual to $(a_1 \mod 2)$ where $\mod 2:H^1(M^4,\Z_4)\rightarrow H^1(M^4,\Z_2)$ is the part of the long exact sequence induced by the short exact sequence $0\rightarrow \Z_2\rightarrow \Z_4\rightarrow \Z_2\rightarrow 0$. Note that $\text{PD}(a_1\mod 2)$ can be chosen to be orientable, because $a_1\mod 2\neq 0$ implies that $a_1\in H^1(M^4,\Z_4)\cong H_3(M^4,\Z_4)$ is not 2-torsion. Moreover, $a_1$ determines the orientation on $\text{PD}(a_1 \mod 2)$ via trivialization of the normal bundle. In particular, the trivialization of the normal bundle allows to induce Spin-structure on $\text{PD}(a_1 \mod 2)$ from Spin-structure on $M^4$. The formula  $\ref{fSPT-Z2Z4-action}$ then defines invariant in terms of the invariant considered in case 1) modulo 4 (only mod 4 value is invariant under the choice of oriented $\text{PD}(a_1 \mod 2)$). Note that equivalently,
\begin{equation}
\int_{M^4}(a_1\mod 2) \cup (a_2\cup \text{ABK}) = f_{\text{PD}(a_1 \mod 2)}(a_2,a_2) \in \Z_4
\end{equation}
with $f$ defined as in the case 2).

4) \footnote{The corresponding cobordism group is $\Omega^4_{\text{Spin}}(B(\Z_4^2))\cong \Z_4^2\times \Z_2$. The presented action corresponds to the generator of $\Z_2$ factor.}
\begin{equation}
\int_{M^4} (a_1\mod 2)\cup (a_2 \mod 2) \cup\text{Arf} \equiv
\text{Arf}( \text{PD}(a_1\mod 2)\cap \text{PD}(a_2 \mod 2)) \in \Z_2.
\end{equation} 
where $\text{Arf}(\Sigma^2)$ denotes $\Z_2$ valued Arf invariant of spin-surface $\Sigma^2$. The spin-structure on $\Sigma^2= \text{PD}(a_1\mod 2)\cap \text{PD}(a_2 \mod 2)$ is induced as follows from the spin structure on $M^4$. As in the case 3), one can first consider a 3-dimensional oriented submanifold $M^3=\text{PD}(a_1\mod 2)\subset M^4$ with induced spin structure. By a similar argument $\Sigma^2 =\text{PD}(a_2|_{M^3} \mod 2)$  is also oriented and gets induced spin-structure from $M^3$.

\subsection{2+1D spin TQFTs from gauging Ising-$\Z_2$ of $\Z_2 \times \Z_2^f$ symmetry}
A 2+1D example is a spin-TQFTs obtained from gauging unitary Ising-$\Z_2$ of $\Z_2\times \Z_2^f$ symmetric fSPTs. 
Before gauging, this represents a class of 2+1D fermionic Topological Superconductor with a $\Z_8$ classification.
The $\Z_2^f$ denotes a fermion number parity symmetry. 
After gauging, the TQFTs are identified in Table 2 of \cite{1612.09298PutrovWang}, matching 
the mathematical classification through cobordism group
$
 \Omega^\text{Spin}_3(B\Z_2)\cong \Z_8
$
 in \cite{Kapustin:2014dxa}. 
 The exactly solvable lattice Hamiltonian constructions for (un-gauged) SPTs \cite{1604.02145Fidkowski}
 and for gauged theory \cite{1605.06125Cheng} have been recently explored.
 Here we follow a TQFT approach following Ref.~\cite{1612.09298PutrovWang}.
 
Given a class $\nu\in \Z_8$, 
the corresponding spin-TQFT partition function reads
\begin{equation}
 Z(M^3,s)=\frac{1}{2}\sum_{a\,\in H^1(M^3,\Z_2)}e^{\frac{\pi i\nu}{4}\text{ABK}[\text{PD}(a),s|_{\text{PD}(a)}]},
 \label{gfSPT-Z2-3D}
\end{equation} 
The partition function is defined on a closed 3-manifold $M^3$ with spin structure $s\in \text{Spin}(M_3)$ with 
a dynamical $\Z_2$ gauge connection $a\in H^1(M^3,\Z_2)$, summed over in the path integral.
As already mentioned,  $\text{ABK}[\ldots]$ is the $\Z_8$ valued Arf-Brown-Kervaire (ABK) invariant of $\text{PD}(a)$, a (possibly non-orientable) surface in $M_3$ which 
represents a class in $H_2(M^3,\Z_2)$ Poincar\'e dual to $a\in  H^1(M^3,\Z_2)$. 
The $s|_{\text{PD}(a)}$ is the $\text{Pin}^-$ structure on $\text{PD}(a)$ 
induced by $s$ as described in 1) above. 
Note that there is no good local realization of ABK invariant via characteristic classes.

\begin{figure}[!h]
\centering
\includegraphics[scale=1.7]{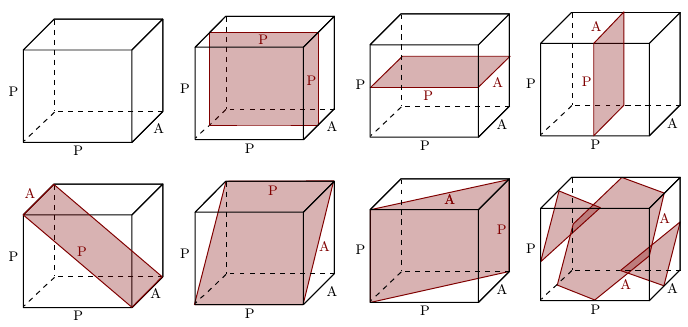}
\caption{Counting GSD on $T^2_o$, that is $\text{Tr}_{\CH_{T^2_\text{o}}}\,1$. 
The shaded 2-tori embedded into a $T^3$ represent Poincar\'e duals to elements of $H^1(M_3,\Z_2)\cong\Z_2^3$. 
The red letters A/P denote Anti-periodic/Periodic boundary conditions on the embedded 2-tori.
}
\label{fig:Z8-GSD}
\end{figure}

To compute GSD on $T^2$ for a spin-TQFT, we have to specify choices of spin structure on the spatial 2-torus $T^2$. 
There are 4 choices corresponding to periodic or anti-periodic (P or A) boundary
conditions along each of the two 1-cycles: (P,P), (A,P), (P,A), (A,A).
It turns out that Hilbert space only depends on the parity 
(i.e. the value of the Arf invariant of $T^2$). 
It is odd for (P,P), and even for (A,P), (P,A), (A,A). This is consistent with the fact that MCG$(T^2)=SL(2,\Z)$ only permutes spin-structures with the same parity. 
We will denote the corresponding two equivalences classes of spin 2-tori as ${T^2_\text{o}}$ and ${T^2_\text{e}}$. 
As described in the beginning of this section, the GSD is determined by the partition function $Z(T^3,s)$, with $M^3=T^3=T^2_{\text{e}|\text{o}}\times S^1_{\text{time}}$.
The time circle $S^1_{\text{time}}$
can have be P or A boundary conditions. 
Consider for example the choice of odd spin structure on $T^2$ and anti-periodic boundary condition on $S^1$. 
Then, as shown in Fig. \ref{fig:Z8-GSD}
\begin{multline}
 \text{Tr}_{\CH_{T^2_\text{o}}}\,1=Z(T^2_\text{o}\times S^1_\text{A})=
\frac{1}{2}\sum_{a\in H^1(T^3,\Z_2)\cong \Z_2^3}e^{\frac{\pi i\nu}{4}\text{ABK}[\text{PD}(a)]}=
\frac{1}{2}\sum_{a\in H^1(T^3,\Z_2)\cong \Z_2^3} e^{\pi i\nu\text{Arf}[\text{PD}(a)]}=\\
=\frac{1}{2}(1+(-1)^\nu+1+1+1+1+1+1)=\left\{\begin{array}{ll} 3, & \nu=1 \mod 2 \\ 4, & \nu=0 \mod 2 \end{array}\right.                                                   
\end{multline} 
where we used the fact that ABK$=4$Arf for oriented surfaces, where Arf is the ordinary Arf invariant of spin 2-manifolds.

Similarly,
\begin{equation}
 \text{Tr}_{\CH_{T^2_\text{o}}}\,(-1)^F=Z(T^2_\text{o}\times S^1_\text{P})
=\frac{1}{2}(1+7(-1)^\nu)=\left\{\begin{array}{ll} -3, & \nu=1 \mod 2 \\ 4, & \nu=0 \mod 2 \end{array}\right.                                                    
\end{equation}
which means that for $\nu=1\mod 2$ all states are femrionic. For the even spin structure on $T^2$ we have:
\begin{equation}
 \text{Tr}_{\CH_{T^2_\text{e}}}\,1=Z(T^2_\text{e}\times S^1_\text{A})
=\frac{1}{2}(7+(-1)^\nu)=\left\{\begin{array}{ll} 3, & \nu=1 \mod 2 \\ 4, & \nu=0 \mod 2 \end{array}\right.                                                    
\end{equation}
\begin{equation}
 \text{Tr}_{\CH_{T^2_\text{e}}}\,(-1)^F=Z(T^2_\text{e}\times S^1_\text{P})
=\frac{1}{2}(7+(-1)^\nu)=\left\{\begin{array}{ll} 3, & \nu=1 \mod 2 \\ 4, & \nu=0 \mod 2 \end{array}\right..                                                    
\end{equation}
So all states are bosonic. The result for all allowed spin structures, 
can be summarized as follows: 
\begin{itemize}
\item For odd $\nu$: 3 fermionic states for ${T^2_\text{o}}$ and 3 bosonic states for ${T^2_\text{e}}$.  
\item For even $\nu$: 4 states, all bosonic, for both ${T^2_\text{o}}$ and ${T^2_\text{e}}$. 
\end{itemize}
We can implement the similar counting for other 2+1D and 3+1D spin TQFTs given in \cite{1612.09298PutrovWang}. 
Notice that at least for 2+1D fermionic topological orders/TQFTs up to some finite states of GSD were classified in \cite{lan2015theory}. 
Our GSD counting can be compared with \cite{lan2015theory}.

So far we have focused on a 2+1D example,
but more 2+1D/3+1D spin TQFT examples are given in
Sec.~\ref{Sec:Dim-Reduce-F} using the dimensional reduction scheme for GSD counting.
We summarize these fermionic TQFTs and GSD data in Table \ref{Table:fTQFT}.




\section{Dimensional Reduction Scheme of Partition Functions and Topological Vacua}

\label{Sec:Dim-Reduce}

\subsection{Bosonic Dimensional Reduction Scheme}

\label{Sec:Dim-Reduce-B}

Here we perform the dimensional reduction of TQFTs described in Sec.~\ref{Sec:Categorification} for explicit bosonic examples, and match with  data computed in Sec.~\ref{Sec:GSD-bTQFT} and \ref{Sec:GSD-bTQFT-higher}.
We write the decomposition in terms of \eqn{eq:C-decompose}, but applicable also to
\eqn{eq:GSD-decompose} and \eqn{eq:Hilbert-decompose}.
The notation $\cC^{d\tD}$ below means a $d+1$D TQFTs.

For 3+1D ${\Z_{2}}$-gauge theory reduction to 2+1D, we can write the continuum field theory form
\bea   \label{eq.Z2toric_3Dfirst}
&& {\cC^{3\tD}_{{\frac{2}{2\pi}}
\int B dA}} =2 {\cC^{2\tD}_{{\frac{2}{2\pi}}
\int B dA}}, 
  \;\;\;\;\; 
\eea
or equivalently in terms of the cocycle $ {\cC^{3\tD}_{1}} =2 {\cC^{2\tD}_{1}}$. The sub-index $1$ means a trivial cocycle.
The $B$ fields represent a 2-form gauge field in $\cC^{3\tD}$,
but represent a 1-form gauge field in $\cC^{2\tD}$.
Here we take the compact $z$-direction (among the $x$-$y$-$z$-$t$ in 3+1D) as the compactification direction, and each sector comes from holonomy around the direction that is
$\oint_z A=0$ or $\pi$ (in terms of the 1-cochain field $\oint_z a=0$ or $1$) respectively.
The resulting sectors of 2+1D $\Z_2$ gauge theories are equivalent in this case.
See Fig.\ref{fig:dim-reduce} for a physical illustration.

For 3+1D twisted ${(\Z_{2})^2}$-gauge theory reduction to 2+1D, we obtain
\footnote{Note that, to be precise, the expressions like $\int A_idA_j$ are of the formal nature, since in general, on a manifold of non-trivial topology, $A_i$ is not globally defined (as there can be non-trivial $U(1)$ bundles.)  One possible way to treat this is to define $A_i$ locally with possible ``jumps'' along codimension-1 loci. More rigorously, it can be treated using Deligne-Beilinson cohomology (see e.g. \cite{Guadagnini:2008bh}). In 3 dimensions (since any 3-manifold is a boundary of some 4-manifold), one can also extend the theory to 4 dimensions with the corresponding term $\int F_iF_j$, $F_i=dA_i$. }

\be \label{eq.Z2toric_3D}
{\cC^{3\tD}_{\frac{2}{2\pi} \underset{i=1}{\overset{2}{\sum}}\int B_i dA_i +c_{122} A_1A_2dA_2}} = 
{\cC^{2\tD}_{\frac{2}{2\pi} \int \underset{i=1}{\overset{2}{\sum}} B_i dA_i}} 
\oplus {\cC^{2\tD}_{\frac{2}{2\pi} \int (\underset{i=1}{\overset{2}{\sum}} B_i dA_i +  A_2dA_2)}}
\oplus 2 {\cC^{2\tD}_{\frac{2}{2\pi} \int(\underset{i=1}{\overset{2}{\sum}} B_i dA_i +  A_1dA_2)}}.  \;\;\;\;\; 
\ee
We can explain this easily by converting those continuous descriptions to the cochain-field theory description with 4-cocycle and 3-cocycles. 
Relevant cocycles are a 4-cocycle $\omega_{4,\tII}^{(12)}=(-1)^{\int {a_1 \cup a_2  \cup a_2  \cup a_2}}$ in 3+1D; and also
3-cocycles 1, $\omega_{3,\tI}^{(1)}=(-1)^{\int {a_1 \cup a_1  \cup a_1 }} \equiv (-1)^{\int (a_1)^3}$ and $\omega_{3,\tII}^{(12)}=(-1)^{\int {a_1 \cup a_2  \cup a_2 }}$ in 2+1D.
Here all the $a_i$ (say $a_1$, $a_2$, etc.) are the $\Z_2$-valued 1-cochain field.
In this section, all these $a_i$, $B$ and $A$ are dynamical fields, which we need to sum over all configurations in the path integral also in $\cC^{d\tD}$
to obtain long-ranged entangled TQFTs 
(instead of short-ranged entangled SPTs).
The above \eqn{eq.Z2toric_3D} can be derived, effectively, as 
\bea
\begin{split}
{\cC^{3\tD}_{a_1 \cup a_2  \cup a_2  \cup a_2}}
&={\cC^{2\tD}_{1}} \oplus {\cC^{2\tD}_{(a_2)^3}} \oplus {\cC^{2\tD}_{a_1 \cup a_2  \cup a_2 }} \oplus {\cC^{2\tD}_{(a_1+ a_2) \cup a_2  \cup a_2 }}\\
&={\cC^{2\tD}_{1}} \oplus {\cC^{2\tD}_{(a_2)^3}} \oplus 2{\cC^{2\tD}_{a_1 \cup a_2  \cup a_2 }}. 
\end{split}
\eea
In the first line, we decompose the 3+1D theory with respect to holonomies around the compactifying $z$-direction (among the $x$-$y$-$z$-$t$ in 3+1D) which are
$(\oint_z a_{1},\oint_z a_{2})=(0,0),(1,0), (0,1), (1,1)$.  Then we obtain the second line by field redefinition, or equivalently a $\mathrm{SL}(2,\mathbb{Z}_2)$ transformation,
sending $a_1+ a_2 \to a_1$ in the last sector. 

For 3+1D twisted ${(\Z_{2})^4}$-gauge theory reduction to 2+1D, we can also use the cochain-field expression to ease the calculation,
\bea \label{eq:A4-decompose-cochain}
\begin{split}
{\cC^{3\tD}_{a_1 \cup a_2  \cup a_3  \cup a_4}}
&={\cC^{2\tD}_{1}} \oplus 4 {\cC^{2\tD}_{a_j \cup a_k \cup a_l}} 
\oplus 6 {\cC^{2\tD}_{{a_j \cup a_k \cup a_l}+{a_j \cup a_k \cup a_{l'}}}} \\
&\;\;\;\;\;\; \oplus 4 {\cC^{2\tD}_{{a_j \cup a_k \cup a_l} + {a_j \cup a_k \cup a_m}+ {a_j \cup a_l \cup a_m}}} 
\oplus {\cC^{2\tD}_{{a_1 \cup a_2 \cup a_3} + {a_1 \cup a_2 \cup a_4}+ {a_1 \cup a_3 \cup a_4} + {a_2 \cup a_3 \cup a_4}}}
\\
&={\cC^{2\tD}_{1}} \oplus 10\;  {\cC^{2\tD}_{a_j \cup a_k \cup a_l}}  \oplus 5 \; {\cC^{2\tD}_{{a_1 \cup a_2 \cup a_3} + {a_1 \cup a_2 \cup a_4}+ {a_1 \cup a_3 \cup a_4} + {a_2 \cup a_3 \cup a_4}}}. 
\end{split}
\eea
In the first line,  we get the each sector in the right hand side from the each holonomy $(\oint_z a_{1},\oint_z a_{2},\oint_z a_{3},\oint_z a_{4})\in \mathbb{Z}_2^4$ around the compactifying $z$-direction. 
We decompose the 16 sectors into a multiplet with multiplicities (1,4,6,4,1),
where the first 1 selects $\vec a^T=(a_{1,z},a_{2,z},a_{3,z},a_{4,z})=(0,0,0,0)$;
The second 4 selects only one element of $\vec a$ as 1 as nontrivial, given by the combinatory ${4 \choose 1}=4$;
The third 16 selects only two elements out of $\vec a$ as 1 as nontrivial, given by the combinatory ${4 \choose 2}=6$;
Similarly, the fourth ${4 \choose 3}=4$ and the fifth ${4 \choose 4}=1$ are selected. 
All these indices $j,k,l,l',m$ given above are dummy but fixed and distinct indices, selected  from
the set $\{1,2,3,4\}$.
In the second line of \eqn{eq:A4-decompose-cochain},
it turns out that we can do a $M \in$ SL$(4,\Z_2)$ transformation in the dimensional reduced sector, among the $\vec a ^T= (a_1,a_2,a_3,a_4)$, 
to redefine the fields through $M \vec a = \vec a' \to \vec a$. 
The second sector and third sector turn out to be the same, via a $M= M_{2 \leftrightarrow 3}$.
The fourth sector and fifth sector turn out to be the same, via a $M= M_{4 \leftrightarrow 5}$.
$$M_{2 \leftrightarrow 3}=\left(
\begin{array}{cccc}
 1 & 1 & 0 & 0 \\
 0 & 1 & 0 & 0 \\
 0 & 0 & 1 & 0 \\
 0 & 0 & 0 & 1 \\
\end{array}
\right), \;\;\;\;
M_{4 \leftrightarrow 5}=
\left(
\begin{array}{cccc}
 1 & 1 & 1 & 1 \\
 0 & 1 & 0 & 0 \\
 0 & 0 & 1 & 0 \\
 0 & 0 & 0 & 1 \\
\end{array}
\right) \in \mathrm{SL}(4,\Z_2).
$$
For example, we see $M_{23}$ can change ${(a_1+a_2)  \cup a_3 \cup a_4}$ to ${a_1 \cup a_3 \cup a_4}$, thus
we can combine the 6 of third sectors into the 4 of second sectors. 
Overall, similar forms of $M= M_{2 \leftrightarrow 3}$ does the job to identify these 10 sectors as
10 equivalent copies of a TQFT written as $10\;  {\cC^{2\tD}_{a_j \cup a_k \cup a_l}}$.
Similarly, we can use the 
similar forms of $M= M_{4 \leftrightarrow 5}$ to
identify the last fourth and fifth sectors, obtaining  5 copies of a TQFT, written as $5\;  {\cC^{2\tD}_{{a_1 \cup a_2 \cup a_3} + {a_1 \cup a_2 \cup a_4}+ {a_1 \cup a_3 \cup a_4} + {a_2 \cup a_3 \cup a_4}}}$.

In terms of continuum gauge field theory, we can rewrite \eqn{eq:A4-decompose-cochain} as 
\bea \label{eq:A4-decompose}
\begin{split}
&{\cC^{3\tD}_{\frac{2}{2\pi} \underset{i=1}{\overset{4}{\sum}}\int B_i dA_i +\frac{1}{\pi^3} A_1A_2A_3 A_4}} 
={\cC^{2\tD}_{\frac{2}{2\pi}\underset{i=1}{\overset{4}{\sum}}\int B_i dA_i}}\\
&\;\;\;\;\oplus
10
{\cC^{2\tD}_{\frac{2}{2\pi} \underset{i=1}{\overset{4}{\sum}}\int B_i dA_i + \frac{1}{\pi^2} A_1 A_2 A_3} }
\oplus
 5 {\cC^{2\tD}_{\frac{2}{2\pi} \underset{i=1}{\overset{4}{\sum}}\int B_i dA_i +
\frac{1}{\pi^2}  (A_1 A_2 A_3 +  A_1 A_2 A_4 + A_1 A_3 A_4  +   A_2 A_3 A_4)  }}.
\end{split}
\eea
In terms of TQFT dimensional decomposition, \eqn{eq:A4-decompose-cochain}/\eqn{eq:A4-decompose} 
is the information we can obtain based on the field theory actions.
What else topological data can we obtain to check the decompositions in \eqn{eq:A4-decompose}? We can consider:
\begin{enumerate}

\item GSD data on ${T^3}$ shows that  
\bea
\begin{split} \label{eq:A4-GSDdecompose-T3}
&\GSD_{T^3,\; 3+1\text{D-TQFT}}= \sum_{b} \GSD_{T^{2},\; 2+1\text{D-TQFT}'(b)}\\
&\Rightarrow {\GSD_{{T^3},\frac{2}{2\pi} \underset{i=1}{\overset{4}{\sum}}\int B_i dA_i +\frac{1}{\pi^3} A_1A_2A_3 A_4}}
= 1576 = 256 + 15 \times (2^2 \times 22).
\end{split}
\eea
The GSD data only distinguish the $b=0$ trivial sector ${\cC^{2\tD}_{\frac{2}{2\pi}\underset{i=1}{\overset{4}{\sum}}\int B_i dA_i}}$ with GSD${}_{T^2}$=256
from the remaining 15 sectors. Each of the remaining 15 sectors has GSD  $=88$,
which is the same as the tensor product ($\Z_2$ gauge theory) $\otimes$ (a {non-Abelian $D_4$ gauge theory}) with trivial DW cocycle in 2+1D \cite{Wang1404.7854}.
Therefore, $\GSD$ cannot distinguish the  second and the last sector  of the decomposition \eqn{eq:A4-decompose}.

\item GSD data on ${\RP^3}$ shows that 
\bea \label{eq:A4-GSDdecompose-RP3}
\begin{split}
 {\GSD_{{\RP^3},\frac{2}{2\pi} \underset{i=1}{\overset{4}{\sum}}\int B_i dA_i +\frac{1}{\pi^3} A_1A_2A_3 A_4}}
= 11& = 1 \times 1+ 4 \times {3 \over 4} + 6 \times {3 \over 4} + 4 \times {1 \over 2} + 1 \times {1 \over 2}.\\
&=1 \times 1 +10 \times {3 \over 4} +5  \times {1 \over 2}.
\end{split}
\eea
In this case, we first compute $Z(\RP^3 \times S^1)=11$ for ${\frac{2}{2\pi} \underset{i=1}{\overset{4}{\sum}}\int B_i dA_i +\frac{1}{\pi^3} A_1A_2A_3 A_4}$.
This data matches with the dimensional reduced 16 sectors of 2+1D TQFTs in terms of their $Z(\RP^3)$ that we also compute. 
In terms of 2+1D TQFTs grouping in \eqn{eq:A4-decompose-cochain} as a multiplet (1,4,6,4,1), the
first sector contributes $Z(\RP^3)=1$, each of the second (4) and third (6) contributes $Z(\RP^3)={3 \over 4}$,
and each of the fourth (4) and fifth (1) contributes $Z(\RP^3)={1 \over 2}$.
\item We can also adopt additional data such as the modular $\cT$ matrix of SL$(2,\Z)$ representation, measuring the topological spin or the self-statistics of anyonic particle/string excitations, in \cite{Wang1404.7854}. The diagonal $\cT$ matrix contains only four distinct eigenvalues, $(1,-1,i,-i)$. We can specify a $\cT$ matrix  by a tuple of numbers containing these eigenvalues,
as $(N_1,N_{-1},N_{i},N_{-i})$.
We find
\bea \label{eq:A4-T-decompose}
\begin{split}
 {\cT^{3D}_{\frac{2}{2\pi} \underset{i=1}{\overset{4}{\sum}}\int B_i dA_i +\frac{1}{\pi^3} A_1A_2A_3 A_4}}  & \text{in terms of }
 (N_1,N_{-1},N_{i},N_{-i})=(836, 580, 80, 80)\\
 &= 1 \times (136, 120, 0, 0) + 10 \times (48, 32, 4, 4) + 5 \times (44, 28, 8, 8).
\end{split}
\eea
The 10 sectors of $\cT^{2D}$ with $(N_1,N_{-1},N_{i},N_{-i})=(48, 32, 4, 4)$ are
 again the same as $\cT^{2D}$ of the 
 ($\Z_2$ gauge theory) $\otimes$ (a {non-Abelian $D_4$ gauge theory}) in 2+1D.
The overall structure of $\cT^{3D}$ decomposition
 agrees with $Z(\RP^3 \times S^1)$ decomposition.
\end{enumerate}
In summary, \eqn{eq:A4-decompose} suggest that there are at most three distinct classes among the 16 sectors of dimensional reduced 2+1D TQFTs, and the distinction among the three is guaranteed by the data of $Z(\RP^3)$ and $\cT$ matrices.

For untwisted gauge theories, one can derive:
\bea
&&\cC^{3\tD}_{\text{3+1D-${D_4}$ gauge}} = 2\cC^{2\tD}_{\text{2+1D-${D_4}$ gauge}} \oplus 2\cC^{2\tD}_{\text{2+1D-$(Z_2)^2$ gauge}} 
\oplus \cC^{2\tD}_{\text{2+1D-$Z_4$ gauge}},\\ 
&&\cC^{3\tD}_{\text{3+1D-${Q_8}$ gauge}} = 2\cC^{2\tD}_{\text{2+1D-${Q_8}$ gauge}} \oplus 3 \cC^{2\tD}_{\text{2+1D-${Z_4}$ gauge}}.
\eea
Each conjugacy class of holonomy round the compactifying circle gives the lower dimensional theory with the maximal subgroup commuting with the holonomy as its subgroup.
These results can be checked by
 the information of \cite{Wang1404.7854, Moradi:2014cfa} and our earlier section's GSD data.

\subsection{Fermionic Dimensional Reduction Scheme}

\label{Sec:Dim-Reduce-F}

Based on Sec.~\ref{Sec:Categorification}'s strategy, we examine the dimensional decomposition of some of the spin TQFTs listed in Sec.~\ref{Sec:fTQFT-GSD}. 
We obtain these spin TQFTs from gauging some global symmetries of fermionic Symmetry-Protected Topological states (fSPTs).

\subsubsection{2+1D $\rightarrow$ 1+1D gauged fSPT reduction: 2+1D $\Z_2^2\times \Z_2^f$ fSPT and its gauged spin-TQFT}
Consider a 2+1D fSPT state with $\Z_2^2\times \Z_2^f$ symmetry and  its partition function
\begin{equation}
 Z_\text{2+1D fSPT}=e^{\frac{\pi}{2} i \int a_1\cup a_2\cup \tilde{\eta}} 
\end{equation} 
where the precise definition of the action is spelled out in point 2) in the beginning of Sec. \ref{Sec:fTQFT-GSD}. 
When implementing the theory on a $M^3=M^2\times S^1$, depending on the spin structure and $\Z_2$ holonomies along $S^1$, 
it reduces to a 1+1D fSPT with $\Z_2^2\times \Z_2^f$ symmetry of one of the 3 following types:

\begin{enumerate}
 \item Trivial. 
\begin{equation}
 Z_\text{1+1D fSPT}=1.
\end{equation} 
Gauged theory has GSD$^\text{I}=4b$ (given by the partition function of $M^2=S^1\times S^1$) independently on the spin structure on $S^1$. 

\item 
\begin{equation}
 Z_\text{1+1D fSPT}=e^{\pi i \int (\sum_{i,j=1}^2\epsilon_{ij}\alpha_i\,a_j)\cup \eta}. 
\end{equation} 
Where $\alpha_i\in \Z_2$ are some parameters not all simultaneously zero, $a_i\in H^1(M^2,\Z_2)$ describe background $\Z_2$ gauge fields and $\epsilon_{ij}$ is the standard anti-symmetric tensor. The formal expression for the action has the following definition:
\begin{equation}
 \int_{M^2} a\cup \eta \equiv \eta[\text{PD}(a)] \equiv  
 \left\{ 
 \begin{array}{cc}
 1, & \text{odd spin structure on PD}[a], \\
 0, & \text{even spin structure on PD}[a].
 \end{array}
 \right.
\end{equation}
where the spin-structure on 1-manifold $\text{PD}[a]$ is induced from the spin structure on $M^2$ in on obvious way\footnote{Equivalently, 
\begin{equation}
\eta[\text{PD}(a)]=q([\text{PD}(a)])/2 
\end{equation}
where $q:H_1(M^2)\rightarrow \Z_4$ is the quadratic enhancement of the intersection form. 
} (cf. beginning of Sec. \ref{Sec:fTQFT-GSD}). Note that 
\begin{equation}
\int_{M^2} (a_1+a_2)\cup \eta = \int_{M^2} a_1\cup \eta+\int_{M^2} a_2\cup \eta+\int_{M^2} a_1\cup a_2.
\end{equation}

 The gauged theory has GSD$^\text{II}_\text{P}=2\times (1f+0b)$ for the odd spin structure on $S^1$ and GSD$^\text{II}_\text{A}=2\times (0f+1b)$ for the even spin structure on $S^1$. 

\item 
\begin{equation}
 Z_\text{1+1D fSPT}=e^{\pi i \int a_1\cup a_2 +(\sum_{i,j=1}^2\epsilon_{ij}\alpha_i\,a_j)\cup \eta}. 
\end{equation} 
Gauged theory, for any $\alpha_i$, has GSD$^\text{III}=1b$ independently on the spin structure on $S^1$. 
\end{enumerate}

Namely, for the even spin strucutre and trivial $\Z_2$ holonomies along $S^1$,  the 2+1D fSPT reduces to trivial (type I) theory on 
$M^2$. For the odd spin strucutre and non-trivial $\Z_2$ holonomies, it reduces to a theory of type II with $\alpha_i=\int_{S^1} a_i$. For the odd spin strucutre and trivial $\Z_2$ holonomies along $S^1$, or the even spin strucure and non-trivial $\Z_2$ holonomies, it reduces to a theory of type III with $\alpha_i=\int_{S^1} a_i$. 

Consider now the gauged 2+1D fSPT on $M^3=T^3$. Let us order the circles in $T^3$ such that the first one is the time circle and the last one is the $S^1$ on which we are doing reduction. The GSD decomposition then reads as follows:
\begin{equation}
 \begin{array}{c} \text{GSD}_{\text{PP}}=\text{GSD}_{\text{P}}^\text{III}+3\,\text{GSD}_{\text{P}}^\text{II}=1b+3\times 2\times (1f+0b)=6f+1b, \\
\text{GSD}_{\text{AP}}=\text{GSD}_{\text{A}}^\text{III}+3\,\text{GSD}_{\text{A}}^\text{II}=1b+3\times 2\times (0f+1b)=7b, \\
\text{GSD}_{\text{PA}}=\text{GSD}_{\text{P}}^\text{I}+3\,\text{GSD}_{\text{P}}^\text{III}=4b+3\times 1b=7b,\\
\text{GSD}_{\text{AA}}=\text{GSD}_{\text{A}}^\text{I}+3\,\text{GSD}_{\text{A}}^\text{III}=4b+3\times 1b=7b.
 \end{array}
 \label{eq:3d-sTQFT-decomp}
\end{equation} 
The decompositions of GSDs can be promoted to the decomposition of the spin-TQFT functor. For odd spin structure on $S^1$:
\begin{multline}
\CC^\text{2D}_{\frac{\pi}{2}\int a_1\cup a_2\cup \tilde\eta}=
\CC^\text{1D}_{\pi\int a_1\cup a_2}\oplus
\CC^\text{1D}_{\pi\int a_1\cup \eta}\oplus
\CC^\text{1D}_{\pi\int a_2\cup \eta}\oplus
\CC^\text{1D}_{\pi\int (a_1+a_2)\cup \eta}=\\
\CC^\text{1D}_{\pi\int a_1\cup a_2}\oplus
3\CC^\text{1D}_{\pi\int a_1\cup \eta}
,
\end{multline}
 For even spin structure on $S^1$:
\begin{multline}
\CC^\text{2D}_{\frac{\pi}{2}\int a_1\cup a_2\cup \tilde\eta}=
\CC^\text{1D}_{0}\oplus
\CC^\text{1D}_{\pi\int a_1\cup a_2+a_1\cup \eta}\oplus
\CC^\text{1D}_{\pi\int a_1\cup a_2+a_2\cup \eta}\oplus
\CC^\text{1D}_{\pi\int a_1\cup a_2+(a_1+a_2)\cup \eta}=\\
\CC^\text{1D}_{0}\oplus
3\CC^\text{1D}_{\pi\int a_1\cup a_2+a_1\cup \eta}
,
\end{multline}
where we used field redefinitions to combine equivalent theories together.
Note that all the summands except the first one give isomorphic Hilbert spaces on $S^1$. This is the reason for the factors of 3 in  (\ref{eq:3d-sTQFT-decomp}).

\subsubsection{3+1D $\rightarrow$ 2+1D gauged fSPT reduction: 3+1D $\Z_4^2\times \Z_2^f$ fSPT and its gauged spin-TQFT}
Consider a 3+1D fSPT state with $\Z_4^2\times \Z_2^f$ symmetry and its partition function
\begin{equation}
 Z_\text{3+1D fSPT}=e^{\pi\int (a_1\mod 2)\cup (a_2 \mod 2) \cup\text{Arf}}
\end{equation} 
where the precise definition of the action is spelled out in point 4) in the beginning of Sec. \ref{Sec:fTQFT-GSD}. 
When putting on $M^4=M^3\times S^1$, depending on the spin structure and $\Z_2$ holonomies along $S^1$, it reduces to a 2+1D fSPT with $\Z_4^2\times \Z_2^f$ symmetry\footnote{The corresponding classifying cobordism group is $\Omega^3_{\text{Spin}}(B(\Z_4^2))\cong \Z_8^2\times \Z_4\times \Z_2^3$. Only the generators of $\Z_2$ subgroups will appear in the decomposition below.} of one of the 3 following types:

\begin{enumerate}
 \item Trivial. 
\begin{equation}
 Z_\text{2+1D fSPT}=1.
\end{equation} 
Gauged theory has GSD$^\text{I}=256b$ independently on spin structure on $T^2$. 

\item 
\begin{equation}
 Z_\text{2+1D fSPT}=e^{\pi i \int (\sum_{i,j}^2\epsilon_{ij}\alpha_i\,a_j \mod 2) \cup \text{Arf}}. 
\end{equation} 
Where $\alpha_i\in \Z_2$ are some parameters not all simultaneously zero and $a_i \mod 2\in H^1(M^3,\Z_2)$ describes background $\Z_2$ gauge fields. Gauged theory has GSD$^\text{II}_\text{PP}=16\times 12f$ for the odd spin structure on $T^2$ and $\text{GSD}^\text{II}_\text{PA}=\text{GSD}^\text{II}_\text{AP}=\text{GSD}^\text{II}_\text{AA}=16\times 12b$ for an even spin structure on $T^2$. 

\item 
\begin{equation}
  Z_\text{2+1D fSPT}=e^{\pi i \int (\sum_{i,j}^2\epsilon_{ij}\alpha_i\,a_j \mod 2) \cup \text{Arf}+(a_1\mod 2)\cup (a_2\mod 2)\cup \eta} 
\end{equation} 
where $\alpha_i\in \Z_2$ are not all simultaneously zero. Gauged theory, for any $\alpha_i$, has GSD$^\text{III}=16\times 9b$ independently on spin structure on $T^2$. 
Note that from the definition \eqn{eq:ABK-def}, we derive
\begin{equation}
\int_{M^3} (b_1+b_2)\cup \text{Arf} = \int_{M^3} b_1\cup \text{Arf}+\int_{M^3} b_2\cup \text{Arf}+\int_{M^3} b_1\cup b_2 \cup \eta.
\end{equation}

\item 
\begin{equation}
Z_\text{2+1D fSPT}=e^{\pi i \int (a_1\mod 2)\cup (a_2\mod 2)\cup \eta}. 
\end{equation} 
Gauged theory has GSD$^\text{IV}_\text{PP}=16\times (6f+1b)$ for the odd spin structure on $T^2$ and $\text{GSD}^\text{IV}_\text{PA}=\text{GSD}^\text{IV}_\text{AP}=\text{GSD}^\text{IV}_\text{AA}=16\times 7b$ for an even spin structure on $T^2$. 

\end{enumerate}

As for 2+1D $\rightarrow$ 1+1D reduction, for the even spin strucutre on $S^1$ and trivial mod 2 holonomies along $S^1$, the 3+1D fSPT reduces to trivial (type I) theory on $M^3$. For the odd spin strucutre and non-trivial mod 2 holonomies it reduces to a theory of type II with $\alpha_i=\int_{S^1} a_i \mod 2$. For the odd spin strucutre and trivial mod 2 holonomies along $S^1$ it reduces to the type IV theory. For the even spin strucure and non-trivial mod 2 holonomies, it reduces to a theory of type III with $\alpha_i=\int_{S^1} a_i\mod 2$. 

Consider now 3+1D fSPTs on $M^4=T^4$. Let us again order the circles in $T^4$ such that the first one is the time circle and the last one is the $S^1$ on which we are doing reduction. The GSD decomposition than reads as follows:
\begin{equation}
 \begin{array}{c}
  \text{GSD}_{\text{(odd)P}}=4\text{GSD}_{\text{(odd)}}^\text{IV}+12\,\text{GSD}_{\text{(odd)}}^\text{II}=4\times 16\times (6f+1b)+12\times 16\times 12f=64\times (42f+1b), \\
  \text{GSD}_{\text{(even)P}}=4\text{GSD}_{\text{(even)}}^\text{IV}+12\,\text{GSD}_{\text{(even)}}^\text{II}=4\times 16\times 7b+12\times 16\times 12b=64\times 43b, \\
  \text{GSD}_{\text{(odd)A}}=4\text{GSD}_{\text{(odd)}}^\text{I}+12\,\text{GSD}_{\text{(odd)}}^\text{III}=4\times 16\times 16b+12\times 16\times 9b=64\times 43b, \\
\text{GSD}_{\text{(even)A}}=4\text{GSD}_{\text{(even)}}^\text{I}+12\,\text{GSD}_{\text{(even)}}^\text{III}=4\times 16\times 16b+12\times 16\times 9b=64\times 43b. \\
 \end{array}
 \label{eq:4d-sTQFT-decomp}
\end{equation} 
where (odd) denotes even PP spin structure on $T^2$ and (even) denotes any of the even spin structures, PA, AP or AA, on $T^2$.

The decompositions of GSDs can be promoted to the decomposition of the spin-TQFT functor. For odd spin structure on $S^1$:
\begin{equation}
\CC^\text{3D}_{\pi\int (a_1\mod 2)\cup (a_2 \mod 2) \cup\text{Arf}}
=
4\CC^\text{2D}_{\pi\int (a_1 \mod 2)\cup (a_1 \mod 2) \cup \eta}\oplus
12\CC^\text{2D}_{\pi\int (a_1\mod 2)\cup\text{Arf}}
\end{equation}
For even spin structure on $S^1$:
\begin{equation}
\CC^\text{3D}_{\pi\int (a_1\mod 2)\cup (a_2 \mod 2) \cup\text{Arf}}
=
4\CC^\text{2D}_{0}\oplus
12\CC^\text{2D}_{\pi\int (a_1\mod 2)\cup\text{Arf}+(a_2\mod 2)\cup\text{Arf}}
\end{equation}
We used field redefinitions to identify equivalent theories.
Note that all the summands except the first one give isomorphic Hilbert spaces on $T^2$.  This is the reason for the factors of 7 in  (\ref{eq:4d-sTQFT-decomp}).

This dimension decomposition method can be applied to all examples given in Table~\ref{Table:fTQFT}.

\section{Long-Range Entangled Bulk/Boundary Coupled TQFTs}
\label{Sec:LRE-bulk-brdy}

Now we consider bulk/boundary coupled TQFT system.
In the work of Ref.~\cite{1705.06728WWW}, 
for a given bulk $d+1$ dimensional $G$-symmetry protected phase characterized by a Dijkgraaf-Witten (DW) cocycle 
$\omega^{d+1}\in \cH^{d+1}(BG,{U(1)})$, a $d$D boundary $K$-gauge theory coupled with the $d+1$D bulk SPT is constructed via a so-called
the \emph{group extension} or \emph{symmetry extension} scheme.
The groups $G$ and $K$ form an exact sequence
$$ 
	1\to K \to H \stackrel{r}{\to} G \to 1,
$$
such that $r^* \omega^{d+1} =1 \in \cH^{d+1}(BH,{U}(1))$ where $r^*$ is the pullback of the homomorphism $r$.
The $r$ is a surjective group homomorphism.
The $H$ is a total group associated to the boundary. To help the readers to remember the group structure 
assignment to the bulk/boundary, we can abbreviate the above group extension as, 
\begin{equation}
\label{eq:exactsequence}
1\to K_{\text{boundary}} \to H_{\text{boundary}} \stackrel{r}{\to} G_{\text{bulk}} \to 1.
\end{equation}
This structure is used throughout Sec.~\ref{Sec:LRE-bulk-brdy}.
In Appendix of \cite{1705.06728WWW}, 
some examples of GSDs are computed for both bulk SRE (ungauged) case and bulk LRE (dynamically gauged) case,
based on the explicit lattice spacetime path integral. Here we examine some examples exposed there, and will argue that when bulk is gauged, some of the boundary degrees of freedom ``\emph{dissolve}" into the bulk.\footnote{Hereby \emph{dissolve}, we mean that the boundary operators can move into the bulk, without any energetic penalty.} In other words,
we will show that after gauging the whole system, 
a \emph{certain} group-\emph{extension} construction in \eqn{eq:exactsequence} is actually equivalent (dual or indistinguishable) to
a group-\emph{breaking} construction also explained in Ref.~\cite{1705.06728WWW} associated to an inclusion $\iota$:
\bea
\label{eq:exactsequence-break}
G'_{\text{boundary}} \stackrel{\iota}{\to} G_{\text{bulk}}.
\eea
where $G'$ is a subgroup that $G$ breaks to, and the inclusion should satisfy
$\iota^* \omega^{d+1} =1 \in \cH^{d+1}(BG',{U}(1))$ where $\iota^*$ is its pullback.
The annotations below $G'$ and $G$ indicate site/link variables are valued in those groups in boundary and bulk respectively as was the case of \eqn{eq:exactsequence}.

\begin{figure}[t]
	\centering
		\includegraphics[width=0.5\linewidth]{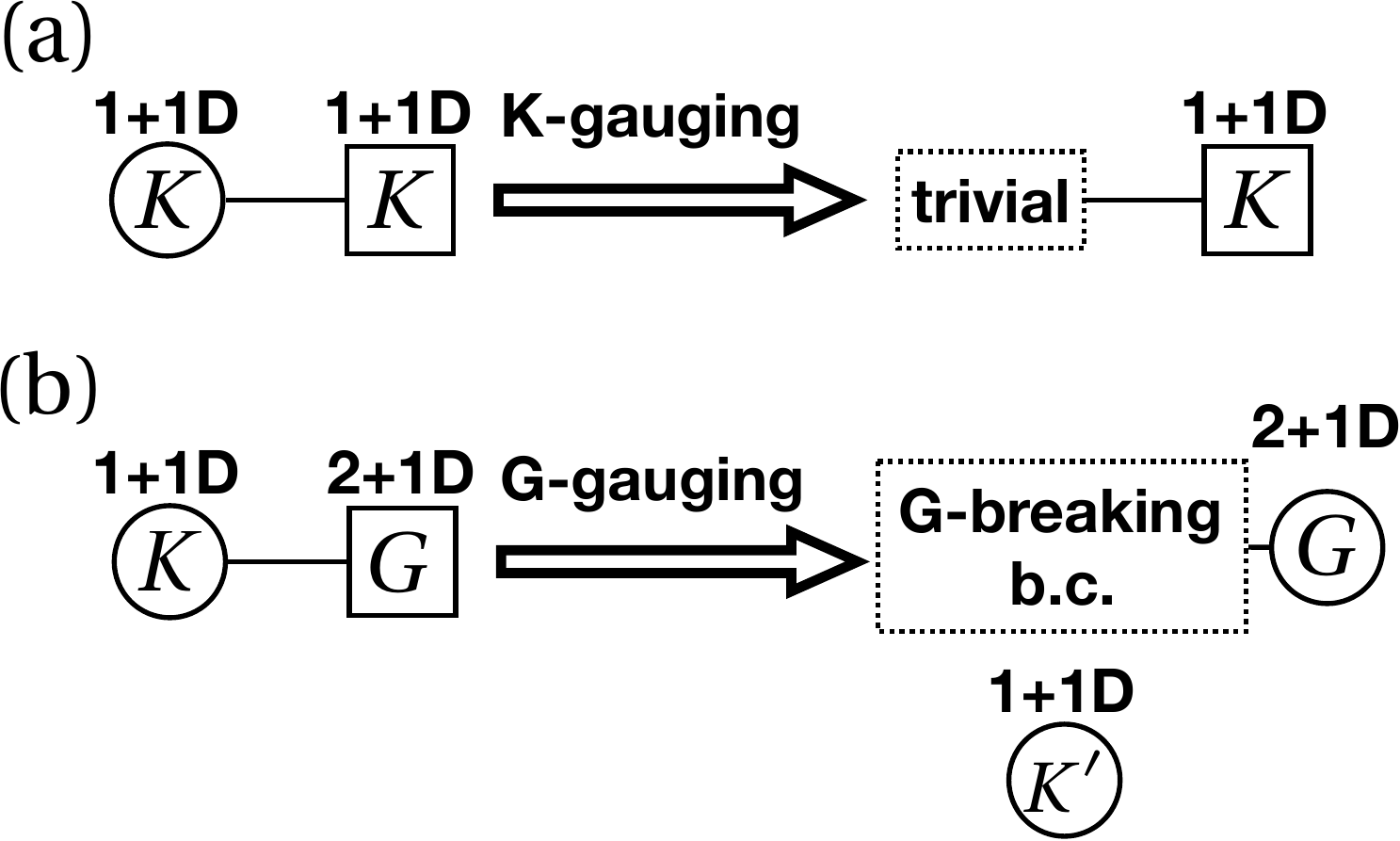}
		\caption{(a) The ``re-gauging'' of a 1+1D (or 2d) abelian finite $K$-gauge theory \cite{Vafa:1989ih, Bhardwaj:2017xp}. 
			In general, a group surrounded by a circle means a gauged symmetry, 
		and a group surrounded by a square means a global symmetry.  
		There is a global non-anomalous $K$ group acting on the theory faithfully, which is shown by the $K$ surrounded by a square. 
		When $K$ is gauged, the whole system becomes trivial, meaning the Hilbert space is 1-dimensional on any topology.	
		(b)  This case is what we focus on in Sec.~\ref{sec:2+1/1+1DZ2}. 
		We start from 1+1D $K$ gauge theory coupled with anomalous $G$ symmetry. 
		The anomalous symmetry is thought to be realized as an SPT phase in 2+1D. 
		After gauging the bulk $G$-symmetry, resulting system would be equivalent to a 2+1D $G$-theory with some boundary condition breaking $G$ into a subgroup $G'$
		and without coupling with a 1+1D system. 
		There can possibly be a decoupled $K'$ gauge theory on the boundary. In the examples, however, we may neglect $K'$ as absence.}
	\label{fig:gaugeeaten} 
\end{figure}

A heuristic reasoning is the following.
This is a generalization of the statement of \cite{Vafa:1989ih, Bhardwaj:2017xp} that a 1+1D gauge theory, 
with an abelian finite gauge group $K$ but without Dijkgraaf-Witten cocycle twist, 
has a global symmetry group isomorphic to $K$, and when the global $K$ is further gauged, the resulting theory is trivial \cite{Vafa:1989ih, Bhardwaj:2017xp}.
In the setup given by \eqn{eq:exactsequence}, the $K$ gauge theory on the boundary is coupled with anomalous $G$-symmetry. Gauging the bulk $G$ symmetry reduces the boundary degrees of freedom as in the pure 1+1D set up. When $K$ is small enough, the boundary degrees of freedom can even be completely gauged away, and no boundary degrees of freedom remain.
Before gauging, the bulk $G$ symmetry have the $G$-preserving boundary condition and coupled with the boundary degrees of freedom. However, when the boundary $K$-gauge theory is gauged away by gauging the bulk $G$, the whole bulk/boundary coupled system should be equivalent to just bulk $G$ symmetry 
with \emph{some boundary condition} without being coupled with 1+1D system, 
possibly accompanied by a decoupled 1+1D system on the boundary.
Namely, we stress the following: \\*
 \emph{There is an equivalence,
only  after gauging,
between 
``a certain bulk/boundary coupled system''
and  ``the bulk system with only some boundary conditions.''}\\*
For such a system to be consistent, the boundary condition should break $G$ into some non-anomalous subgroup $G'$. This discussion is summarized in Figure \ref{fig:gaugeeaten}.

In the rest of this section, we show how the above scenario occurs in more detail in two examples,
in Sec.~\ref{sec:2+1/1+1DZ2} and Sec.~\ref{sec:2+1/1+1DZ23}.
We will then compute partition functions on the $I^1 \times S^1$ topology\footnote{The $I^1$ is an interval.
The $I^1 \times T^{d-1} = (I^1_x \times  T^{d-2})\times  S^1_{\text{time}}$ 
can be regarded as
(an annulus or cylinder) $\times$ (a torus topology) in space, then $\times$ (a compact time).} 
in two ways: using the explicit 
lattice path-integral model coming from \cite{1705.06728WWW}'s \eqn{eq:exactsequence} 
and using the non-trivial boundary condition on $G$ with no boundary degrees of freedom.
Furthermore, we will try to generalize the discussion to 3+1D/2+1D system, in Sec.~\ref{sec:3+1/2+1DZ24}.
However, we will see that a certain exotic type of boundary condition of the bulk theory occurs after the bulk gauging. The complete understanding of the higher dimensional case remained for a future work.

We clarify that in the discussions of Sec.~\ref{Sec:LRE-bulk-brdy}, when we state 
``\emph{breaking}'' this means breaking the (gauge/global) symmetry with respect to the \emph{electric} sector (instead of the \emph{magnetic} sector), and when we state
``\emph{preserving}'' this means preserving the (gauge/global)  symmetry with respect to the electric sector (instead of the \emph{magnetic} sector), too.

We will also use the language of \cite{1705.06728WWW}  summarized in Table \ref{tab:system} throughout Sec.~\ref{Sec:LRE-bulk-brdy}.
\begin{table}[!h]
	\centering
	\begin{tabular}{l l l }
	\hline
System & \begin{minipage}[c]{1.8in}
$d+1$D Bulk/ $d$D Boundary\\ \;\;\; Entanglement property; \end{minipage} &  Group Extension Construction\\
	  \hline
		\hline
		 System (i) & SRE/SRE SPT/Symmetry&   $1 \to K_{\text{boundary}}^{\text{global sym}} \to H_{\text{boundary}}^{\text{global sym}} \stackrel{r}{\to} G_{\text{bulk}}^{\text{global sym}} \to 1$ \\
		\hline
		System (ii) & SRE/LRE  SPT/SET(TQFT)& 
		$1 \to K_{\text{boundary}}^{\text{gauge}} \to H_{\text{boundary}}^{\text{total}} \stackrel{r}{\to} G_{\text{bulk}}^{\text{global sym}} \to 1$ \\
		\hline
		System (iii) & LRE/LRE TQFT/TQFT&  $1 \to K_{\text{boundary}}^{\text{gauge}} \to H_{\text{boundary}}^{\text{gauge}} \stackrel{r}{\to} G_{\text{bulk}}^{\text{gauge}} \to 1$ \\
		\hline
	\end{tabular}
	\caption{We re-examine the $d+1$D Bulk/ $d$D Boundary coupled system based on a group \emph{extenssion} construction, developed in \cite{1705.06728WWW}, 
	in terms of more field theoretic understandings in Sec.~\ref{Sec:LRE-bulk-brdy}.
	The system we analyze the most is System (iii)'s LRE/LRE Bulk/Boundary TQFT.
	LRE/SRE stands for Long/Short Ranged Entangled. 
	SPT/SET stands Symmetry Protected/Enriched Topological states.
	We will especially comment about the gauging process (say, from System (i) to (ii), or (ii) to (iii)), 
	and especially focus on the issue still left open: 
	The \emph{boundary conditions} and some of their \emph{dualities} to \emph{breaking} construction \eqn{eq:exactsequence-break}, 
	after gauging $G_{\text{bulk}}$.
	}
	\label{tab:system}
\end{table}

\subsection{2+1/1+1D LRE/LRE TQFTs: 
Gauging an extension construction is dual to a gauge-breaking construction
}
\label{sec:2+1/1+1DZ2}

Let us start from the easiest case as a warm up, where the bulk is a 2+1D $G=\mathbb{Z}_2$ gauge theory. Namely, the bulk is the $\mathbb{Z}_2$ gauge theory of field $A$ (represented by a $\Z_2$-valued 1-cochain) 
with the unique non-trivial cocycle $(-1)^{\int A \cup A  \cup A}$.
\footnote{Most of the discussion in this subsection does not rely on that the bulk $\mathbb{Z}_2$ gauge field has a non-trivial DW action.
Here we assume a non-trivial DW action just because non-trivial DW terms will be important in the rest of the section.}
This cocycle can be cancelled by a boundary cochain when the boundary has a $K=\mathbb{Z}_2$ gauge theory, as shown in \cite{1705.06728WWW}.
In this case the sequence \eqn{eq:exactsequence} is
\begin{equation}
	1\to {\mathbb{Z}_2^K}_{\text{boundary}}\to {\mathbb{Z}_4^H}_{\text{boundary}} \to{\mathbb{Z}_2^G}_{\text{bulk}}\to 1.
	\label{eq:Z2Z4Z2}
\end{equation}
We use the upper indices $G$ to denote the group for the bulk, and the indices $K$ and $H$ to denote the groups for the boundaries following \eqn{eq:exactsequence} and \cite{1705.06728WWW}.
In \cite{1705.06728WWW}, the GSD on $D^2$ is computed to be $Z(D^2\times S^1) = 1$ when both bulk and boundary are dynamical. 
This hints that the boundary $\mathbb{Z}_2^K$ degrees of freedom are actually absent 
when the bulk $\mathbb{Z}_2^G$ is gauged. 
Below we aim to show that, \emph{only} after gauging $\mathbb{Z}_2^G$, this extension construction \eqn{eq:Z2Z4Z2} becomes equivalent to the breaking 
construction \eqn{eq:exactsequence-break} (also formulated in \cite{1705.06728WWW}) as
\begin{equation} 	\label{eq:0Z2}
	 1_{\text{boundary}}^{G'} \to{\mathbb{Z}_2^G}_{\text{bulk}}.
\end{equation}
We use the upper indices $G'$ to denote the preserved group for the boundary as \eqn{eq:exactsequence-break}.

In brief, it can be explained as follows. On boundary, 
there is a vortex operator $\phi(x)$ localized at a point $x$. For the boundary $\mathbb{Z}_2^K$ gauge theory to be coupled with the bulk $\mathbb{Z}_2^G$ symmetry, the operator should be shifted under the bulk $\mathbb{Z}_2$ transformation:
$
	\phi\to \phi + \lambda,
$
where $\lambda$ is a $\mathbb{Z}_2$ valued parameter of the bulk symmetry transformation. 
Then, we have an operator invariant under the bulk $\mathbb{Z}_2$ transformation 
\begin{equation}
	\exp(\mathrm{i} \pi(\int_{x_1}^{x_2}A - \phi(x_1) + \phi (x_2))),
	\label{eq:Z2line}
\end{equation}
where $A$ is the bulk $\mathbb{Z}_2^G$ field, $x_1$ and $x_2$ are the boundary points.
When the bulk $\mathbb{Z}_2^G$ is gauged, the boundary vortex operator $\phi$ is gauged out and therefore there is no longer a $\mathbb{Z}_2^K$ degeneracy on boundary,  and the bulk electric electric $\mathbb{Z}_2$ Wilson line can end on the boundary. 
Thus, the whole system is indistinguishable to just a  bulk $\mathbb{Z}_2^G$ gauge theory with boundary condition breaking the electric $\mathbb{Z}_2^G$, without any additional degrees of freedom on the boundary.

We give a more explicit explanation in the following.
Before gauging the  bulk $\mathbb{Z}_2^G$, the partition function of the full spacetime, with bulk ${M^3}$ of 2+1D $\mathbb{Z}_2^G$-SPTs 
and boundary $({\partial M})^2$ of 1+1D $\mathbb{Z}_2^K$-gauge theory, is
\bea \label{eq:Z2SPTZ2gaugeZ}
Z_A=(-1)^{\int_{M^3} A \cup A  \cup A} \sum_{{\alpha\in C^1( (\partial M)^2, \Z_2),}\atop{\phi \in C^0( (\partial M)^2, \Z_2)}}
 (-1)^{\int_{\partial M} \phi \delta \alpha +  \alpha \cup A + \phi A \cup A} ,
\eea
where $\phi$ and $\alpha$ are $\Z_2$-valued 0-cochain and 1-cochain fields respectively.
We denote all such $\Z_n$ valued $m$-cochain fields on the spacetime manifold $\CM$ in the cochain $C^m( \CM, \Z_n)$.
The $Z_A$ depends on the background  $\mathbb{Z}_2^G$ field $A$, and only $\phi$ and $\alpha$ are dynamical here.
Under their gauge transformations, 
$A \to A+\delta \lambda$, 
$\phi  \to \phi+ \lambda$ and $\alpha  \to \alpha+ \lambda  \delta \lambda$ with
$\lambda$ is an integral 0-cochain ($\lambda \in C^0(M, \Z_2)$), 
similar to the gauge-invariant calculation done in \cite{Kapustin1403.1467},
we find the $Z_A$ is gauge invariant.

After gauging the  bulk $\mathbb{Z}_2^G$, we propose the partition function of the full spacetime, 
with bulk ${M^3}$ of 2+1D $\mathbb{Z}_2^G$-gauge theory with boundary (of 1+1D $\mathbb{Z}_4^H$-gauge theory including also $\mathbb{Z}_2^K$ gauge sector), 
in continuum field description, is 
\begin{equation} \label{eq:Z2Z4Z2gaugeZ}
Z=\cN^{-1} \sum_A Z_A=  \int [DA][DB][D \phi] \; e^{ i (\frac{1}{2\pi}\int\limits_{M^3} (2 BdA+ AdA) +\frac{1}{2\pi} \int\limits_{\partial M} (2\phi  d B +2 B A + \phi  d A ))}.
\end{equation}
Here we use continuum field notations, where $A$ and $B$ are 1-form gauge fields, and $\phi$ becomes a 0-form scalar. The whole  partition function  $Z$ is gauge invariant,
under $A \to A+d \eta_A$, $B \to B+ d \eta_B$ and $\phi \to \phi -\eta_A$,
where $\eta_A$/$\eta_B$ are locally 0-forms.
The $\alpha$ as a 1-cochain field in \eqn{eq:Z2SPTZ2gaugeZ} is related to the B as the continuum 1-form field in \eqn{eq:Z2Z4Z2gaugeZ}.
We give several remarks in order to explain the gauging $\mathbb{Z}_2^G$ process:
\begin{enumerate}
\item \emph{SRE/SRE bulk/boundary}:
Starting from System (i) SRE/SRE bulk/boundary in Table \ref{tab:system}, as shown in \cite{1705.06728WWW},
this is a $\mathbb{Z}_2^G$-SPTs in a bulk, while it has a $\mathbb{Z}_4^H$-symmetry extended boundary. All global symmetries are preserved and unbroken.

\item 
 \emph{SRE/LRE bulk/boundary $\Rightarrow$ SRE/(SRE+LRO) bulk/boundary}:
After gauging the $\Z_2^K$ on the boundary, we arrive System (ii)'s SRE/LRE bulk/boundary in Table \ref{tab:system}, whose partition function is $Z_A$ in \eqn{eq:Z2SPTZ2gaugeZ}. 
In Sec.~3.3 of Ref.~\cite{1705.06728WWW}, it is found that the two holonomies of $\Z_2^K$ (or two ground states on a disk $D^2$ for this System (ii)) has different $\Z_2^G$-symmetry charge. 
The trivial holonomy of $\Z_2^K$ has a trivial (no or even) $\Z_2^G$ charge.
The non-trivial holonomy of $\Z_2^K$ has an odd $\Z_2^G$ charge. 
We find this fact can be understood as \eqn{eq:Z2SPTZ2gaugeZ}'s $Z_A$  has the $\Z_2^K$-holonomy $\int \alpha$ coupled to the 
$\Z_2^G$-background field in $(-1)^{\int_{\partial M} \alpha \cup A}$.\footnote{
When the bulk-$\Z_2^G$ is not gauged and therefore treated as an $\Z_2^G$-SPT state, 
the interpretation of the operator \eqn{eq:Z2line} is different. 
In that case, if the probe operator $\int A$ end on the boundary, it changes the boundary vacuum to a different state.}
Such a $\Z_2^K$-gauge theory turns out to develop 
$\Z_2^G$-\emph{spontaneous global symmetry breaking} (SSB) \emph{long-range order} (LRO) \cite{1705.06728WWW}.
Thus, it turns out that this SRE/LRE bulk/boundary by design turns into an SRE/SRE bulk/boundary,
because the $\Z_2^G$-SSB boundary has a gapped edge, which has LRO (but no Goldstone modes) but is SRE.

\item \emph{LRE/LRE bulk/boundary}:
After gauging the $\Z_2^G$ in the bulk (the boundary $\Z_2^K$ is also gauged), 
we arrive System (iii)'s LRE/LRE bulk/boundary in Table \ref{tab:system}, whose partition function we propose as $Z$ in \eqn{eq:Z2Z4Z2gaugeZ}. 
By massaging \eqn{eq:Z2Z4Z2gaugeZ}, we obtain\footnote{
Note that the 
$\phi$ is only defined on the boundary
${\partial M}$, but an arbitrary extension into the bulk give an unique action.
}
\bea \label{eq:Z2Z4Z2gaugeZ-break}
Z= \int [DA][DB][D \phi] \; e^{ i(\frac{2}{2\pi}\int\limits_{M^3}  (A+d\phi)dB+ \frac{1}{2\pi}\int\limits_{M^3}  (A+d\phi)d(A+d\phi))}.
\eea
From this expression of the partition function we can make several physical observations and predictions listed in  below.

(1). When we gauge bulk's $\Z_2^G$, both $A$, $\phi$ and $\alpha$ (cochain fields of \eqn{eq:Z2SPTZ2gaugeZ}) become dynamical.
This yields $A+ \delta \phi$ having no gauge transformation on the boundary, thus $B$ integration implies
\begin{equation}
	(A+\delta\phi)|_{\partial M}=0.
	\label{eq:bdry-cond-Z4Z2Z2}
\end{equation}
The dynamical vortex field $\phi$ (at the open ends of $A$) becomes \emph{deconfined} on the boundary.
This can be viewed as the $\Z_2$ electric charge particle (the $e$ anyon) becomes \emph{deconfined} and \emph{condensed}
on the boundary.
By anyon \emph{condensed} on the boundary, we mean that there can be nontrivial expectation value 
\bea 
\langle \exp(i \phi) \rangle  \neq 0,
\eea
for the ground state(s), since the $\phi$ are freely popped up and absorbed into the boundary.
Thus, gauging bulk's $\Z_2^G$ causes however the $\mathbb{Z}_2^G$ \emph{gauge symmetry broken on the boundary}.

(2) We can (and later will) also read the boundary condition directly from the cochain fields in \eqn{eq:Z2SPTZ2gaugeZ}. 
The $\Z_2^G$-SPT partition function indicates the following boundary condition after gauging $\Z_2^G$:
$$
{A  |_{\partial M}=0, \;\;\; A \cup A |_{\partial M}=0},
$$
in terms of cohomology. 
For example, integrating out $\alpha$ in \eqn{eq:Z2SPTZ2gaugeZ} 
forces $A$ to be exact on the boundary.
The first condition is equivalent to \eqn{eq:bdry-cond-Z4Z2Z2}, 
while the second condition automatically holds at the path integral after imposing the first condition.

(3).  After gauging $\mathbb{Z}_2^G$, we expect 
all boundary operators can be \emph{dissolved} into the bulk.
Which means the apparent boundary operator
$
	(-1)^{\int_{C} \alpha},
$
where $C$ is a 1-cycle in $\partial M$,
should be identified with the magnetic $\mathbb{Z}_2^G$ line operator in the bulk, since the electric $\mathbb{Z}_2^G$ is broken on the boundary as we saw.
Thus we can physically understand the conversion from ${\int \alpha}$ (1-cochain field) to ${\int B}$ (1-form magnetic $\Z_2$ field), from
\eqn{eq:Z2SPTZ2gaugeZ}'s $Z_A$ to \eqn{eq:Z2Z4Z2gaugeZ}'s $Z$, \emph{only after gauging} the $\mathbb{Z}_2^G$.
This agrees with the fact that the GSD is $Z(D^2\times S^1)=1$ in \cite{1705.06728WWW}.

(4). As an additional check, we consider GSD on $I^1\times S^1$.
From \cite{1705.06728WWW}, the lattice computation shows $Z(I^1 \times T^2)=2$. 
This is consistent with the alternative description of the symmetry breaking boundary condition, 
since the $\int A$ line can have a nontrivial $\mathbb{Z}_2$ value along the interval $I^1$, 
while cannot have nontrivial holonomy along the spatial $S^1$. 
(This coincides with the gauge symmetry breaking boundary condition explored in \cite{1212.4863.WangWen}, e.g. Table II).
See also Table \ref{tab:GSDZ2}
\end{enumerate}

\begin{table}[!h]
	\centering
	\begin{tabular}{|c|c|c|c|}
	\hline
	\multicolumn{4}{c}{LRE/LRE 2+1D bulk/1+1D boundary coupled TQFTs}  \\
		\hline
		\hline
		 $Z(M^2 \times S^1)$ & $Z(T^3)$ & $Z(D^2 \times S^1)$  & $Z(I^1\times T^2)$\\
		\hline
		spatial topology & $T^2$ & $D^2$ & $I^1\times S^1$\\
		\hline
		GSD & 4 & 1  & 2\\
		\hline
	\end{tabular}
	\caption{
	The GSDs of the LRE/LRE bulk/boundary TQFT theory of dynamically gauged $(-1)^{\int A \cup A  \cup A}$ SPT.
	The boundary theory is constructed via \eqn{eq:Z2Z4Z2}  in \cite{1705.06728WWW}.
	Our Sec.~\ref{sec:2+1/1+1DZ2} can explain the GSD data in terms of field theoretic description.}
	\label{tab:GSDZ2}
\end{table}

In summary, the LRE/LRE bulk-boundary coupled TQFT systems given by the exact sequence \eqn{eq:Z2Z4Z2}, 
\emph{only} when $\mathbb{Z}_2^G$ is gauged, 
is equivalent to just $\mathbb{Z}_2^G$ breaking boundary condition, namely, the \emph{double semion condensation} (see \cite{1212.4863.WangWen}), when both bulk and boundary groups are gauged. The gauge symmetry breaking condition is given in \cite{1705.06728WWW} \eqn{eq:0Z2} as
$
1\to\mathbb{Z}_2^G. 
$
In 2+1D/1+1D LRE/LRE bulk/boundary, the underlying physics of \eqn{eq:Z2Z4Z2} and \eqn{eq:0Z2}  coincides with the \emph{double semion condensation}. For example, we can write the bulk gauge theory as a twisted $\Z_2$ gauge theory
or $\Z_2$ double-semion topological order,
$
{\int \frac{2}{2\pi}
 B dA} + \frac{1}{2\pi}A dA,
$
then the \emph{double semion condensation} can be achieved by $\left. A \right\vert_{\partial M}=0$ boundary condition in \eqn{eq:bdry-cond-Z4Z2Z2}. 
We can gap the boundary by turning on the cosine sine-Gordon term 
\bea
g \int dt dx \cos(2 \phi)
\eea 
at a strong coupling $g$, where the scalar field $\phi(x,t)$ is the same vortex operator mentioned above.
Under SL$(2,\Z)$ field redefinition, we can rewrite the bulk theory as
$
{\int \frac{1}{4\pi}
 \Big(\begin{smallmatrix} 2 & 0 \\
 0& -2 \end{smallmatrix}} \Big)_{IJ}\, A_I' dA_J',
$
then we can gap the boundary by the cosine term of vortex field $ \phi'_I$ of $A_I'$, \cite{1212.4863.WangWen}
\bea
g \int  dt dx \cos(2 ( \phi'_1 + \phi'_2))
\eea
at a strong coupling $g$.
See the comparison of the physical setup of 
 \emph{double semion condensation}, or precisely the condensation 
of semion $s$ and anti-semion $\bar{s}$ in \cite{1212.4863.WangWen}.

\subsection{2+1/1+1D LRE/LRE TQFTs: 
Gauging an extension construction is dual to a partially gauge-breaking construction
}

\label{sec:2+1/1+1DZ23}

We would like to generalize the argument in the previous subsection into a more nontrivial case, 
namely 2+1/1+1D coupled system associated to the following exact sequence:
\begin{equation}
     1\to (\mathbb{Z}_2)^K_{\text{boundary}} \to (D_4\times \mathbb{Z}_2)^H_{\text{boundary}} \to (\mathbb{Z}_2^3)^G_{\text{bulk}} \to1,
     \label{eq:D4exact}
\end{equation}
where the leftmost $(\mathbb{Z}_2)^K$ goes into the order-8 non-Abelian dihedral group $D_4$, and the $\mathbb{Z}_2$ in the middle just become one of the factors of 
$(\mathbb{Z}^3_2)^G$.
Again the upper index $G$ denotes the group for the bulk, and the indices $K$ and $H$ denote the groups for the boundaries following \eqn{eq:exactsequence} and \cite{1705.06728WWW}.
The bulk  $\mathbb{Z}^3_2$ gauge fields $A_1,A_2,A_3$ have the exponentiated action
\begin{equation}
	(-1)^{\int_M A_1\cup A_2 \cup A_3}.
	\label{eq:Z23action}
\end{equation}
To cancel the anomaly induced by the bulk, the boundary dynamical $\mathbb{Z}_2$ cochain fields $\alpha$ and $\phi$ have the coupling 
\begin{equation}
	Z_A=(-1)^{\int_M A_1\cup A_2 \cup A_3} 
 \sum_{{\alpha\in C^1( (\partial M)^2, \Z_2),}\atop{\phi \in C^0( (\partial M)^2, \Z_2)}}
	(-1)^{\int_{{\partial}M} ( \phi \delta \alpha + \alpha\cup A_1 + \phi A_2\cup A_3)}.
	\label{D4action}
\end{equation}
This indicates the boundary conditions
\begin{equation}
	A_1=0, \quad A_2\cup A_3=0,
	\label{eq:boundary1}
\end{equation}
for the LRE/LRE system.

Let us compute the partition function on $I^1\times T^2$ 
with the boundary condition \eqn{eq:boundary1}, 
and compare it with the result from the method of the Appendix of \cite{1705.06728WWW}.
To be precise, the boundary condition ``$A_1=0$" means that the field $A_1$ is an element of the relative cohomology $H^1(M,\partial M; \mathbb{Z}_2)$, while keeping $A_{2,3}$ to be inside $H^1(M;\mathbb{Z}_2)$.
The second condition of \eqn{eq:boundary1} will be imposed by the path-integral, as we will see.
Then, the partition function is
\begin{equation}
	Z(I^1\times T^2) =  \cN^{-1} \sum_{A} Z_A=  \cN^{-1} \sum_{(A_1,A_2,A_3)} (-1)^{\int_{I\times T^2} A_1\cup A_2 \cup A_3},
	\label{eq:PartitionIT2}
\end{equation}
where $(A_1,A_2,A_3)$ runs through $H^1(M,\partial M;\mathbb{Z}_2)\oplus H^1(M;\mathbb{Z}_2)^{\oplus 2}$ with $M=I\times T^2$ as said, 
and $\cN$ is a normalization constant which is to be determined. 
Note that $A_1\cup A_2 \cup A_3$ defines an element of $H^3(M,\partial M)$ so that it can be integrated over the fundamental class $[M] \in H_3(M,\partial M)$.
In this expression, only the first condition of \eqn{eq:boundary1} is imposed by hand, 
while the summation over $A_1$ with the sign acts as a projection (times an integer) imposing the second condition.

Let us first compute the partition function \eqn{eq:PartitionIT2} up to the normalization constant $\cN$.
It is convenient to pack the holonomy data $(A_1,A_2,A_3)$ into a $3\times3$ matrix
\begin{equation}
	H_{ij}= \int_\text{$i$-th direction} A_j \in \mathbb{Z}_2,
\end{equation}
where the first direction is $I^1$ and the second and third directions are $S^1$'s in $T^2$ of $M=I^1 \times T^2$.
The partition function up to the normalization constant can be computed by
\begin{equation}
	\cN Z(I^1\times T^2)  = \sum_{H}(-1)^{\mathrm{det}H} = 20,
	\label{eq:Hsum}
\end{equation}
where the summation is constrained by 
the conditions 
$A_1\in H^1(M,\partial M;\mathbb{Z}_2)$ and 
$A_2,A_3 \in H^1(M;\mathbb{Z}_2)^{\oplus 2}$, which mean $H_{21}=H_{31}=H_{12}=H_{13}=0$.
In the sum of \eqn{eq:Hsum}, contribution from the configurations which do not satisfy the second equation of \eqn{eq:boundary1} on boundary is automatically canceled.
For example, take a configuration given by
\begin{equation}
	H
	=
	\begin{pmatrix}
		H_{11} & 0 & 0 \\	
		0 & 1 & 0\\
		0 & 0 & 1
	\end{pmatrix}.
\end{equation}
With this configuration,  $\int_{T^2} A_2\cup A_3 =1 $ which does not satisfy the second equation of \eqn{eq:boundary1}.
However, summation over $H_{11}$ yeilds
\begin{equation}
	\sum_{H_{11}=0,1}(-1)^{\mathrm{det}\,\mathrm{diag}(H_{11},1,1)}=0.
\end{equation}
In this way, the summation over $H_{11}$ in \eqn{eq:Hsum} is essentially just projecting out the configurations which does not satisfy the second equation of \eqn{eq:boundary1}, and therefore the partition function $Z$ counts the number of configurations satisfying \eqn{eq:boundary1} up to some constant.

The normalization constant $\cN$ should be the number of residual gauge transformations 
which fixes the cohomology classes $(A_1,A_2,A_3)$.
Such gauge transformations are given by elements of $H^0(M,\partial M;\mathbb{Z}_2)\oplus H^0(M;\mathbb{Z}_2)^{\oplus 2} =  \mathbb{Z}_2^{\oplus 2}$, therefore the normalization constant $\cN$ is $4$.
Then, the partition function can be computed to be 
\begin{equation}
	Z(I^1\times T^2) = 5.
\end{equation}
Our independent computation matches exactly with the lattice model computation \cite{1705.06728WWW} 
based on the exact sequence \eqn{eq:D4exact}, see Table~\ref{tab:GSDZ23}.

\begin{table}[!h]
	\centering
	\begin{tabular}{|c|c|c|c|}
		\hline
			\multicolumn{4}{c}{LRE/LRE 2+1D bulk/1+1D boundary coupled TQFTs}  \\
                 \hline
                 \hline
		 $Z(M^2 \times S^1)$ & $Z(T^3)$ & $Z(D^2 \times S^1)$  & $Z(I^1\times T^2)$\\
		\hline
		spatial topology & $T^2$ & $D^2$ & $I^1\times T^1$\\
		\hline
		GSD & 22 & 1  & 5\\
		\hline
	\end{tabular}
	\caption{
	The GSDs of the LRE TQFT theory of dynamically gauged \eqn{eq:Z23action}.
	The boundary theory is constructed via \eqn{eq:D4exact}.
	Our Sec.~\ref{sec:2+1/1+1DZ23} can explain the GSD data in terms of field theoretic description.}
	\label{tab:GSDZ23}
\end{table}

\begin{figure}[!h]
	\centering
	\includegraphics[width=.7\linewidth]{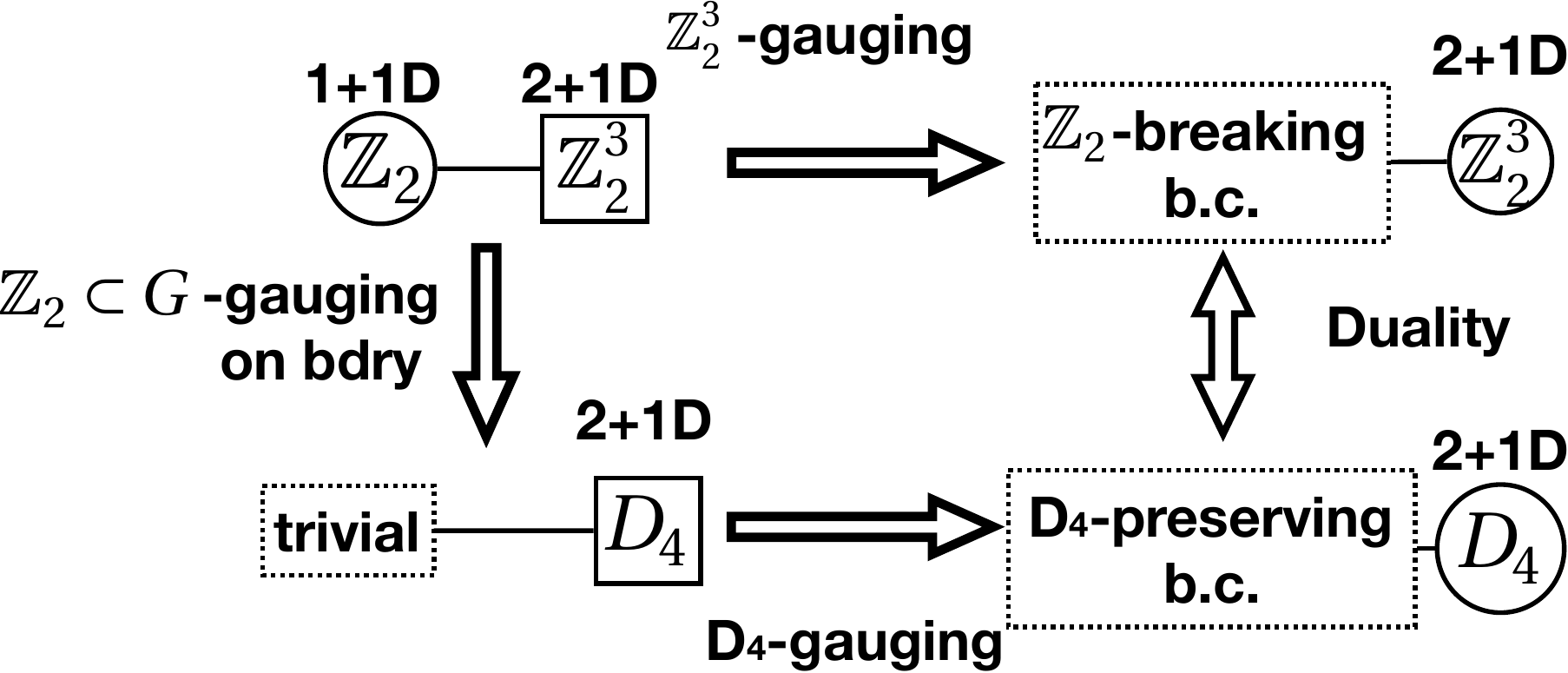}
	\caption{The interpretation of the boundary condition \eqn{eq:boundary1}. The downward arrow is due to \cite{Bhardwaj:2017xp}. The boundary condition \eqn{eq:boundary1} breaks one of $\mathbb{Z}_2\subset (\mathbb{Z}_2^3)^G$, and preserves (the electric parts of) the other two. On the other hand, we can do the same gauging in two steps. First we gauge one $\mathbb{Z}_2$ only on boundary, getting the trivial theory coupled with non-anomalous $D_4$, and then we gauge every symmetry realized in the system. In this way, we get $D_4$ gauge theory in the bulk with $D_4$ preserving boundary condition, which should be dual to the \emph{twisted} $(\mathbb{Z}_2^3)^G$ gauge theory in the bulk with
	$\Z_2$-breaking boundary condition.}
	\label{fig:Z23duality}
\end{figure}

The $\mathbb{Z}_2^3$ gauge theory with the action is known to be equivalent to the $D_4$ gauge theory (the order 8 dihedral group)\cite{deWildPropitius:1995cf}. 
The boundary condition \eqn{eq:boundary1} can be understood as the $D_4$ preserving boundary condition. Then, the partition function 
$Z(I^1\times T^2)=5$ exactly counts the number of $D_4$ holonomies up to conjugacy around the $S^1$ 
inside a time slice $I^1\times S^1$.
This observation is in line with a result in \cite{Bhardwaj:2017xp}.
In \cite{Bhardwaj:2017xp}, it was shown that when a $\mathbb{Z}_2$ subgroup of $\mathbb{Z}_2^3$ 
symmetry with anomaly \eqn{eq:Z23action} of a 1+1D theory is dynamically \emph{gauged}, the resulting theory has a non-anomalous $D_4$ symmetry.
We can divide the bulk $(\mathbb{Z}_2^3)^G$ gauging into two parts, as depicted in 
Fig.~\ref{fig:Z23duality}, one is gauging of the $\mathbb{Z}_2$ subgroup on boundary, and the other is gauging of the rest of the symmetry.
When the first $\mathbb{Z}_2\in G$ global symmetry is gauged on boundary, the $\mathbb{Z}_2^K$ gauge theory on the boundary is gauged away, and the resulting system is the trivial theory coupled with non-anomalous $D_4$ symmetry due to the result of \cite{Bhardwaj:2017xp}.
Then, gauging the rest of the global symmetry merely results in the bulk $D_4$ gauge theory with $D_4$ preserving boundary condition.
{This is consistent with the fact that the $D_4$ gauge theory in $2+1$D is dual to the twisted $\mathbb{Z}_2^3$ gauge theory\cite{deWildPropitius:1995cf}.}
See also Figure \ref{fig:Z23duality}.

In summary, the group extension construction \eqn{eq:D4exact} of coupled bulk/boundary (after gauging $G$) system
is equivalent/dual to the \emph{partially gauge breaking} construction as \eqn{eq:exactsequence-break} into 
\bea
(\mathbb{Z}_2^2)^{G'}_{\text{boundary}} \to (\mathbb{Z}_2^3)^G_{\text{bulk}}.
\eea

\subsection{3+1/2+1D LRE/LRE TQFTs: 
Comment on constructions of gauging an extension, and 1-form breaking v.s. ``fuzzy-composite'' breaking
}

\label{sec:3+1/2+1DZ24}

The system described by the action \eqn{D4action} can be generalized to $3+1$D $\mathbb{Z}_2^{4}$ gauge theory whose  exponentiated action is
\begin{equation}
	(-1)^{\int A_1\cup A_2\cup A_3\cup A_4}
	\label{eq:Z24action}
\end{equation}
where $(A_1,A_2,A_3,A_4) \in H^1(M;\mathbb{Z}_4)^{\oplus 4}$ are $\mathbb{Z}_2^4$ cocycle fields.
Correspondingly, we consider the following new exact sequence, following \eqn{eq:exactsequence}:
\begin{equation}
	1\to (\mathbb{Z}_2)^K_{\text{boundary}} \to (D_4\times \mathbb{Z}_2^2)^H_{\text{boundary}}\to (\mathbb{Z}_2^4)^G_{\text{bulk}}\to 1,
	\label{eq:exactD4Z22}
\end{equation}
with an order-8 non-Abelian dihedral group $D_4$.
Namely, the anomaly described by \eqn{eq:Z24action} can be cancelled by a $\mathbb{Z}_2^K$ gauge theory on a 2+1D boundary.
GSDs of this bulk-boundary coupled system when both bulk and boundary are gauged for some topologies, 
computed by the lattice model described in \cite{1705.06728WWW}, now we compute the new data and list it in Table \ref{tab:GSDZ24}.

\begin{table}[!h]
	\centering
	\begin{tabular}{|c|c|c|c|c|}
	\hline
     \multicolumn{5}{c}{LRE/LRE 3+1D bulk/2+1D boundary coupled TQFTs}  \\
		\hline
		\hline
		 $Z(M^3 \times S^1)$ & $Z(T^4)$ & $Z(D^3 \times S^1)$ & $Z(D^2\times T^2)$ & $Z(I^1\times T^3)$\\
		\hline
		spatial topology & $T^3$ & $D^3$ & $D^2\times S^1$ & $I^1\times T^2$\\
		\hline
		GSD & 1576 & 1 & 50 & 484\\
		\hline
	\end{tabular}
	\caption{The GSDs of the LRE TQFT theory of dynamically gauged 
	\eqn{eq:Z24action} on spatial topologies $M_{\text{time slice}}^3$, which are equal to partition functions 
	$Z(M^3\times S^1)$, computed by lattice path integral construction  \cite{1705.06728WWW}. 
	On each boundary of $M_{\text{time slice}}^3$, there is a $\mathbb{Z}_2^K$ gauge theory and coupled with the bulk through the exact sequence \eqn{eq:exactD4Z22}
	via $1\to \mathbb{Z}_2^K \to (D_4\times \mathbb{Z}_2^2)^H\to (\mathbb{Z}_2^4)^G\to 1$.
	Here both the bulk/boundary are LRE/LRE coupled TQFTs.
	Our Sec.~\ref{sec:3+1/2+1DZ24} can explain the GSD data in terms of field theoretic description. 
	}
	\label{tab:GSDZ24}
\end{table}

For the boundary $\mathbb{Z}_2^K$ gauge theory to cancel the anomaly \eqn{eq:Z24action},
there should be the following coupling:
\begin{equation}
	(-1)^{\int_{\partial M} (\alpha \cup A_1\cup A_2 + \beta \cup A_3 \cup A_4)},
	\label{eq:D4Z2Z2boundarycoupling}
\end{equation}
where $\alpha$ is the boundary $\mathbb{Z}_2$ gauge field, $\beta$ is its magnetic dual.
Integrating $\alpha$ and $\beta$ out, 
we get the following boundary conditions 
\begin{equation} 	\label{eq:boundaryZ24}
	A_1\cup A_2 =0, \quad A_3\cup A_4=0.
\end{equation}
The partition function on $M$ with this boundary condition would be counted by
\begin{equation}
	Z(M) = \cN(M) \sum_{(A_1,A_2,A_3,A_4)} (-1) ^{\int_M{ A_1\cup A_2\cup A_3\cup A_4}}
\end{equation}
where summation should be taken over some cohomology group which precisely realizes the boundary condition implicated by \eqn{eq:boundaryZ24},  and $\cN(M)$ is the normalization factor counting the residual gauge transformations.
Unfortunately, the precise cohomology group which actually realize the above condition is harder to determine. 
We don't have an exact answer for $\cN(M)$ yet.
Nonetheless, we can at least count the possible $\mathbb{Z}_2^4$ holonomies consistent with \eqn{eq:boundaryZ24}. 

For $M=I^1 \times T^2 \times S^1$, we can parametrize possible holonomies by a 3 by 4 matrix
\begin{equation}
	H_{ij}= \int_{S_i^1} A_j \in \mathbb{Z}_2,
\end{equation}
where $S_i^1$ is the $i$th 1-cycle of $T^3$.
We define 2 by 2 submatrices $H(a,b;c,d)$ of $H$ by taking $a$th and $b$th rows and $c$th and $d$th columns of $H$.
The boundary condition \eqn{eq:boundaryZ24} forces that 
\begin{equation}
	\mathrm{Det}(H(a,b;c,d))=0
	\label{eq:Hconst}
\end{equation}
for $(c,d)=(1,2)$ or $(c,d)=(3,4)$.
The number of matrices $H$ with elements $0$ or $1$ satisfying \eqn{eq:Hconst} is 484. 
Since we are yet to determine the physics affected by the $I^1$ direction and the normalization constant $\cN(M)$,  
this is not a 
complete computation.\footnote{With condition \eqn{eq:boundaryZ24} on the boundary, 
we do not expect that the action \eqn{eq:Z24action} can become $-1$.}
However, the fact that the number of possible holonomies around $T^3$ with the condition \eqn{eq:boundaryZ24} coincides with the GSD computed by the lattice computation, 
it suggest that the condition \eqn{eq:boundaryZ24} is physically sensible in some way.
Similarly, for the topology $M=D^2\times T^2$, we can count the holonomies around $T^2$ satisfying \eqn{eq:boundaryZ24} to be 100, which is different from $Z(D^2\times T^2)$ calculated by the lattice model by a factor of two, which is controlled by $\cN(M)$. 

\begin{figure}[t]
	\centering
	\includegraphics[width=.6\linewidth]{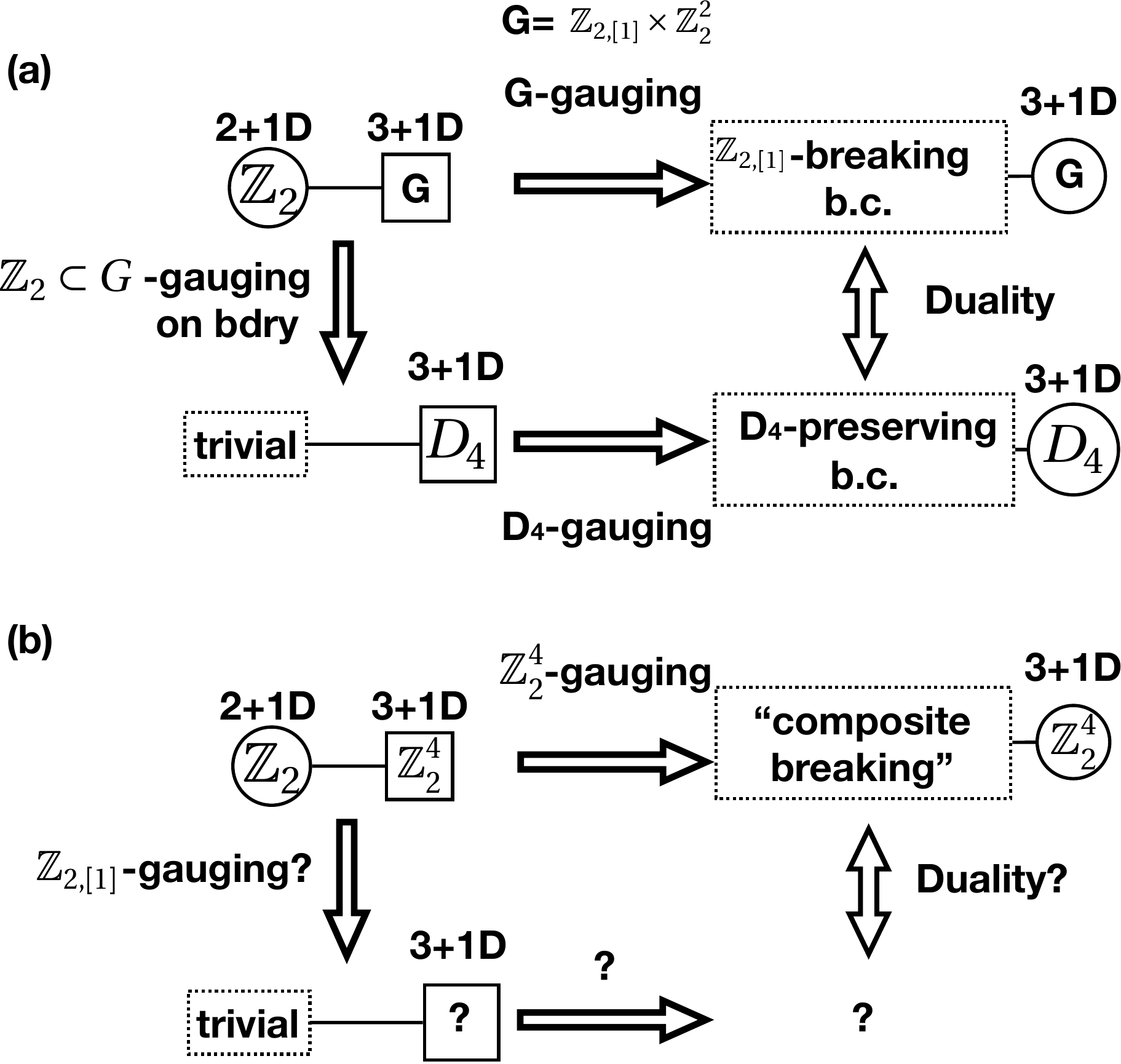}
	\caption{(a) A straightforward generalization of Figure \ref{fig:Z23duality}. In one higher dimension, one $\mathbb{Z}_2$ of $(\mathbb{Z}_2^3)^G$ in 2+1D/1+1D case is replaced by higher form $\mathbb{Z}_2$ symmetry here in 3+1D/2+1D, see \eqn{eq:1-form-Z2-Z23}. 
	(b) Currently we do not know precisely how to generalize from the previous Fig.~\ref{fig:Z23duality} associated to eqn. \ref{eq:D4exact} 
	to the present case associated to \eqn{eq:exactD4Z22}.
	 A possible relationship with the case of (a) is discussed in the main text. The downord arrow should be $\mathbb{Z}_{2,[1]}$-gauging on the boundary 
	in some sense to get a trivial theory out of the $\mathbb{Z}_2$ gauge theory in 2+1D, but the precise relation of $\mathbb{Z}_{2,[1]}$ to the $(\mathbb{Z}_2^4)^G$ is left open for
	future investigation. }
	\label{fig:Z24duality}
\end{figure}  

Mysteriously the situation here is \emph{not} an obvious generalization of what was studied in the previous section. Rather, a straightforward generalization of the situation \eqn{eq:D4exact} is 
that the bulk theory has a $\mathbb{Z}_{2,[1]}\times \mathbb{Z}_{2}^2$ symmetry, where $\mathbb{Z}_{2,[1]}$ is a 1-form $\mathbb{Z}_2$ symmetry, with an action
\begin{equation} \label{eq:2-form-anomaly}
	(-1)^{\int_{M^4} A^{(2)} \cup A_3\cup A_4},
\end{equation}
where $ A^{(2)}$ is the 2-form gauge field of $\mathbb{Z}_{2,[1]}$ symmetry and $a_{1,2}$ are 1-form gauge fields of $\mathbb{Z}_2^2$. See Figure \ref{fig:Z24duality}(a).
A boundary $\mathbb{Z}_2$ gauge field $\alpha$ can be coupled with the field $A^{(2)}$ through
\begin{equation}
	(-1)^{\int_{\partial M} \alpha \cup A^{(2)}+ \beta \cup A_3\cup A_4}.
	\label{eq:alphaA2}
\end{equation}
The recent paper \cite{Tachikawa:2017gyf} shows that when $\mathbb{Z}_{2,[1]}$ of $\mathbb{Z}_{2,[1]}\times (\mathbb{Z}_2^2)^G$ with an anomaly \eqn{eq:2-form-anomaly} is gauged in 2+1D, the resulting theory has a non-anomalous $D_4$ symmetry. Therefore the whole picture of Figure \ref{fig:Z23duality} can be lifted in this case, by lifting one $\mathbb{Z}_2$ symmetry into $\mathbb{Z}_{2,[1]}$. In particular, a breaking boundary condition
\begin{equation} \label{eq:1-form-Z2-Z23}
	\mathbb{Z}_2^2 \to \mathbb{Z}_{2,[1]}\times (\mathbb{Z}_2^2)^G
\end{equation}
should be dual to the $D_4$ preserving boundary condition.

What was studied in this subsection is more involved. 
Instead of \eqn{eq:alphaA2}, we have the coupling
 \eqn{eq:D4Z2Z2boundarycoupling}.
Still, we can observe a similarity between the \eqn{eq:D4Z2Z2boundarycoupling} and \eqn{eq:alphaA2}. Namely, $A_1\cup A_2$ plays the roll of $A^{(2)}$ in \eqn{eq:alphaA2}.
Thus, one might somehow find relations between these $\mathbb{Z}_2^2$ fields and a ``composite" $\mathbb{Z}_{2,[1]}$ field. One might regard the boundary condition \eqn{eq:boundaryZ24} relating to a boundary condition breaking this ``composite" $\mathbb{Z}_{2,[1]}$. See Figure \ref{fig:Z24duality}(b).

The boundary condition \eqn{eq:boundaryZ24} suggests that, on the boundary, some composite strings, composed of particles charged under the $(\mathbb{Z}_2^4)^G$ symmetries, are condensed, and thus this boundary condition might have some novel feature.
Since the dimensionality of the composite strings (1+1D) and their composed particles (0+1D) is different,
here we introduce the new concept of the ``\emph{fuzzy composite} object''
composed by one-lower-dimensional object.
We add \emph{fuzzy} to emphasize the dimensionality differences between the two objects.
The condensation of such ``\emph{fuzzy composite} objects'' on the boundary can be understood as the
 ``\emph{fuzzy composite-breaking}'' boundary condition.

Investigating this boundary condition in detail, in particular constructing a microscopic model (other than the lattice Hamiltonian and lattice path integral given in Ref.~\cite{1705.06728WWW})
to realize the physical mechanism
at the microscopic level on the lattice will be interesting.

\section{Conclusions}

\label{Sec:conclude}

Below we conclude with remarks on long-range entanglements and entanglement entropy,
the generalization of topological boundary conditions,
and the potential application to strongly-coupled gauge theories, and quantum cosmology.

\subsection{Remarks on Long-Range Entanglement and Entanglement Entropy with Topological Boundaries}

It is well known that the long range entanglement (LRE) can be partially captured by the topological entanglement entropy (TEE) \cite{0510092KitaevPreskill,0510613LevinWen}. 
In 2+1D, the topological entanglement entropy is the constant part of the entanglement entropy (EE), and one can extract the TEE by computing a linear combination of entanglement entropy as suggested\cite{0510092KitaevPreskill,0510613LevinWen}.  
For discrete gauge theories with gauge group $G$, when the entanglement cut does not wrap around the spatial cycle, the topological entanglement entropy is $-\log |G|$. 
For instance, the $(\mathbb{Z}_N)^{k}$ gauge theory has the TEE $=-k\log n$. 
Notice that the value of the TEE is independent of the twisting parameter (i.e. cocycle) of twisted gauge theories, hence one is not likely to distinguish different Dijkgraaf-Witten (DW) theories with the same gauge group using the TEE 
of the ground state wavefunctions on a closed manifold. 

In the following, we consider two generalizations to obtain a richer structure of the entanglement entropy. 

One generalization is to go to 3+1 dimensions and consider the Walker-Wang twisted theory. As was discussed in \cite{1108.4038GroverTurnerVishwanath} and \cite{1710.01747YZhengBernevig}, for discrete gauge theories of Walker-Wang type \cite{WalkerWang1104.2632} with gauge group $\mathbb{Z}_N$ 
and twisting parameter $p$ (namely, $\int BF +BB$ in Sec.~\ref{sec:BB-GSD}), the topological entanglement entropy across a torus (which does not wrap around the spatial cycle) is $-\log \gcd(N,2p)$, 
which depends on the twisting parameter. Hence the TEE can probe the twisting level $p$ of Walker-Wang model. This fact can be understood as follows: The genuine line and surface operators of the Walker-Wang model coincide 
with the line and surface operators of $\Z_{\gcd(n,2p)}$ gauge theory, see \cite{Gaiotto:2014kfa} and Sec.~\ref{sec:BB-GSD}. The TEE of the $\mathbb{Z}_N$ 
Walker-Wang model with twists is precisely the TEE of the $\mathbb{Z}_{\gcd(N,2p)}$ 
ordinary gauge theory without twists, i.e.  $-\log |G_{\text{eff}}|=-\log \gcd(N,2p)$.

\begin{figure}[h!]
	\centering
	\includegraphics[width=.9\textwidth]{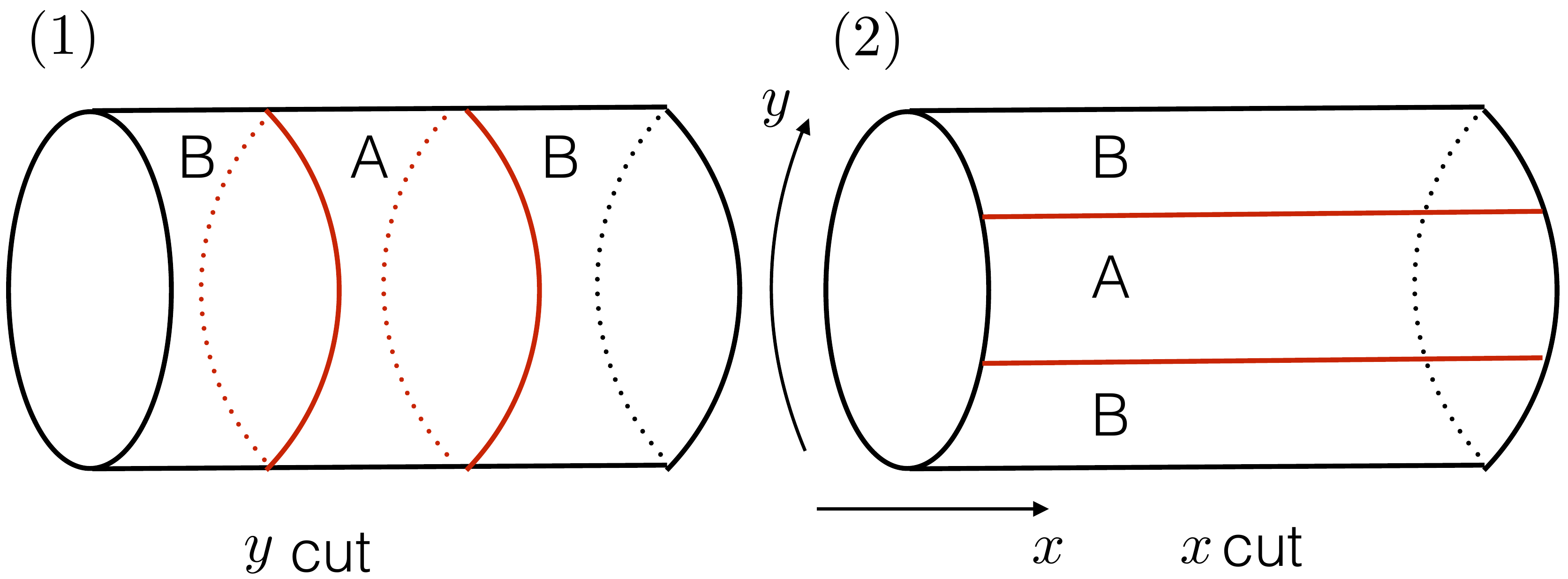}
	\caption{In the left panel, the entanglement cuts wrap around the $y$-cycle, which we denote the $y$-cut. 
	In the right panel, the entanglement cuts extend along $x$ direction, and end at the two boundaries, which we denote the $x$-cut.}
	\label{pic}
\end{figure}

Another generalization is to consider the entanglement entropy on a spatial manifold with boundaries where the entanglement cut wraps around spatial cycles. We consider the $\mathbb{Z}_2$ gauge theories (i.e., $\Z_2$-toric code model and $\Z_2$-double semion [a twisted $\mathbb{Z}_2$] model) on a cylinder geometry with two boundaries, 
as shown in Fig.~\ref{pic}. Let us first focus on the left panel and discuss the $\Z_2$-toric code model. {In the toric code model, there are four types of anyons $\{1, e, m, \epsilon\}$. Let us denote $|W_x, T_x\rangle$ to be the eigenstate of the line operators $W_x=e^{i\oint_{x}A_xdx}$ (i.e., $e$-line in the $x$ direction) and $T_x=e^{i\oint_x B_xdx}$ (i.e., $m$-line in the $x$ direction).  A generic ground state is a linear combination of $|W_x, T_x\rangle$, 
\begin{eqnarray}
|\psi\rangle=c_1 |1\rangle+c_e|e\rangle+c_m|m\rangle+c_\epsilon|\epsilon\rangle
\end{eqnarray}
where we label $|1\rangle\equiv |0, 0\rangle, |e\rangle=|1, 0\rangle, |m\rangle=|0, 1\rangle, |\epsilon\rangle=|1,1\rangle$, and the coefficients are properly normalized $|c_1|^2+|c_e|^2+|c_m|^2+|c_\epsilon|^2=1$. Following the computation in \cite{FradkinLeigh0802.3231}, we can derive that the entanglement entropy of $|\psi\rangle$ is a linear combination of the entanglement entropy of $|1\rangle, |e\rangle, |m\rangle, |\epsilon\rangle$,  
denoted as $S^{y}(|\psi\rangle)=-\sum_{i=1, e, m, \epsilon}|c_i|^2\log |c_i|^2+\sum_{i=1, e, m, \epsilon}|c_i|^2S^{y}(|i\rangle)$ where the super-script $y$ indicates the direction of the entanglement cut. 
In the following, we are only interested in the subleading (topological) part of the entanglement entropy, i.e., 
\begin{eqnarray}\label{EE}
S^{y}_{\mathrm{topo}}(|\psi\rangle)=-\sum_{i=1, e, m, \epsilon}|c_i|^2\log |c_i|^2+\sum_{i=1, e, m, \epsilon}|c_i|^2S^{y}_{\mathrm{topo}}(|i\rangle)
\end{eqnarray}
Because $e$ and $m$ are both self and mutual bosons, they can separately condense on the boundaries. Let us denote $a|b$ as $a$-condensation on the left boundary and $b$-condensation on the right boundary. There are 3 types of boundary conditions on a cylinder: the $e|e$, $m|m$ and $e|m$. When both boundaries are $e$-condensation, i.e, the $e|e$ boundary condition, there are two distinct sectors, with odd/even number of $e$-lines across the entanglement cut respectively. The generic ground state is 
\begin{eqnarray}
|\psi, e|e\rangle=c_1|1\rangle+c_e|e\rangle
\end{eqnarray}
where we have explicitly shown the boundary condition. 
Notice that there are no $|m\rangle$ and $|\epsilon\rangle$ in the expansion because the $m$-particle and $\epsilon$-particle cannot end on the boundary. Therefore the $m$-line and the $\epsilon$-line must cross the entanglement cut twice. For each sector, the even/odd-ness of the $e$-line crossing the entanglement cut is fixed, hence $S^y_{\mathrm{topo}}(|1\rangle)=S^y_{\mathrm{topo}}(|e\rangle)=-\log 2$. According to Eq.~\eqref{EE}, we have 
\begin{eqnarray}
\begin{split}
S^{y}_{\mathrm{topo}}(|\psi, e|e\rangle)&= -|c_1|^2\log |c_1|^2-|c_e|^2\log |c_e|^2-\log 2\\&=-|c_1|^2\log |c_1|^2-(1-|c_1|^2)\log (1-|c_1|^2)-\log 2
\end{split}
\end{eqnarray}
When $|c_1|^2=|c_e|^2=\frac{1}{2}$, the entanglement entropy $S^{y}_{\mathrm{topo}}(|\psi, e|e\rangle)$ is maximized, $S^y_{\mathrm{MaxES}, \mathrm{topo}}(|\psi, e|e\rangle)=0$. When $|c_1|^2=0, |c_e|^2=1$ or $|c_1|^2=1, |c_e|^2=0$, the entanglement entropy is minimized, $S^y_{\mathrm{MinES}, \mathrm{topo}}(|\psi, e|e\rangle)=-\log 2$. We can further consider other boundary conditions and the entanglement cuts along the $x$ direction, and the results are summarized in Table~\ref{TCEE}(a). Furthermore, we also consider the entanglement entropy of the double semion model as shown in Table~\ref{DSEE}(b). In the double semion model, there are four types of anyons $\{1, s, \bar{s}, b\}$, where the only nontrivial boson is $b$. Hence there is only one type of boundary condition, i.e., $b$ condensation on both boundaries, which we denote as $b|b$. From the data in Table~\ref{TCEE}(a) and \ref{DSEE}(b), we have the following observations:
	\begin{table}
		\centering
	(a) \begin{tabular}{| c | c | c | c |}
		\hline
		b.c & $e|e$ & $m|m$ & $e|m$ \\ \hline
		$S_{\mathrm{MaxES}}^y$ & 0 & 0 & $-\log 2$ \\ \hline
		$S_{\mathrm{MinES}}^y$ & $-\log 2$ & $-\log 2$ & $-\log 2$ \\ \hline
		$S_{\mathrm{MaxES}}^x$ & 0 & 0 & $0$ \\ \hline
		$S_{\mathrm{MinES}}^x$ & $-\log 2$ & $-\log 2$ & $0$ \\ 
		\hline
	\end{tabular}\;\;\;\;\;\;\;\;\;
	(b)
		\begin{tabular}{| c | c |}
			\hline
			b.c & $b|b$ \\ \hline
			$S_{\mathrm{MaxES}}^y$ & 0  \\ \hline
			$S_{\mathrm{MinES}}^y$ & $-\log 2$ \\ \hline
			$S_{\mathrm{MaxES}}^x$ & 0 \\ \hline
			$S_{\mathrm{MinES}}^x$ & $-\log 2$ \\ 
			\hline
		\end{tabular}
	\caption{(a) Maximal and minimal entanglement entropy of the $\Z_2$-toric code associated with various boundary conditions. (b) 
	{Maximal and minimal entanglement entropy of the $\Z_2$-double semion model associated with various boundary conditions.}}
	\label{TCEE} 		\label{DSEE}
	\end{table}
\begin{enumerate}
	\item The maximal and minimal entanglement entropy depend on the boundary condition. In particular,  when the types of boundary conditions are the same, $S_{\mathrm{MaxES}}-S_{\mathrm{MinES}}=\log 2$. However, when the types of boundary conditions are different, $S_{\mathrm{MaxES}}-S_{\mathrm{MinES}}=0$. 
	\item When the types of boundary conditions differ on two sides, the entanglement entropy is sensitive to whether the cut is in the $x$-direction or the $y$-direction. This enables us to use the entanglement entropy to probe the boundary conditions. 
\end{enumerate}
We can implement the above approach to other examples studied in Sec.~\ref{Sec:LRE-bulk-brdy}.
In particular, given a bulk LRE system, we can design various boundary conditions (by group \emph{extension} or by 0-form/higher-form \emph{breaking}) on different boundaries.
By generalizing the above analysis, we expect that EE is sensitive to not only the bulk but also the boundary/interface conditions.
We leave a systematic analysis of the interplay between other boundary/interface conditions and the long range entanglement for future work.

\subsection{More Remarks}

\begin{enumerate}

\item \emph{Counting extended} (\emph{line/surface}) \emph{operators}:
For 3+1D bosonic Abelian-$G$ TQFTs (with or without cocycle twists) we studied, 
the number of distinct types of \emph{pure} line operators (viewed as the worldline of particle excitations)
and 
the number of distinct types of \emph{pure} surface operators (viewed as the worldsheet of string/loop excitations)
are the same, equal to the number of group elements in $G$ (thus the order of group $|G|$).
By a \emph{pure} surface operator, we mean that the particular surface operator 
does not have additional lower dimensional line operators attached; and vice versa.\footnote{
The \emph{pure} operators here \cite{1602.05951,Lan1704.04221} are not equivalent to the \emph{genuine} operators
defined in \cite{Gaiotto:2014kfa}.  The \emph{genuine} operators in $d$-dimensions mean
that those operators do not require their attachment to higher dimensional objects.
}
In a spacetime picture, a \emph{pure} surface operator is associated to the worldsheet of \emph{pure} 
string/loop excitations (of a constant time slice) that does not have additional particles attached.
For 3+1D bosonic non-Abelian-$G$ TQFTs, with or without cocycle twists, 
we however expect 
that the number of distinct line/surface operators is related to the number of representation/conjugacy classes of $G$.\footnote{However, 
when the same TQFT has two (or more) descriptions of different gauge groups (so called the duality),
the counting of electric/magnetic operators can be different.
For example, in 2+1D, (1) the $(\Z_2)^3$-gauge theory with type-III cocycle $(-1)^{\int A_1 \cup A_2 \cup A_3}$ in Sec.~\ref{sec:A3-GSD} 
is equivalent to (2) the order-8 $D_4$ gauge theory \cite{deWildPropitius:1995cf, 1405.7689}.
In terms of (1), there are 8 group elements, thus which implying 
8 pure electric $e$-operator of $A$ (the trivial line operator $1$ included),
8 pure magnetic $m$-operator of $B$ (the trivial line operator $1$ included)
and some dyonic operators mixing $e$ and $m$.
In contrast, in (2), as a non-Abelian $D_4$, there are
5 pure electric $e$-operator (the trivial line operator $1$ included) related to representation of $D_4$,
5 pure magnetic $m$-operator of $B$ (the trivial line operator $1$ included)
related to conjugacy classes of $D_4$,
and some dyonic operators mixing $e$ and $m$.
Therefore, two dual descriptions of the same theory may give rise to different countings of $e$ and $m$ operators.
Nonetheless, the total number of distinct extended operators are the same: There are 22 distinct line operators in both cases \cite{deWildPropitius:1995cf, 1405.7689, He1608.05393}.
}
In both case, we find that the number of pure line and pure surface operators are equivalent.
One simple argument \cite{1602.05951,Lan1704.04221} is that the number of ground states on the spatial manifold $S^2 \times S^1$ in the Hilbert space must be
spanned by (1) the eigenstates obtained from inserting all possible pure line operators along $S^1$ into $D^3 \times S^1$ or 
(2) the eigenstates obtained from inserting all possible pure surface operators along $S^2$ into $S^2 \times D^2$, because TQFT assigns a state-vector 
in a Hilbert space to an open manifold,  in the spacetime path integral picture.
Here we use the fact that $\partial (D^3 \times S^1)= \partial (S^2 \times D^2)= S^2 \times S^1$ and the gluing along their boundary produces
$(D^3 \times S^1) \sqcup (S^2 \times D^2) =S^4$ \cite{1602.05951}.
It is obviously that the operators along $S^1$ must be 1-lines,
and the operators along $S^2$ must be 2-surfaces where the loop excitations created by this surface can be shrunk to nothing into the vacuum.
Thus, they correspond to pure line/surface operators creating pure particle/string excitations. 
Because the GSD on $S^2 \times S^1$ from two derivations is the same,
the numbers of linear-independent pure line operators equal that of pure surface operators.

For fermionic spin-TQFTs (with or without twists from gauging the cobordism topological terms), 
the line/surface operators in general will
have additional labels compared to the bosonic case. In particular,
for line operators, one can introduce a 0+1D fermionic SPT (labelled by
$\Z_2$) supported on the line operator. Note that, as often required
in QFT, the line operators should be equipped with framing
(trivialization of the normal bundle) in order to be well defined.
Then, the spin structure on the spacetime manifold induces a spin
structure on the support of the line operator. For a non-trivial
choice (i.e. $1\in \Z_2$) of the 0+1D SPT, the expectation value of the
line operator will be multiplied by $\pm 1$ for even/odd induced spin
structure. As in the bosonic case, the surface operators can be
understood in terms of drilling out a tabular neighborhood of the
operator and considering the theory on the resulting manifold with
boundary with some condition on the holonomies on the boundary. For
fermionic theories, one has to also choose a spin-structure on this
manifold with boundary. This choice corresponds to the extra label
assigned to the surface operator.\footnote{One can non-canonically
identify spin structures on the space-time manifold $M$ with elements
of $H^1(M,\Z_2)$. Then by considering the relevant part of the
Mayer-Vietoris sequence
\begin{equation}
\ldots \rightarrow H^1(M,\Z_2) \rightarrow  H^1(M\setminus
\Sigma,\Z_2)\oplus H^1(\Sigma,\Z_2) \rightarrow H^1(\Sigma\times
S^1,\Z_2) \cong H^1(\Sigma,\Z_2) \oplus \Z_2\rightarrow \ldots,
\end{equation}
one can see that (for connected $\Sigma$) that the set of
spin-structures on the compliment is given by (also non-canonically)
$\text{Spin}(M\setminus \Sigma) \cong H^1(M\setminus
\Sigma,\Z_2)=H^1(M,\Z_2)\oplus \Z_2$.}

\item  \emph{Fuzzy-composite object/breaking and  extended operators}:
Earlier in Introduction \eqn{eq:composite-bdry} and in Sec.~\ref{sec:3+1/2+1DZ24}'s \eqn{eq:boundaryZ24}, we introduce a new mechanism to obtain 
a peculiar gapped  topological boundary condition: $\left. {A_{i} \cup  A_{j}} \right\vert_{\Sigma^3}=0$ that could be viewed as the
condensation of a \emph{fuzzy-composite} string formed by two different particles (associated to the ends of two different line operators).
We term it as the condensation of the \emph{fuzzy-composite object} associated to the open ends of \emph{a set of extended operators}. More generally, 
we could anticipate that in higher spacetime dimensions, say ${M^{d+1}}$,
there could be other general topological boundary conditions on ${\Sigma^{d}}=\partial{M^{d+1}}$  
as
\bea
\left. {A_{i} \cup  A_{j}} \cup B_{k} \cup \dots  \right\vert_{\Sigma^{d}}=0,
\eea
in terms of the condensation of the composite object from a set of extended operators of \emph{different dimensionality} 
(1-form, 2-form fields, etc., or 1-cochain, 2-cochain fields, etc.). Its further detailed study is left for the future.

\item \emph{Boundary/Interface Deconfinement}: 
We had discussed in Sec.~\ref{Sec:LRE-bulk-brdy} (See also \cite{1705.06728WWW}),
gauging dynamically the bulk of SRE/LRE Bulk/Boundary coupled TQFTs to obtain the LRE/LRE Bulk/Boundary coupled TQFTs in Sec.~\ref{Sec:LRE-bulk-brdy}.
We notice that the former system has a SRE bulk (e.g. SPT state) thus naturally has only \emph{non-fractionalized} excitations in the bulk.\footnote{
Readers can find many other examples of surface topological orders on the boundary of SRE SPT states in this informative recent review \cite{Senthil1405.4015}
and References therein. The original idea is from \cite{SenthilBF}'s observation
of quantum disordering the symmetry defects to restoring the broken symmetry as
 topologically ordered boundary. Their approach is rather different from our constructions. 
\label{footnote:Senthil}}
The later system has a LRE bulk (e.g. topologically ordered state) thus can have also \emph{deconfined fractionalized} excitations even in the bulk.
However, we stress that the important ingredient, for both cases, 
is that the \emph{deconfined fractionalized} excitations happen on the boundary/interface, without much energy penalty. 
In Sec.~\ref{Sec:LRE-bulk-brdy}, we find that the \emph{deconfined fractionalized} excitations indeed condense on the boundary/interface.
On the lattice scale, the energy cost for having \emph{deconfined} excitations in the SRE bulk is $\Delta E \to \infty$ (i.e. impossible),
while that in the LRE bulk costs only $\Delta E \simeq \# J$ (some order of lattice coupling $J$, see Fig.~\ref{fig:vacua}).
But having  \emph{deconfined} excitations on the boundary/interface is $\Delta E \simeq 0$, if 
the ground state is obtained from extended operators ending on the boundary.

We note that there are some recent interests to study the deconfined domain walls 
\cite{1501.06773PoppitzSulejmanpasic, 1608.09011Unsel, 1706.05731KomargodskiUnsel}
(See also footnote \ref{footnote:Senthil}, and Reference therein \cite{Senthil1405.4015}), 
where the bulks of systems are however confined without fractionalized excitations 
but only the boundary harbors deconfined excitations. 
Our work potentially could help to understand such systems systematically and quantitatively.

\item
Besides the generic mixed (gauge/global) symmetry-breaking/extension construction of topological interfaces in
Ref.~\cite{1705.06728WWW}, 
there are many other recent work and applications on related issues.
For example, we can study the quantum code or topological quantum computation with boundary\cite{9811052BravyiKitaev, 1509.03626Yoshida,
1609.02037ICongChengWang, 1703.03564ICongChengWang, 1707.04564ICongChengWang,
1707.05490ICongChengWang, 1710.07197ICongWang}. One can construct
Hamiltonian models for gapped boundaries \cite{1706.00650HuWanWu, 1706.03329HuLuoPankovichWan, 1706.03611BullivantHuWan, 1706.09782HuWanWu}.
For LRE/LRE topological bulk/boundary coupled states,
there are applications to LRE fractional topological insulators and SETs  \cite{1311.0767.CW.Levin, 1511.01502Fidkowski, 1701.08828GYCho, 1706.00429GYCho}.
One can also consider entanglement entropy involving the topological interfaces, this is analyzed recently in 2+1D case\cite{1705.09611.Fliss.XDWen}.
There are other formal aspects of studying on boundaries and
surface defects in the categorical set-up (e.g. \cite{{1104.5047KitaevKong}, 1307.3632Fuchs, 1501.01885Fuchs} and References therein).

\item \emph{Tunneling between topological quantum vacua}: 
We had discussed our interpretations of 
tunneling between topological quantum vacua in Sec.~\ref{Sec:tunnel-1}, Sec.~\ref{Sec:tunnel-2}.
The tunneling rate $P$ is determined by the probability to create a pair
of excitations and anti-excitations out of the vacuum and then winding the
pair along a non-contractible spatial cycle. By dimensional analysis,
the tunneling rate $P$ is about $P \sim f [(E_T /\Delta), (a/L)]$. It is a function
$f$ proportional (up to some power) to the energy fluctuation $E_T$ (quantum or thermal) but
anti-proportional to the energy gap $\Delta=E_n-E_0$ between excited states and
ground states. It is also anti-proportional to the system size $L$ over
the lattice constant $a$ (or the Planck scale).
In TQFT, the energy gap $\Delta$ and system size $L/a$ is usually assumed
to send to the infinite limit. Thus, $P \sim 0$, it is not possible simply
based on the TQFT alone to obtain further detailed calculations of
the tunneling rate. In other words, 
the $P \sim 0$ also guarantees the robustness of fault-tolerant topological quantum computation \cite{Kitaev2003}:
The ground state (and data) is topologically robust against local perturbations, 
unless the artificial manual process dragging the excitations along a non-contractible spatial cycle.
However, it is possible to consider some lattice
models or TQFT coupling to other massive QFT with some physical
tunneling rate $P$.

The original motivation of our systems is inspired by long-range entangled condensed matter and strongly correlated electron systems with intrinsic topological orders. 
These systems are fully quantum and highly entangled.
In contrast, in a different discipline, most of the set-up and analysis in cosmology on vacuum-tunneling 
is somehow semi-classical, for example,
S.~Coleman's study on the fate of the false vacuum \cite{1997ColemanPRD}  to 
a more recent work\cite{1312.1772Turok}, mostly in a semiclassical theory, 
and references therein \cite{coleman1988aspects}.
We anticipate (or at least speculate)
the potential use of 
 topological quantum vacua tunneling,
 through the extended operators (for example, in terms of cosmic strings or higher dimensional analogs),
 in quantum cosmology.

\end{enumerate}


\section{Acknowledgements}

The authors thank Zhenghan Wang and Edward Witten for conversations.
KO and JW thank Yuji Tachikawa for helpful discussions, especially during the stay in IPMU, and also for conversations about his recent paper.
JW thanks Meng Cheng, Clay Cordova and Xueda Wen for informing their recent works, and thanks Ling-Yan Hung for the past conversations on entanglements. 
JW gratefully acknowledges the Corning Glass Works Foundation Fellowship and NSF Grant PHY-1314311.
Partial work of JW is performed in 2015 at Center for Mathematical Sciences and Applications at Harvard University.
KO gratefully acknowledges support from IAS and NSF Grant PHY-1314311.
PP gratefully acknowledges the support from Marvin L. Goldberger Fellowship and the DOE Grant DE-SC0009988. 
ZW gratefully acknowledges support from NSFC grants 11431010, 11571329.
MG thanks the support from U.S.-Israel Binational Science Foundation. 
YZ thanks H.~He and C.~von Keyserlingk for collaboration on a related project, and thanks B.~A.~Bernevig and Physics Department of Princeton University for support.
HL thanks H.~Wang and K.~Zeng for conversations.
This work is also supported by NSF Grant PHY-1306313, PHY-0937443, DMS-1308244, DMS-0804454, DMS-1159412 and Center for Mathematical Sciences and Applications at Harvard University.

\bibliographystyle{arXiv_new.bst}
\bibliography{DualStringGSD_ref,JCW-ref_12-2017} 


\end{document}